%Paper: hep-th/9412181
%From: FLOREAN@TRIESTE.INFN.IT
%Date: Tue, 20 Dec 1994 17:06:23 +0100 (WET)

\hfuzz=2pt
\def\t{T}
\def\l{L}
\def\q{{\hat q}}
\def\tr{{\rm tr}}

\def\thetacl{\theta_{\rm (cl)}{}}
\def\acl{A_{\rm (cl)}{}}
\def\fcl{F_{\rm (cl)}{}}
\def\mcl{M_{\rm (cl)}{}}
\def\xcl{X_{\rm (cl)}{}}
\def\ncl{N_{\rm (cl)}{}}

\def\btheta{\bar\theta}

\def\igamma{{\mit\Gamma}}

\font\titlefont=cmbx10 scaled\magstep1
\magnification=\magstep1
\null
\rightline{SISSA 204/94/EP}
\rightline{\tt hep-th/9412181}
\vskip 1.5cm
\centerline{\titlefont THE RENORMALIZATION GROUP FLOW}
\smallskip
\centerline{\titlefont OF THE DILATON POTENTIAL}
\vskip 1.5cm
\centerline{\bf R. Floreanini \footnote{$^*$}{\tt florean@ts.infn.it}}
\smallskip
\centerline{Istituto Nazionale di Fisica Nucleare, Sezione di
Trieste}
\centerline{Dipartimento di Fisica Teorica, Universit\`a di Trieste}
\centerline{Strada Costiera 11, 34014 Trieste, Italy}
\bigskip\smallskip
\centerline{\bf R. Percacci \footnote{$^{**}$}{\tt
percacci@tsmi19.sissa.it}}
\smallskip
\centerline{International School for Advanced Studies, Trieste,
Italy}
\centerline{via Beirut 4, 34014 Trieste, Italy}
\centerline{and}
\centerline{Istituto Nazionale di Fisica Nucleare,
Sezione di Trieste}
\vskip 1.8cm
\centerline{\bf Abstract}
\smallskip
\midinsert
\narrower\narrower\noindent
We consider a scalar-metric gauge theory of gravity with independent
metric, connection and dilaton. The role of the dilaton is to
provide the scale of all masses, via its vacuum expectation value.
In this theory, we study the renormalization group flow of the
dilaton potential, taking into account threshold effects at the
Planck scale. Due to the running of the VEV of the dilaton
all particles that would naively seem to have masses larger than
Planck's mass, may actually not propagate. This could solve the
problem of unitarity in these theories.
\endinsert
\bigskip
\vskip 1cm
\vfil\eject

\leftline{\bf 1. Introduction}
\medskip
\noindent
The general ideas that go under the name of ``effective field
theory'' are playing an increasingly important role in elementary particle
physics. The variety of physical phenomena is divided into
energy ranges, whose boundaries usually coincide with the masses of
various particles. In each range one has an effective field
theory. Very often the theories describing neighbouring energy
ranges are of the same type, the only difference being
that one particle which is
present in the higher energy range has been ``integrated out'' in
the lower energy range.
In other cases, however, the description of the physics in one
energy range is quite different from that in the
next energy range. It is clearly desirable to
relate all these descriptions, but this has not always been
possible so far.

It is widely believed that Einstein's General Relativity
is also an effective theory, valid for energies
lower than Planck's energy [1,2]. This does not mean
that it can only be treated classically; it only means that any quantum
calculation in Einstein's theory will have a natural cutoff at
the Planck scale. From this point of view,
the nonrenormalizability of Einstein's theory is not a problem.

Just below Planck's energy corrections due to higher derivative
terms are expected to appear.
The most general action with at most four derivatives is
$$
S(g_{\mu\nu})=
\int d^4x\,\sqrt{\det g}\Bigl[
\Lambda+{1\over16\pi G}R
+\alpha_1 R_{\mu\nu\rho\sigma}R^{\mu\nu\rho\sigma}
+\alpha_2 R_{\mu\nu}R^{\mu\nu}
+\alpha_3 R^2 \Bigr]\ ,\eqno(1.1)
$$
where $R_{\mu\nu}{}^\rho{}_\sigma$ denotes the curvature tensor
constructed with the metric $g_{\mu\nu}$.
As emphasized in [1] the gravitational action (1.1)
can be treated as a quantum field theory using the rules of
``chiral perturbation theory'', which were devised in the context
of a theory of mesons [3].
If regarded as ``fundamental'', the theory (1.1) is renormalizable
[4,5],
but another problem appears: the terms quadratic in curvature
contain higher derivatives of the metric and therefore
violate unitarity. Again, this is not a problem in the ``effective theory''
picture: the ghosts
have masses of the order of Planck's mass, and therefore will
never be excited if one remains at lower energies.
At energies exceeding the Planck energy, some ``new physics''
is expected to appear.

As mentioned above, there is no compelling reason to believe that the
theory describing the physics above the Planck energy will be of the same
type of the theory (1.1): it may be described by a completely different
set of degrees of freedom. Nevertheless,
in this paper we will conservatively assume that the ``new physics''
can still be described by a field theory. Motivated by the success
of the gauge principle in explaining all other known forces of nature
we consider a gauge theory of gravity, with independent
metric and connection. We will assume that the action is at most
quadratic in curvature and torsion.
These theories have a long history [6];
from a particle physicist's point of view their most attractive
feature is perhaps that they present a gravitational analogue of
the Higgs phenomenon [7]: the vierbein behaves like a Higgs field and
when it acquires a nonvanishing, constant, vacuum expectation value,
its kinetic term (torsion squared) becomes a mass term for the
connection. The mass is obviously of the order of Planck's mass,
so below Planck's energy the connection degrees
of freedom cannot be excited. Yet the connection itself does not
vanish: it becomes the Levi--Civita connection, whose components in a
coordinate frame are the Christoffel symbols. In this way the theory we
will consider gives rise naturally to the action (1.1) as an effective
theory below the Planck scale. The descriptions of physics above
and below Planck's energy are easy to relate in this case.
The occurrence of the Higgs phenomenon is often related to
unification, and indeed the theory can be generalized so as to
unify gravity with all other interactions (assuming that these
are already put together in a GUT). The vierbein then appears as the
order parameter that breaks the symmetry between gravity and the
rest [7]. This is a true unification,
in the sense in which the word is used in particle physics,
and does not require higher dimensional spacetimes.

The problem with these gauge theories of gravity is that there seems
to be no one of them which is at the same time unitary and
renormalizable. This is the same dilemma that one faces in (1.1):
without the higher derivative terms the theory is not renormalizable,
and with them it is not unitary.
However, this time the problem cannot be avoided because we are not aware
of any higher mass scale which might act as a cutoff and allow these
theories to be regarded as effective field theories.
Any theory that purports to describe physics above the
Planck scale had better be consistent on its own.

Our point of view is that we do not know enough on these theories
to draw any conclusion. The reason is, obviously, that they are very
complicated. In the general case we only have a tree level analysis
of their propagators [8]. A more rigorous study of the spectrum
would be desirable, but for the moment has been done only for Einstein's
theory [9]. It is quite possible that the quantum propagators
are very different from what they seem to be at the tree level.
For example, some degrees of freedom may be confined.
To the best of our knowledge, no one has performed
any quantum calculation so far.

Since the ghosts and/or tachyons would have masses of the order
of Planck's mass, the issue of unitarity could be resolved
in a rather drastic way if particles with such masses
were forbidden to propagate.
In this paper we elaborate on our earlier proposal for a mechanism
in which this could happen.
There are two main ingredients in this proposal:
the dilaton, and the renormalization group.
The dilaton is a scalar field, coupled to the metric, connection and to
other matter fields in such a way that all masses are equal to its
vacuum expectation value (VEV), multiplied by some
dimensionless coupling constant.
This field is closely associated to the (quantum mechanical) breaking
of Weyl invariance in a manner which has been discussed in detail in
[10,11] and is reviewed in Appendix A. In the present work,
we compute the renormalization group flow of the dilaton potential.
In particular, we are interested in the running of the VEV of
the dilaton, since this gives the dominant contribution to
the running of the masses. (Dimensionless coupling constants are
expected to run only logarithmically.)
The tentative conclusion of our analysis will be that the position of
the pole of a particle with mass comparable or larger than Planck's mass
can be shifted to exponentially large energies.
The sole remnant at lower energy would be the graviton.
(In the generalized gauge theory of gravity alluded to above,
one would also have an unbroken GUT sector).

This paper will be organized as follows: in Section 2 we will
describe in detail the gauge theory of gravity that
we are going to consider. In Section 3 we discuss the
linearization of the theory around flat space and define the
effective potential for the dilaton.
In section 4 we define the average effective potential.
In section 5 we write the renormalization group equations
for the average
effective potential and study their solutions.
In Section 6 we draw our conclusions.

\bigskip
\goodbreak
\leftline{\bf 2. Lagrangians}
\smallskip
\noindent
We consider a theory of gravity with
independent metric $g_{\mu\nu}$, connection
$\igamma_\lambda{}^\mu{}_\nu$ and dilaton $\rho$.
The connection is required to be metric:
$$
\partial_\lambda g_{\mu\nu}
-\igamma_\lambda{}^\tau{}_\mu g_{\tau\nu}
-\igamma_\lambda{}^\tau{}_\nu g_{\mu\tau}
=0\ ,\eqno(2.1)
$$
but can have nonvanishing torsion
$T_\lambda{}^\mu{}_\nu=
\igamma_\lambda{}^\mu{}_\nu-\igamma_\nu{}^\mu{}_\lambda$.
The curvature of the connection will be denoted
$R_{\mu\nu}{}^\rho{}_\sigma=
\partial_\mu \igamma_\nu{}^\rho{}_\sigma
-\partial_\nu \igamma_\mu{}^\rho{}_\sigma
+\igamma_\mu{}^\rho{}_\tau \igamma_\nu{}^\tau{}_\sigma
-\igamma_\nu{}^\rho{}_\tau \igamma_\mu{}^\tau{}_\sigma$.
It is antisymmetric in $\mu$, $\nu$, and using (2.1) one can
show that
$R_{\mu\nu\alpha\beta}=g_{\alpha\rho}R_{\mu\nu}{}^\rho{}_\beta$
is also antisymmetric in $\alpha$, $\beta$. It has no other symmetry
property.

The most general diffeomorphism invariant action which
is at most quadratic in the derivatives of the fields is
$$
\eqalign{
&S(g,\igamma,\rho)=
\int d^4x\,\sqrt{\det g}\Bigl[
{1\over2}b_0\partial_\mu\rho\partial^\mu\rho
+2b_1\rho\partial_\mu\rho T^\mu
+V(\rho)+g_0 \rho^2 R\cr
&
\qquad\qquad\qquad\qquad
+a_1 \rho^2 T_{\mu\nu\rho}T^{\mu\nu\rho}
+a_2 \rho^2 T_{\mu\nu\rho}T^{\mu\rho\nu}
+a_3 \rho^2 T^{\rho}T_{\rho}\cr
&
+g_1 R_{\mu\nu\rho\sigma}R^{\mu\nu\rho\sigma}
+g_2 R_{\mu\nu\rho\sigma}R^{\mu\rho\nu\sigma}
+g_3 R_{\mu\nu\rho\sigma}R^{\rho\sigma\mu\nu}
+g_4 R_{\mu\nu}R^{\mu\nu}
+g_5 R_{\mu\nu}R^{\nu\mu}
+g_6 R^2 \Bigr]\cr}\eqno(2.2)
$$
where $R_{\mu\nu}=R_{\rho\mu}{}^\rho{}_\nu$,
$R=g^{\mu\nu}R_{\mu\nu}$
and $T_\nu=T_\nu{}^\mu{}_\mu$.
Indices are raised and lowered with $g$.
The couplings of the dilaton can be understood as due to
the quantum mechanical breaking of Weyl symmetry in a
classically Weyl invariant theory. This point is
discussed in Appendix A, but is not necessary for what follows.

There are some special choices of coefficients that should be pointed out
since they will be useful later. The first of these
can be expressed by the identity
$$\eqalign{
{1\over 768\pi^2}\int d^4x\, & \sqrt{\det g}\Bigl[
R_{\mu\nu\rho\sigma}R^{\rho\sigma\mu\nu}
-4R_{\mu\nu}R^{\nu\mu}
+R^2 \Bigr]\cr
&={1\over 128\pi^2}\int d^4x{1\over\sqrt{\det g}}
\varepsilon^{\alpha\beta\gamma\delta}\varepsilon^{\mu\nu\rho\sigma}
R_{\alpha\beta\mu\nu}R_{\gamma\delta\rho\sigma}=\chi\ ,}
\eqno(2.3)
$$
($\varepsilon^{1234}=1$)
which is an integral representation of the Euler invariant $\chi$.
With this choice of action, the theory is topological.
Another useful identity is
$$
R=R(g)+{1\over4}T_{\mu\nu\rho}T^{\mu\nu\rho}
+{1\over2}T_{\mu\nu\rho}T^{\mu\rho\nu}
-T^{\rho}T_{\rho}+2\nabla_\mu T^\mu
,\eqno(2.4)
$$
where $R(g)$ denotes the Ricci scalar of the metric $g_{\mu\nu}$
and $\nabla_\mu$ is the covariant derivative with respect to the
Levi-Civita connection. Using this formula one can replace
$R$ by $R(g)$ in (2.2) at the cost of redefining the constants
$a_i$ and $b_1$.
In particular, if we choose the only nonzero coefficients to be
$b_0$, $g_0$, $a_1=-{1\over4}g_0$, $a_2=-{1\over2}g_0$,
$a_3=g_0$ and $b_1=2g_0$, (2.2) becomes the action for
a nonminimally coupled scalar field:
$$
\int d^4x\,\sqrt{\det g}\,
\left[{1\over2}b_0\,\partial_\mu\rho\partial^\mu\rho
+g_0\, R(g)\rho^2\right]
\ .\eqno(2.5)
$$
For $\rho=\ $constant, this is Einstein's action.

Finally, we mention two alternative ways of writing the torsion
squared and curvature squared terms. In discussing the linearized
theory it is useful to write them in the more compact form
$$
G^{\mu\nu\alpha\beta\rho\sigma\gamma\delta}
R_{\mu\nu\alpha\beta}\, R_{\rho\sigma\gamma\delta}
+\rho^2 H^{\mu\alpha\nu\rho\beta\sigma}
T_{\mu\alpha\nu}T_{\rho\beta\sigma}
\ ,\eqno(2.6)
$$
where
$$
\eqalignno{
G^{\mu\nu\alpha\beta\rho\sigma\gamma\delta}=&
\,{\rm SYMM}\bigl\{
g_1\, g^{\mu\rho}g^{\nu\sigma}
g^{\alpha\gamma}g^{\beta\delta}
+g_2\, g^{\mu\rho}g^{\gamma\nu}
g^{\alpha\sigma}g^{\beta\delta}
+g_3\, g^{\gamma\mu}g^{\delta\nu}
g^{\alpha\rho}g^{\beta\sigma}\cr
&\qquad\qquad+g^{\alpha\mu}g^{\gamma\rho}
(g_4\, g^{\nu\sigma}g^{\beta\delta}
+g_5\, g^{\delta\nu}g^{\beta\sigma}
+g_6\, g^{\nu\beta}g^{\sigma\delta})\bigr\}\ ,&(2.7a)\cr
H^{\mu\alpha\nu\rho\beta\sigma}=&
\,{\rm SYMM}\bigl\{g^{\mu\rho}(a_1\, g^{\alpha\beta}g^{\nu\sigma}
+a_2\, g^{\alpha\sigma}g^{\beta\nu}
+a_3\, g^{\alpha\nu}g^{\beta\sigma})\bigr\}\ .&(2.7b)\cr}
$$
The prefix ``SYMM'' indicates that one has to take the proper
combinations so that $G$ is antisymmetric in
$(\mu,\nu)$, $(\rho,\sigma)$, $(\alpha,\beta)$, $(\gamma,\delta)$
and symmetric
under the simultaneous interchange of $\mu\nu\alpha\beta$ and
$\rho\sigma\gamma\delta$
and similarly $H$ is antisymmetric in $(\mu,\nu)$, $(\rho,\sigma)$
and symmetric under the simultaneous interchange of $\mu\alpha\nu$ and
$\rho\beta\sigma$.

The other way of writing the action is based on the decomposition
of the torsion and curvature tensors into irreducible parts
with respect to the Lorentz group. This is discussed in Appendix B.

\bigskip
\goodbreak
\leftline{\bf 3. Linearization}
\smallskip
\noindent
In this section we set up the formalism for computing,
at one loop, the Euclidean effective action $S_{\rm eff}(g,\igamma,\rho)$.
We will not ultimately do the calculation (this was done for example
in [11]), but what we describe here is a preliminary material
for the definition of the average effective action, that will be
given in the next section.
We will restrict ourselves to a flat background
$$
g_{\mu\nu}=\delta_{\mu\nu}\ \ ,\qquad\qquad
\igamma_\lambda{}^\mu{}_\nu=0\ \ ,\qquad\qquad
\rho={\rm constant}\ .\eqno(3.1)
$$
In this case we can write
$S_{\rm eff}(g,\igamma,\rho)=\int d^4x\,V_{\rm eff}(\rho)$,
where $V_{\rm eff}$ is the effective potential for the dilaton.
The first step in this calculation is to linearize the action
around the background (3.1).
We define the fluctuation fields $\delta \igamma_\lambda{}^\mu{}_\nu$,
$\delta g_{\mu\nu}$ and $\delta\rho=\sigma$ as the deviations of
$\igamma$, $g$ and $\rho$ from their background values (3.1).
In principle there is a total of $64+10+1$ fields.
However, the condition (2.1) implies that
$$
\partial_\lambda\delta g_{\mu\nu}-\delta \igamma_{\lambda\mu\nu}
-\delta \igamma_{\lambda\nu\mu}=\,0\ .\eqno(3.2)
$$
(From now on indices are raised and lowered with $\delta_{\mu\nu}$.)
These are $40$ constraints, reducing the number of independent fields to
$35$.

We define $\omega_{\lambda\mu\nu}=\delta \igamma_{\lambda[\mu\nu]}$
and $\varphi_{\mu\nu}={1\over2}\rho\delta g_{\mu\nu}$.
The rescaling of $\delta g_{\mu\nu}$ is convenient for dimensional reasons:
with this definition all the fluctuations have dimension of mass.
This redefinition is legitimate here since we are assuming $\rho$
to be constant; it is related to the choice of functional measure
in the path integral. We will see later that it does not have any
effect on the renormalization group equations.
After having written out the linearized action in terms of
$\delta \igamma$, $\delta g$ and $\sigma$, one can use (3.2)
to eliminate the symmetric part of $\delta \igamma$ in favor of $\varphi$:
$\delta
\igamma_{\lambda(\mu\nu)}={1\over\rho}\partial_\lambda\varphi_{\mu\nu}$.
At this point the linearized Euclidean action is a quadratic form
which can be written, after Fourier transforming
(we use $\partial_\mu\to iq_\mu$)
$$
S^{(2)}(\varphi,\omega,\sigma;\rho)
=\ {1\over2}\int d^4q \sum\limits_{A,B}
\ \Phi_A(q)\cdot{\cal O}_{[AB]}\cdot\Phi_B(-q)\ ,\eqno(3.3)
$$
where the indices $A$, $B$ label the three types of fields
$\Phi_1=\omega$, $\Phi_2=\varphi$ and $\Phi_3=\sigma$
and the dots stand for contraction over the tensor indices.
When written out explicitly in terms of the components
of the fields, ${\cal O}$ is a $35\times 35$ matrix.
The components of this matrix are listed below:
$$
\eqalignno{
&{\cal O}_{[\omega\omega]}{}_{\mu\alpha\beta}{}^{\rho\gamma\delta}=
\,8 G_\mu{}^\nu{}_{\alpha\beta}{}^{\rho\sigma\gamma\delta}q_\nu q_\sigma
+8  H_{\mu\alpha\beta}{}^{\rho\gamma\delta}\rho^2
+2 g_0\rho^2\delta_\beta^\gamma\left(\delta_{\alpha\mu}\delta^{\delta\rho}-
\delta_\alpha^\rho\delta_\mu^\delta\right)
\ ,&(3.4a)\cr
&{\cal O}_{[\varphi \omega]}{}_{\alpha\mu}{}^{\rho\gamma\delta}=
\,-8i\rho H^\sigma{}_{\alpha\mu}{}^{\rho\gamma\delta}q_\sigma
+2ig_0\rho\left(\delta_{\alpha\mu}\delta^{\delta\rho}q^\gamma
-\delta_\mu^\delta\delta_\alpha^\rho q^\gamma
+\delta_\mu^\delta\delta^{\gamma\rho}q_\alpha\right)
\ ,&(3.4b)\cr
&{\cal O}_{[\varphi\varphi]}{}_{\alpha\mu}{}^{\beta\rho}=
\,8H_{\mu\alpha}{}^{\nu\rho\beta\sigma}q_\nu q_\sigma
+{V\over\rho^2}(\delta_{\alpha\mu}\delta^{\beta\nu}
-2\delta_\alpha^\rho\delta^\beta_\mu)
\ ,&(3.4c)\cr
&{\cal O}_{[\sigma\omega]}{}^{\rho\gamma\delta}=
2i\rho(b_1-2g_0)\delta^{\rho\gamma}q^\delta
\ , &(3.4d)\cr
&{\cal O}_{[\sigma\varphi]}{}^{\mu\nu}=
\,2b_1(\delta^{\mu\nu}q^2-q^\mu q^\nu)
+{1\over\rho}{dV\over d\rho}\delta^{\mu\nu}
\ ,  &(3.4e)\cr
&{\cal O}_{[\sigma\sigma]}=
\,b_0 q^2+{d^2V\over d\rho^2}
\ .&(3.4f)\cr}
$$
For the purpose of clarity we have omitted to indicate explicitly
symmetrizations and antisymmetrizations on the r.h.s.
(for example $(3.4b)$ should be symmetrized in
$\alpha$, $\mu$ and antisymmetrized in $\gamma$, $\delta$).
When $V=0$, this linearized action is invariant
under linearized gauge transformations.
Let $x'^\mu=x^\mu-v^\mu$ be an infinitesimal
coordinate transformation. The variations of the fields are
$$
\delta \igamma_\lambda{}^\mu{}_\nu=\partial_\lambda\partial_\nu v^\mu\ ,\qquad
\delta g_{\mu\nu}=\partial_\mu v_\nu+\partial_\nu v_\mu\ ,\qquad
\delta\rho=0\ . \eqno(3.5)
$$
There follows that the fields
$$
\omega_{\lambda\mu\nu}={1\over2}q_\lambda(q_\mu v_\nu-q_\nu v_\mu)
\ ,\qquad
\varphi_{\mu\nu}={i\over2}\rho\, (q_\mu v_\nu+q_\nu v_\mu)\ ,\qquad
\sigma=0 \eqno(3.6)
$$
are null vectors for the operator ${\cal O}$.
This can be verified by explicit calculations.
To make ${\cal O}$ invertible, we add to the
linearized action the following gauge-fixing term:
$$
S_{\rm GF}=
{1\over2\alpha}\int d^4x\,
\partial_\mu\varphi^{\mu\nu}\partial^\rho\varphi_{\rho\nu}\ .
\eqno(3.7)
$$
The ghost contribution has the form
$$
S_{\rm ghost}(\bar d,d)
=\ {1\over2}\int d^4q\,
\bar d^\mu {{\cal O}_{[dd]}}_\mu{}^\nu d_\nu
\ ,\eqno(3.8)
$$
where, $\bar d$, $d$, are
anticommuting ghost fields and
$$
{{\cal O}_{[dd]}}_{\mu}{}^{\nu}=
{1\over2}\rho\,(\delta_\mu^\nu q^2+q_\mu q^\nu)\ .\eqno(3.9)
$$
Apart from the overall factor $\rho$, which can be eliminated by
a redefinition of the measure and is irrelevant, this operator
is field-independent. It can be neglected in what follows.

To compute the one-loop effective action one now needs to calculate
the functional determinant of the operator $\cal O$ appearing in the
previous formulas.
The determinant of ${\cal O}$ on the 35 dimensional space
spanned by the fields is very hard to compute as it stands.
One can partially diagonalize
these operators in blocks corresponding to spin and parity.
This is because ${\cal O}$ is a Lorentz covariant wave operator
and therefore does not mix fields with different spin and parity.
There are two modes with spin-parity $2^+$, coming from $\omega$ and
$\varphi$, one $2^-$ mode from $\omega$, two $1^+$ modes
coming from $\omega$, three $1^-$ modes,
two from $\omega$ and one from $\varphi$,
four $0^+$ of which one comes from $\omega$, two from $\varphi$
and one from $\sigma$,
and finally one $0^-$ mode from $\omega$.
One counts indeed
$2\times5+1\times5+2\times3+3\times3+4\times1+1\times1=35$.

The total linearized quadratic action, including the gauge-fixing, ghost
and potential terms, can be rewritten as
$$
S^{(2)}={1\over 2}\int d^4q\ \Phi_A(-q)\cdot
a_{ij}^{AB}(J^{\cal P})\, P_{ij}^{AB}(J^{\cal P})\cdot \Phi_B(q)\ ,
\eqno(3.10)
$$
where $P_{ij}^{AB}(J^{\cal P})$ are spin projection operators
[12,8,10] that we list in Appendix C,
and $a_{ij}^{AB}(J^{\cal P})$ are coefficient matrices,
representing the inverse propagators of each set of fields with
definite spin and parity.
For $V=0$ these matrices are given by
$$
\eqalignno{
&a(2^+)=\left[
\matrix{G_1q^2+B_1\rho^2
     & -i\sqrt2 B_1|q|\rho \cr
        i\sqrt2 B_1|q|\rho
     & B_2 q^2 \cr}
                              \right]\ ,&(3.11a)\cr
&\null\cr
&a(2^-)=G_2 q^2+B_1\rho^2\ , &(3.11b)\cr
&\null\cr
&a(1^+)=\left[
\matrix{G_3q^2+B_3\rho^2
        & -\sqrt 2 B_4\rho^2
                              \cr
          -\sqrt 2 B_4\rho^2
        &  B_5\rho^2
                    \cr}
                              \right]\ , &(3.11c)\cr
&\null\cr
&a(1^-)=\left[
\matrix{G_4q^2+B_6\rho^2
       & \sqrt2 B_7\rho^2
       & i\sqrt2 B_7|q|\rho
        \cr
         \sqrt2 B_7\rho^2
       & B_8\rho^2
       & iB_8|q|\rho
        \cr
         -i\sqrt2 B_7|q|\rho
       & -iB_8|q|\rho
       & B_8 q^2
                      \cr}
                               \right]\ ,&(3.11d)\cr
&\null\cr
&a(0^+)=\left[
\matrix{  G_5q^2+B_9\rho^2
        & -i \sqrt2 B_9|q|\rho
        & 0
        & -i\sqrt6 B_{11} |q|\rho
 \cr
          i\sqrt2 B_9|q|\rho
        & B_{10}q^2
        & 0
        & \sqrt3 B_{12}q^2
 \cr
          0
        & 0
        & 0
        & 0
  \cr
          i\sqrt6 B_{11} |q|\rho
        & \sqrt3 B_{12}q^2
        & 0
        & b_0q^2
                                  \cr}\right]\ , &(3.11e)\cr
             &\null\cr
&a(0^-)=G_6q^2+B_{13}\rho^2\ , &(3.11f)\cr}
$$
where
$$
\eqalign{
&G_1=4g_1+2g_2+4g_3+g_4+g_5\ ,\cr
&G_2=4g_1+g_2\ ,\cr
&G_3=4g_1-4g_3+g_4-g_5\ ,\cr
&G_4=4g_1+g_2+2g_4\ ,\cr
&G_5=4g_1+2g_2+4g_3+4g_4+4g_5\ ,\cr
&G_6=4g_1-2g_2\ ,\cr
&B_1=2a_1+a_2+g_0\ ,\cr
&B_2=4a_1+2a_2\ ,\cr
&B_3=6a_1-5a_2-g_0\ ,\cr
&B_4=2a_1-3a_2-g_0\ ,\cr}
\qquad\qquad
\eqalign{
&B_5=4a_1-2a_2\ ,\cr
&B_6=2a_1+a_2+2a_3-g_0\ ,\cr
&B_7=a_3-g_0\ ,\cr
&B_8=2a_1+a_2+a_3\ ,\cr
&B_9=2a_1+a_2+3a_3-2g_0\ ,\cr
&B_{10}=4a_1+2a_2+6a_3\ ,\cr
&B_{11}=b_1-2g_0\ ,\cr
&B_{12}=2b_1\ ,\cr
&B_{13}=8a_1-8a_2-2g_0\ ,\cr
&B_0=b_0\ .\cr}
\eqno(3.12)
$$
We observe that if we did not redefine the fluctuation of
the metric and worked with $\delta g$, the only effect on the
coefficient matrices would be to multiply by $\rho/2$ the
second row and column of $a(2^+)$, the third row and column of
$a(1^-)$ and the second and third row and column of $a(0^+)$.
This would change the determinants of these matrices by an overall
power of $\rho^2$, which, as we shall see, does not affect the
renormalization group equations.

There are a few checks that one can make on these matrices.
First we observe that the matrices $a(1^-)$ and $a(0^+)$ are degenerate.
The proportionality of the last two rows and columns of $a(1^-)$ and the
vanishing of the third row and column of $a(0^+)$ are direct consequences
of the diffeomorphism invariance.

If we take the only nonzero coefficients to be $g_3$,
$g_5=-4g_3$, $g_6=g_3$, corresponding to the action (2.3),
the coefficient matrices are identically zero;
this is because the corresponding
action is a topological invariant (actually zero, since we expand
around flat space).
If we take the only nonzero coefficients to be $g_0$,
$a_1=-{1\over4}g_0$, $a_2=-{1\over2}g_0$, $a_3=g_0$, $b_0$ and
$b_1=2g_0$,
corresponding to the Lagrangian (2.5),
the coefficient matrices reduce to
$$
\eqalignno{
&a(2^+)=\left[
\matrix{0
     & 0\cr
        0
     & -2g_0 q^2 \cr}
                              \right]\ ,&(3.12a)\cr
&\null\cr
&a(0^+)=\left[
\matrix{ 0
        & 0
        & 0
        & 0
 \cr
          0
        & 4g_0 q^2
        & 0
        & 4\sqrt3g_0 q^2
 \cr
          0
        & 0
        & 0
        & 0
  \cr
          0
        & 4\sqrt3 g_0 q^2
        & 0
        & B_0q^2
                                  \cr}\right]\ .&(3.12b)\cr
           \cr}
$$
Note that for $g_0=b_0/12$ the second and fourth rows and columns
of $a(0^+)$ are proportional, and the matrix has rank one.
This is because
in this case the action is Weyl invariant (see Appendix A).
If we freeze $\rho=$constant, the last row and column of $a(0^+)$
can be suppressed and we are left with
the familiar coefficient matrices of Einstein's theory [12].

The contribution of the potential to the inverse propagators is
$$
\eqalignno{
&a(2^+)_{\rm Pot}=\left[
 \matrix{  0
        & 0 \cr
          0
        & -{2\over\rho^2}V \cr}
                      \right]\ , &(3.14a)\cr
&\null\cr
&a(1^-)_{\rm Pot}=\left[
\matrix{ 0
       & 0
       & 0
           \cr
         0
       & 0
       & 0
           \cr
         0
       & 0
       & -{2\over\rho^2}V
                            \cr}\right]\ ,&(3.14b)\cr
             &\null\cr
&a(0^+)_{\rm Pot}=\left[
\matrix{  0
        & 0
        & 0
        & 0
    \cr
          0
        & {1\over\rho^2}V
        & {\sqrt3\over\rho^2}V
        & {\sqrt3\over\rho}{d V\over d\rho}
    \cr
        0
        & {\sqrt3\over\rho^2}V
        & -{1\over\rho^2}V
        & {1\over\rho}{d V\over d\rho}
    \cr
          0
        & {\sqrt3\over\rho}{d V\over d\rho}
        & {1\over\rho}{d V\over d\rho}
        & {d^2 V\over d\rho^2}
                               \cr}   \right]\ . &(3.14c)\cr
                   \cr}
$$
Note that these matrices do not have the degeneracies of
$(3.11d,e)$ or $(3.12b)$. This is because flat space (with $\rho=$constant)
is a solution of the field equations only if $V=0$.
Finally the contribution of the gauge fixing terms is
$$
\eqalignno{
&a(1^-)_{\rm GF}=\left[
\matrix{ 0
       & 0
       & 0
       \cr
         0
       & 0
       & 0
       \cr
         0
       & 0
       & {1\over2\alpha}q^2
       \cr}
                \right]\ ,&(3.15a)\cr
             &\null\cr
&a(0^+)_{\rm GF}=\left[
\matrix{  0
        & 0
        & 0
        & 0
    \cr
          0
        & 0
        & 0
        & 0
    \cr
          0
        & 0
        & {1\over\alpha}q^2
        & 0
    \cr
          0
        & 0
        & 0
        & 0\cr}
               \right]\ . &(3.15b)\cr}
$$

With these results, the usual one-loop effective action is equal to the
sum over spin $J$ and parity $\cal P$ of the logarithms of the
determinants of the matrices $a$.
These are polynomials in $q^2$, $\rho^2$, $V$ and its derivatives
of dimension up to eight.
Taking into account the multiplicity of these contributions,
the one-loop effective potential is
$$
V_{\rm eff}(\rho)={1\over2}\sum\limits_{J,{\cal P}}(2J+1)
\int {d^4q\over(2\pi)^4}
\ln\left( {\det a(J^{\cal P})(\rho)\over
\det a(J^{\cal P})(\rho_0)}\right)
\ .\eqno(3.16)
$$
We have normalized the effective action with the determinant of
a free field, which is obtained by linearizing the action around
a fixed constant field $\rho_0$. It is natural to choose $\rho_0$
as the minimun of $V_{\rm eff}$ itself, in which case
$V_{\rm eff}(\rho_0)=0$.

Given the previous explicit form for the matrices $a(J^{\cal P})$,
one can now compute the expression for $V_{\rm eff}(\rho)$, using standard
renormalization techniques. As explained in the next section, we shall follow
instead a different strategy: we shall derive the equation that describes
the renormalization group flow of $V_{\rm eff}$. This allows a more
accurate
discussion of the scale-dependence of the parameters that characterize
the effective potential.

\bigskip
\goodbreak

\leftline{\bf 4. Average effective potential}
\medskip
\noindent
To study the renormalization group flow of the effective potential,
we shall use ideas originally introduced by Wilson [13].
One begins from some action $S_{k_1}$ which is supposed to describe
accurately the physics at some momentum scale $k_1$.
Physics at a lower momentum scale $k_2$ is then described
by an effective action $S_{k_2}$ which is obtained by functionally
integrating $\exp(-S_{k_1})$
over all fluctuations of the fields with momenta between $k_1$ and
$k_2$. The procedure is then iterated: the effective action at
scale $k_3<k_2$ is obtained by functionally integrating
$\exp(-S_{k_2})$ from $k_2$ to $k_3$, and so on.
The flow of $S_k$ with $k$ is the renormalization group flow.
Each step of the integration should not cover too large a range of momenta.
In this way one can accurately compute the effective action at some low
momentum scale $k$, starting from a high momentum scale $\Lambda$.
Note that this is not the same as performing the functional integral
from $\Lambda$ to $k$ in one step, because the action is updated
at each step of the integration. In the following we will
refer to this updating as ``the renormalization group improvement''.

Originally this idea was applied to spin fields on the lattice,
but it was subsequently adapted to the continuum, where it was used to
clarify and simplify the notion of renormalizability [14], and was also
applied to gauge theories [15].
The particular implementation of this idea that we shall use here
has been discussed in [16,17].
One can start from the usual definition of the effective action
$S_{\rm eff}$,
defined as the Legendre transform of $W=-\ln Z$, where $Z$ is the
partition function. The effective action has a perturbative expansion
in Feynman diagrams, which correspond to integrals of certain
functions of the momenta. By introducing some kind of infrared cutoff $k$
in the propagators one obtains a new effective action $S_k$,
which depends on the scale $k$.

A way of implementing this idea in the path integral formalism
is to add to the classical action $S$ a term $\Delta S_k$ that constrains
the average of the field $\phi$
in spheres of radius $\approx k^{-1}$ centered around the
point $x$ to be equal to a predetermined function
$\bar\phi(x)$ (one can apply this
discussion to the theory we are interested in by replacing the
generic notation $\phi$ with $\igamma$, $g$ and $\rho$).
The average of a field $\phi$ around a point $x$ is defined
by the convolution $f_k\phi(x)=\int d^4y\sqrt{\det g} f_k(x-y)\phi(y)$,
or, when $g$ is flat, in Fourier space, by
$f_k\phi(q)=f_k(q)\phi(q)$, where $f_k$ is the function
$$
f_k(q)=\exp\left(-a\left({q^2\over k^2}\right)^b\right)\ ,\eqno(4.1)
$$
with $a$ and $b$ are constants. This function interpolates between
a gaussian, for $b=1$, and a step function for $b\to\infty$.

In [10] we discussed this procedure in the context of a gauge
theory of gravity where only the coefficients $g_1$ and $a_1$ were
assumed to be nonzero. The specific choice of the term $\Delta S_k$
that we used there had the disadvantage that some of the propagators
were not well defined. This was not important in [10] because these
terms did not contribute to the quantities that we computed there.
However, it could be a problem in more general calculations.
For this reason we shall use here a simpler definition:
we will assume that the term $\Delta S_k$ is defined
in such a way that the only effect it has on the linearized action
\vfill\break\noindent
is to replace in the operators ${\cal O}$ the term $q^2$ with the function
$$
P_k(q)={q^2\over 1-f_k(q)^2}\ ,\eqno(4.2)
$$
(and $|q|$ by $\sqrt{P_k(q)}$).
The function $P_k(q)$ approaches exponentially fast the function
$q^2$ for $q^2>k^2$, but goes to a constant ($b=1$) or diverges
($b>1$) for $q^2\to 0$.
Thus, replacing $q^2$ by $P_k$ in the propagators suppresses the
modes with $q^2<k^2$ and therefore has an effect similar to
putting an IR cutoff at momentum $k$.
This definition of scale-dependent effective action $S_k$ is
equivalent, at least at one-loop order, to the one given in [17].

It is quite clear that the considerations that were made in [17]
for nonabelian gauge theories can be extended in a rather
straightforward way to the case of gravity.
There is one point, however, that requires some extra care: it is
the definition of the absolute normalization of the potential.
In the presence of gravity, the value of the potential at the
minimum is interpreted as a cosmological constant.
It affects the field equations, and therefore cannot be shifted
arbitrarily. Furthermore, we see from (3.14) that it corresponds
also to the mass of the graviton.
In principle, the value of the potential at the minimum could
depend on the scale, but one has to make sure that at least in the
extreme IR limit this value be zero, to ensure that the graviton
be massless.

We take the following definition of the scale dependent effective potential
for $\rho$:
$$
V_k(\rho)={1\over2}
\sum\limits_{J,{\cal P}}(2J+1)\int {d^4q\over(2\pi)^4}
\ln\left({\det a_k(J^{\cal P})(\rho)
\over\det a_k(J^{\cal P})(\rho_k)}\right)\ ,
\eqno(4.3)
$$
where $a_k(J^{\cal P})$ are obtained from the matrices $a(J^{\cal P})$
given in (3.11,14,15) by replacing $q^2$ with the function
$P_k(q)$ and $\rho_k$ is defined to be the minimum of $V_k$.
Note that with this definition $V_k(\rho_k)=0$ for all $k$, so that
the cosmological constant is actually zero at all scales.
In this connection see also [18].

Finally, we observe that since the determinants $\det a_k(J^{\cal P})$
are functions of $\rho^2$, it is consistent to assume that
$V_k$ is a function of $\rho^2$.
It will be convenient to define $V'={dV\over d(\rho^2)}$.
Then in (3.14$c$) we can write
${1\over\rho}{d V\over d\rho}=2V'$
and ${d^2V\over d\rho^2}=4\rho^2 V''+2 V'$.

\bigskip
\goodbreak
\leftline{\bf 5. Renormalization group flow}
\medskip
\noindent
The average effective potential obeys a renormalization group
equation that is obtained from (4.3) by taking its derivative with respect
to $k$. It has the form
$$
k{dV_k\over dk}=F(V_k,V_k',V_k'')\ ,
\eqno(5.1)
$$
where the functional $F$ comes from the r.h.s. of (4.3):
$$
F(V,V',V'')={1\over2}
\sum\limits_{J,{\cal P}}(2J+1)\int {d^4q\over(2\pi)^4}
{1\over \det a_k(J^{\cal P})}\
k{d\over dk}\left(\det a_k(J^{\cal P})\right)\ ,
\eqno(5.2)
$$
with
$$
k{d \over dk}(\det a_k)=
{\partial (\det a_k)\over \partial P_k}\,
k{\partial P_k\over \partial k}\ .\eqno(5.3)
$$
After taking these derivatives, we substitute in $F$
the classical potential $V$ with the effective potential $V_k$:
this is the ``renormalization group improvement''.
This substitution gives (5.1), a differential equation
for the function $V_k(\rho)$.
Notice that thanks to the behaviour of $P_k$ and its $k$-derivative in
(5.3), the integral in (5.2) is actually dominated by a finite range of
momenta and does not need an ultraviolet regularization.

Although derived in the context of a one loop approximation,
this renormalization group improved equation has a validity that
goes beyond one loop [19,13,14]. Clearly one cannot follow the evolution
of the whole function $V_k$, so some other kind of approximation
becomes necessary. In the following we shall study only the
first few terms of the Taylor expansion of $V_k$.
As explained in the introduction, we are interested in the
spontaneously broken phase, with a nonzero VEV of $\rho$ at $k=0$.
Thus, we parametrize $V_k$ by the
position of its minimum, $\rho_k$,
and the quartic coupling at the minimum $\lambda_k$:
$$
V'_k(\rho_k)=0\ \ ,\qquad
\lambda_k=V''_k(\rho_k)\ .\eqno(5.4)
$$
In terms of these parameters the potential reads
$$
V_k(\rho)={1\over 2}\lambda_k(\rho^2-\rho_k^2)^2\ ,\eqno(5.5)
$$
which is the Taylor expansion of $V_k$ as function of $\rho^2$
around its minimum.

The equations governing the evolution in $k$ of $\rho^2_k$
and $\lambda_k$ can be obtained by differentiating the definitions (5.4).
This leads to the following set of coupled partial differential equations:
$$\eqalignno{
k{\partial\rho^{\, 2}_k\over\partial k}=&\,\gamma(k) k^2\ ,&(5.6b)\cr
k{\partial\lambda_k\over\partial k}=&\,\beta(k)\ ,&(5.6c)}
$$
with
$$
\eqalignno{
\gamma(k)=&-{1\over\lambda k^2}\,
\left(k{\partial\over\partial k} V'_k\right)(\rho_k)\,
={1\over32\pi^2}\int dx x\,{\cal R}_\gamma(P_k,\rho_k^2)\,
k{\partial P_k\over\partial k}\ ,&(5.7b)\cr
\beta(k)=&
\left(k{\partial\over\partial k}V_k''\right)(\rho_k)=
{1\over32\pi^2}\int dx x\,{\cal R}_\beta(P_k,\rho_k^2)\,
k{\partial P_k\over\partial k}\ .&(5.7c)}
$$
In the definition of $\beta(k)$ we are neglecting a term containing
$V_k'''(\rho_k)$, which takes into account the $k$ dependence of the
point of definition of $\lambda_k$ and is a signal of the presence
of operators of higher dimension in the potential (``irrelevant''
operators). Consistent with our approximation, we will neglect
the effect of these terms.
The functions $\beta$ and $\gamma$ have the general form shown
in the r.h.s. of (5.7), where $x=q^2$ and ${\cal R}$ are rational
functions of dimension $k^{-6}$. These functions can be computed
explicitly from the representation (4.3) of $V_k$, by taking
derivatives with respect to $\rho^2$ and $k$, and using (5.3).
The general expressions are complicated and not particularly
illuminating, so we will not give them here.
It is not possible to find a solution of the system of p.d.e.'s
(5.6-7) in closed form. However, analytic solutions can be
obtained in some asymptotic limit.

Let us assume first that $\rho_k^2$ is small with respect to $k$.
One would expect this to describe the behaviour of the theory in the
regime when $k$ is large compared to the Planck mass.
In simpler systems, this approximation indeed reproduces the results
of perturbative calculations of beta functions at momenta much
larger than the characteristic mass of the theory.
Because of the factor $k{\partial P_k\over\partial k}$,
the integrals are dominated by the region $x\approx k^2$.
In this region, $P_k$ is itself of order $k^2$, so that $\rho_k^2$
is small with respect to $P_k$. One can therefore expand the
functions ${\cal R}$ in powers of $\rho_k^2$.
In the function ${\cal R}_\gamma$ the dominant term is a constant
(independent of $\rho_k$) but the function ${\cal R}_\beta$
has a pole for $\rho_k\to 0$, coming from the
contribution of the spins $2^+$, $1^-$ and $0^+$.
The beta functions reduce to the following:
$$
\gamma(k)=\gamma_0+O\!\left(\rho_k^2\over k^2\right)\ \ ;
\qquad
\beta(k)=\beta_{-1}{k^2\over\rho_k^2}+\beta_0
+O\!\left(\rho_k^2\over k^2\right)\ ,\eqno(5.8)
$$
where
$$
\eqalignno{
&\gamma_0={1\over32\pi^2}{1\over\lambda_k}
\biggl[5{B_1B_2-2B_1^2\over G_1B_2}
+5{B_1\over G_2}
+3{B_3B_5-2B_4^2\over G_3B_5}
+3{B_6B_8-2B_7^2\over G_4B_8}\cr
&\qquad
+{6\lambda_k(B_{10}-2B_{12})\over B_0B_{10}-3B_{12}^2}
+{B_9\over G_5}
-{2B_0B_9^2-12B_9B_{11}B_{12}+6B_{10}B_{11}^2
\over G_5(B_0B_{10}-3B_{12}^2)}
+{B_{13}\over G_6}\biggr]I_{-2}\ ,\quad &(5.9a)\cr
&\beta_{-1}={1\over32\pi^2}\left[{10\over B_2}
-{B_0\over B_0B_{10}-3B_{12}^2}+13\alpha\right]\lambda_kI_{-2}\
.&(5.9b)\cr
&\beta_0={1\over 32\pi^2}
\biggl[10{(B_1B_2-2B_1^2)^2+4\lambda_kB_1^2
\over G_1^2B_2^2}
+10{B_1^2\over G_2^2}
+6{(B_3B_5-2B_4^2)^2\over G_3^2B_5^2}\cr
&\qquad
+6{(B_6B_8-2B_7^2)^2\over G_4^2B_8^2}
+{24(2B_0B_{10}+3B_{10}^2-12B_{10}B_{12}+7B_{12}^2)-16\alpha B_{10}
\over(B_0B_{10}-3B_{12}^2)^2}\lambda_k^2\cr
&\qquad
+{432(B_{10}B_{11}-B_9B_{12})^2
-4(B_0B_9-3B_{11}B_{12}+12B_{10}B_{11}-12B_9B_{12})^2
\over G_5(B_0B_{10}-3B_{12}^2)^2}\lambda_k &(5.9c)\cr
&\qquad
+{2(2B_0B_9^2-B_0B_9B_{10}+6B_{10}B_{11}^2-12B_9B_{11}B_{12}+3B_9B_{12}^2)^2
\over G_5^2(B_0B_{10}-3B_{12}^2)^2}
+2{B_{13}^2\over G_6^2}\biggr]I_{-3}\ .\qquad \cr}
$$
The positive dimensionless constants $I_n$ are defined by
$$
\eqalign{
&I_{-3}=
\int dx x P_k^{-3}k{\partial P_k\over\partial k}=1\ ,\cr
&I_{-2}=\ k^{-2}
\int dx x P_k^{-2}k{\partial P_k\over\partial k}
={2\over(2a)^{1/b}}\,\Gamma(1+1/b)\ ,\cr
&I_{0}=\ k^{-6}
\int dx x k{\partial P_k\over\partial k}
={2\over(2a)^{3/b}}\,\Gamma(1+3/b)\,\zeta(3/b)\ .\cr}
\eqno(5.10)
$$
Infrared convergence of $I_0$ requires that $b<3$.

Note that in this approximation the dominant term in $\beta(k)$
is the one coming from the pole, unlike other known examples
where the dominant term is the constant $\beta_0$ [16].
This peculiarity can be traced to the presence of the
undifferentiated potential $V$ in the inverse propagators,
which is characteristic of gravity (cfr. (3.14)).
The situation would be different if we allowed a nonzero
cosmological constant.

Treating $\gamma_0$, $\beta_{-1}$ and $\beta_0$ as constants
and neglecting integration constants one finds that
$\rho_k^2={1\over2}\gamma_0 k^2$ and
$\lambda_k=2{\beta_{-1}\over\gamma_0}\ln k^2$.
This is the behaviour that one would expect on dimensional grounds.
In this calculation one neglects the running of the coupling constants.
This is a reasonable approximation if one considers
the behaviour of the theory over a range of momenta
which is not too large.
One could take into account the running of $\lambda_k$,
for which the evolution equation is known, but the result would
not be very significant: the other coupling constants
are also expected to run logarithmically, and their (unknown)
contribution could easily overwhelm the one coming from $\lambda_k$.

Assuming that all the couplings appearing in $\gamma_0$ run
logarithmically, the true solution for $\rho_k^2$ would deviate
from the one given above by a sublogarithmic correction.
The validity of the approximation $\rho_k^2\ll k^2$ hinges
on the sign of these corrections. If $\rho_k^2$ grows
slower than $k^2$, the approximation could be justified.
This was the case in the calculation we did in [10].
However, it may not be generally true.

Given that, in general, the validity of the approximation
$\rho_k^2\ll k^2$ is questionable, it would be desirable to say something
on the large--$k$ behaviour without making this assumption.
One general conclusion that can be drawn with reasonable confidence is
that $\rho_k^2$ will be proportional to $k^2$, up to (at most)
logarithmic corrections.
To see this consider again the general form (5.7) of the
functions $\gamma(k)$ and $\beta(k)$.
As mentioned before, the $x$-integration is cut off exponentially
for $x>k^2$, and as a power for $x<k^2$,
so if we are only interested in the dominant behaviour
of the integral, we can replace $P_k$ by $k^2$.
Assume further that $\rho_k^2=ck^2$ for some constant $c$.
Then ${\cal R}(k^2,ck^2)$ is a constant that can be taken out
of the $x$-integration and the functions $\gamma(k)$ and $\beta(k)$
become simply constants. The equation for $\rho_k^2$ becomes an
algebraic equation that implicitly determines the constant $c$,
so the assumption $\rho_k^2=ck^2$ is justified a posteriori.

Let us now consider the opposite limit: $k^2\ll\rho_k^2$.
This is the limit
$k\to 0$, when $\rho_0\not=0$.
In this case we retain in each sum the term with the highest power
of $\rho_k^2$. The contributions of the different spins are not
all of the same order in $k^2/\rho_k^2$. Keeping only the
leading terms, the beta functions reduce to
$$
\gamma(k)=\gamma_2{k^4\over\rho_k^4}+O\!\left({k^6\over\rho_k^6}\right)\ \ ;
\qquad
\beta(k)=\beta_1{k^2\over\rho_k^2}+O\!\left({k^4\over\rho_k^4}\right)\ .
\eqno(5.11)
$$
where
$$
\eqalignno{
&\gamma_2={1\over32\pi^2}{1\over\lambda_k}
\Biggl[5{G_1B_2\over B_1B_2-2B_1^2}
+5{G_2\over B_1}
+3{G_3B_5\over B_3B_5-2B_4^2}
+3{G_4B_8\over B_6B_8-2B_7^2} \cr
&\quad -{G_5B_{10}\over B_9}
+{6G_5B_9(B_{10}B_{11}-B_9B_{12})\over(B_9B_{10}-2B_9^2)^2}
+{3B_0(2B_9-B_{10}-4B_{11}+2B_{12})\over 8\lambda_k(2B_9-B_{10})} &(5.12a)
\cr
&\quad +9{(2B_9-B_{10}-4B_{11}+2B_{12})
   (B_9B_{12}^2-4B_9B_{11}B_{12}+2B_{10}B_{11}^2)
                 \over 8\lambda_k B_9(2B_9-B_{10})^2}
+{G_6\over B_{13}}\Biggr]I_0\ ,\qquad\ \cr
&\beta_1=
{1\over32\pi^2}\left[{10\over B_2-2B_1}
-{5\over 2B_9-B_{10}}+15\alpha\right]
I_{-2}\lambda_k \ .&(5.12b)}
$$
In this regime the running of
$\lambda_k$ and $\rho_k^2$ is damped by powers of
$k^2/\rho_k^2$ and stops for $k\to 0$.
The solutions for small $k$ are
$$
\eqalignno{
\rho_k^2=&\,\rho_0^2
\left[1+{\gamma_2\over6}{k^6\over\rho_0^6}\right]\ ,&(5.13a)\cr
\lambda_k=&\,\lambda_0
\left[1+{\beta_1\over2}{k^2\over\rho_0^2}\right]\ .&(5.13b)\cr}
$$

For generic values of the parameters $g_i$, $a_i$, $b_i$,
all modes except for the graviton are massive,
with masses of the order of Planck's mass. One would expect that
these modes can be neglected when describing the physics
below the Planck scale.
Thus, the running of the dilaton potential at low energies
should be derivable entirely from the action (2.5).
This can be easily checked using the coefficient matrices
given in  (3.12). The renormalization group
equations become for small $k$
$$
\eqalignno{
k{\partial\rho^{\, 2}_k\over\partial k}=&
{3\over32\pi^2}{12g_0-B_0\over 8\lambda_k^2}I_0{k^6\over\rho_k^4}
\ ,&(5.14a)\cr
k{\partial\lambda_k\over\partial k}=&
{15\over32\pi^2}\lambda_k\left(\alpha-{1\over8 g_0}\right)
I_{-2}{k^2\over\rho_k^2}
\ .&(5.14b)}
$$
These have again the solutions (5.16).
Note that the coefficient $\gamma_2$ vanishes when
$g_0=b_0/12$, in which case the action (2.5) is conformally invariant.
(See Appendix A).

\bigskip
\goodbreak
\leftline{\bf 6. Concluding remarks}
\smallskip
\noindent
One attractive feature of the theory we have considered here is
that one can easily describe the transition to Einstein's theory at
sufficiently low energies. Indeed, for generic values of the
coupling constants $g_i$ and $a_i$, all components of the
connection $\igamma_\mu{}^\nu{}_\rho$ are massive, with squared masses of the
form $\rho_0^2 B/G$, where $B$ and $G$ are appropriate combinations
of the coupling constants $a_i$ and $g_i$. As mentioned in the
introduction, this is due to the occurrence of a Higgs phenomenon.
The dilaton becomes massive too, with
a mass squared equal to $\lambda_0\rho_0^2$.
At energies lower than the VEV $\rho_0$, all these particles
decouple, leaving the graviton as the only remnant.
Its effective dynamics is given by the second term in (2.5).
If we assume that the Einstein term of the low energy world
comes entirely from this source, we see that
$\rho_0$ has to be of the order of the Planck mass, $m_P$.

The results of the preceding section can now be summarized as
follows: at scales $k$ much larger than $m_P$, the
coupling constant $\lambda_k$ runs logarithmically and the VEV
$\rho_k$ runs
quadratically, as one would expect on dimensional grounds.
These results are the reflection of the logarithmic and
quadratic divergences that one would encounter if one tried
to remove the ultraviolet cutoff from the theory.
On the other hand for scales $k$ below the Planck mass the
running is suppressed, and both $\lambda_k$ and $\rho_k$ tend
to constants for $k\to 0$.

When we said that there are massive particles in the theory
we used in the mass formula the VEV of $\rho$ at the
scale $k=0$. This is the naive procedure that one would
follow, at tree level, with a classical potential.
However, the running of the VEV of the dilaton affects
the position of the pole of the propagators.
Let us see this first in the case of the dilaton itself.
{}From the parametrization (5.5) of the potential $V_k$,
the running mass $m^2(k)={d^2 V_k\over d\rho^2}\big|_{\rho_k}$
is equal to $\lambda_k\rho_k^2$.
The physical mass of the particle, {\it i.e.} the position
of the pole of the propagator, is defined implicitly
by the condition
$$
m^2_{\rm phys}=m^2(k)\big|_{k^2=m^2_{\rm phys}}\ .\eqno(6.1)
$$
This is equivalent to the statement, which was made in
[10], that in a propagator of the form $(q^2-m^2(k))^{-1}$
the pole is to be found by evaluating the running mass at
$k=|q|$.
Equation (6.1) is solved graphically, by finding the intersection
of the plots of the l.h.s. $y=m_{\rm phys}^2$ and the r.h.s.
$y=\lambda_k\rho_k^2\big|_{k^2=m_{\rm phys}^2}$, as functions of
the independent variable $m_{\rm phys}^2$.
The l.h.s. is represented by the straight line at 45$^0$.
The r.h.s. is the plot of the running of $\rho_k$ that we have
computed in the previous section, multiplied by a factor
$\lambda_0$, up to logarithmic factors that we neglect for the moment.
As we have seen, it starts flat at $\lambda_0\rho_0^2$ and grows
linearly for large $m_{\rm phys}^2$.
If $\lambda_0\ll 1$, the intersection occurs in the region
where the r.h.s. is constant, and therefore the mass squared is
$\lambda_0\rho_0^2\ll m^2_P$.
On the other hand if $\lambda_0$ is of the order or bigger than
one, the intersection is shifted to higher energies.
Exactly where it occurs depends crucially on the logarithmic
corrections. If the leading logarithmic factors in the r.h.s.
appear with a negative power, $m^2_{\rm phys}$ is of the order
$\exp(\lambda_0)\rho_0^2$. If they appear with a positive power,
there may be no intersection at all. In this case, the dilaton
would disappear completely from the spectrum.
Note that an anomalous dimension of $\rho$ would give a power
correction to the running of the mass, so it would be even more
important than the logarithmic corrections in the previous
considerations.

This discussion can be repeated for the connection
$\igamma_\lambda{}^\mu{}_\nu$,
which potentially carries dangerous ghost or tachyon states.
The running of $\rho_k$ could eliminate these states from the spectrum.
That such a mechanism could exist was suggested in [20],
but no concrete support for this idea had been given until now.
In order to draw some definite conclusion one would have to
compute the beta functions for the couplings $g_i$ and $a_i$
for large $k$, find the ultraviolet fixed point, if there is any,
and evaluate (5.9) at that point.

The same discussion can be repeated also for any other matter field.
In this theory the matter is supposed to be coupled to gravity
in such a way that the masses of all particles are proportional
to the VEV of the dilaton. For example, in the case of a scalar
field $\phi$, the action would be
$$
S(\phi,g,\rho)={1\over2}\int d^4x\sqrt{\det g}
\left[g^{\mu\nu}\partial_\mu\phi\partial_\nu\phi
+h\rho^2\phi^2\right]\ ,\eqno(6.2)
$$
where $h$ is a dimensionless coupling constant.
The reason why the known particles have masses much smaller than
Planck's mass would be the smallness of the coupling constant $h$.
For such particles the intersection of the curves
given by the l.h.s. and r.h.s. of (6.1)
occurs in the region where the VEV of the dilaton
is constant, so the poles are exactly where one would expect
to find them.
In GUT theories the expected masses are only a few orders
of magnitude smaller than the Planck scale, so that the
renormalization group corrections envisaged here may
not be negligible.
In the case of a scalar field, this is discussed in a separate
publication [21].

Finally, we mention that the definition (4.3) of the scale dependent
effective potential is not the only possibility.
One could choose a different
normalization of $V_k(\rho)$ such that its minimum
(the cosmological constant) is actually $k$-dependent.
These alternatives could be of relevance for example in cosmological
problems.

\bigskip
\centerline{\bf Acknowledgements}

R.P. wishes to thank G. Veneziano for the kind hospitality and support
at the Theory Division of CERN, where part of this work was done.
We are also grateful to I. Oda and M. Reuter for several discussions, and to
F. Hehl for calling our attention on Ref. [23].
\vfil
\eject

\leftline{\bf Appendix A: Broken Weyl invariance}
\smallskip
\noindent
Let us consider the behaviour of the action (2.2) under Weyl
scalings
$$
g_{\mu\nu}\to\Omega^2g_{\mu\nu}\ \ ,\qquad\qquad
\rho\to\Omega^{-1}\rho\ \ ,\qquad\qquad
\igamma_\lambda{}^\mu{}_\nu\to \igamma_\lambda{}^\mu{}_\nu
+\delta^\mu_\nu\Omega^{-1}\partial_\lambda\Omega\ ,
\eqno(A.1)
$$
where $\Omega$ is a general function of the position.
There is also another possible way in which
a Weyl transformation could act on the connection: it is defined
by requiring that $\igamma$ transforms like the Christoffel symbols of
$g$. This alternative transformation
acts trivially on the torsion tensor, whereas $(A.1)$ acts in a
nontrivial way. We have chosen the transformation $(A.1)$ because
it can be generalized to a local $GL(4)$ transformation.

It is easy to see that the curvature tensor $R_{\mu\nu}{}^\rho{}_\sigma$
is inert under $(A.1)$, so the curvature squared terms are invariant
(as in Yang--Mills theory).
The remaining terms can be made
invariant by choosing
$$
b_0=6(2a_1+a_2+3a_3)\ ,\qquad
b_1=2a_1+a_2+3a_3\ ,\qquad
V(\rho)={1\over2}\lambda\rho^4\ ,\eqno(A.2)
$$
where $\lambda$ is an arbitrary constant.
Note in particular that with the choices leading to (2.5)
this gives $g_0=b_0/12$.

It is sometimes convenient to write the action in a different form.
Define the combinations
$$
\tilde g_{\mu\nu}=\rho^2g_{\mu\nu}\ \ ,\qquad\qquad
\tilde\igamma_\lambda{}^\mu{}_\nu=\igamma_\lambda{}^\mu{}_\nu
+\delta^\mu_\nu\rho^{-1}\partial_\lambda\rho\ ,\eqno(A.3)
$$
which by construction are inert under the transformation $(A.1)$.
We also define the curvature of $\tilde\igamma$,
$\tilde R_{\mu\nu}{}^\rho{}_\sigma=
\partial_\mu\tilde\igamma_\nu{}^\rho{}_\sigma
-\partial_\nu\tilde\igamma_\mu{}^\rho{}_\sigma
+\tilde\igamma_\mu{}^\rho{}_\tau\tilde\igamma_\nu{}^\tau{}_\sigma
-\tilde\igamma_\nu{}^\rho{}_\tau\tilde\igamma_\mu{}^\tau{}_\sigma$,
and the torsion of $\tilde\igamma$,
$\tilde T_\lambda{}^\mu{}_\nu
=\tilde\igamma_\lambda{}^\mu{}_\nu-\tilde\igamma_\nu{}^\mu{}_\lambda$.
The following relations hold:
$$
\eqalignno{
\tilde R_{\mu\nu}{}^\rho{}_\sigma=&\,R_{\mu\nu}{}^\rho{}_\sigma\ ,&(A.4)\cr
\tilde T_\lambda{}^\mu{}_\nu=&\,
T_\lambda{}^\mu{}_\nu+\delta^\mu_\nu\rho^{-1}\partial_\lambda\rho
-\delta^\mu_\lambda\rho^{-1}\partial_\nu\rho\ ,&(A.4b)\cr
\partial_\lambda \tilde g_{\mu\nu}
-\tilde\igamma_\lambda{}^\tau{}_\mu \tilde g_{\tau\nu}
-\tilde\igamma_\lambda{}^\tau{}_\nu \tilde g_{\mu\tau}
=&\,
\rho^2\left(\partial_\lambda g_{\mu\nu}
-\igamma_\lambda{}^\tau{}_\mu g_{\tau\nu}
-\igamma_\lambda{}^\tau{}_\nu g_{\mu\tau}\right)\ .&(A.4c)\cr
}
$$
Note from $(A.4c)$ that if $\rho\not=0$, metricity of one connection guarantees
the metricity of the other.

With the relations $(A.2)$, the action (2.3) can be written
$S(g,\igamma,\rho)=\tilde S(\tilde g,\tilde\igamma)$,
where
$$
\eqalign{
&\tilde S(\tilde g,\tilde\igamma)=
\int d^4x\,\sqrt{\det\tilde g}\Bigl[
{1\over2}\lambda+g_0 \tilde R
+a_1 \tilde T_{\mu\nu\rho}\tilde T^{\mu\nu\rho}
+a_2 \tilde T_{\mu\nu\rho}\tilde T^{\mu\rho\nu}
+a_3 \tilde T^{\rho}\tilde T_{\rho}\cr
&
+g_1 \tilde R_{\mu\nu\rho\sigma}\tilde R^{\mu\nu\rho\sigma}
+g_2 \tilde R_{\mu\nu\rho\sigma}\tilde R^{\mu\rho\nu\sigma}
+g_3 \tilde R_{\mu\nu\rho\sigma}\tilde R^{\rho\sigma\mu\nu}
+g_4 \tilde R_{\mu\nu}\tilde R^{\mu\nu}
+g_5 \tilde R_{\mu\nu}\tilde R^{\nu\mu}
+g_6 \tilde R^2 \Bigr]\cr}\eqno(A.5)
$$
where $\tilde R_{\mu\nu}=\tilde R_{\rho\mu}{}^\rho{}_\nu$,
$\tilde R=\tilde g^{\mu\nu}\tilde R_{\mu\nu}$
and $\tilde T_\nu=\tilde T_\nu{}^\mu{}_\mu$. In this action
indices are raised and lowered with $\tilde g$.
This is the most general action quadratic in $\tilde T$ and $\tilde R$,
and it is invariant under the transformations $(A.1)$ in a trivial
way, since all quantities entering in this action are inert under
those transformations.
It is obtained from (2.2) by choosing the conformal gauge in which
$\rho$ is constant, and rescaling the fields by factors $\rho$.
Conversely, (2.2) with the relations $(A.2)$ is obtained from $(A.5)$
by using the definitions $(A.3)$. Note that the field $\rho$, which
we call the dilaton in this paper, may be called the conformal
factor of the metric $\tilde g$ if one started from the action $(A.5)$.
In the literature on conformal gravity the two actions
(2.3)-$(A.2)$ and $(A.5)$ are said to be written
in the Jordan and Einstein conformal frames, respectively.
They are completely equivalent at the classical level.

Let us now consider the quantization of the theory in the
Weyl-invariant case. We note first of all that when
$(A.2)$ holds, $B_{11}=B_9$ and
$B_{12}=B_{10}$. Thus at the linearized level the new gauge invariance
manifests itself in the proportionality of the second and fourth
rows and columns in $a(0^+)$. Therefore, one has to
fix the gauge also for Weyl transformations.
Assuming that this has been done, it is easy to see that there exists
a quantization procedure that preserves Weyl invariance.
One has to use the form $(A.5)$ for the action,
and define the functional measure by means of the metric $\tilde g$.
In this case the effective action can be written
again as a functional of $\tilde g$ and $\tilde\igamma$ alone, and therefore
is automatically invariant under $(A.1)$ [22].

One is naturally inclined to preserve as much as possible
the classical symmetries in the quantization process, so this choice
of measure may seem to be the only sensible one.
However, this is not the case. Other choices are possible and, from a
certain point of view, may even be more natural: if we interpret the
metric $g$ as the one defining the geometry of spacetime, then
it is natural to use $g$ rather than $\tilde g$ in the definition of the
functional measure, and this leads to a quantum theory in which Weyl
invariance is broken.

In a concrete calculation, the definition of the measure reflects itself
in the definition of the cutoff. We have shown in [10,11],
that if one starts from a Weyl invariant theory and defines the
UV momentum cutoff as $g^{\mu\nu}q_\mu q_\nu<\Lambda^2$
(rather than $\tilde g^{\mu\nu}q_\mu q_\nu<\Lambda^2$), then the effective
action $S_{\rm eff}(g,\igamma,\rho)$ will not have Weyl invariance anymore.

In this paper we have only discussed the renormalization group
flow of the effective action, which does not necessitate the
explicit introduction of an UV cutoff. Still, Weyl invariance
cannot be maintained. This can be seen as follows.
Suppose we study the small fluctuations of the gravitational field
around flat space at some energy scale $k_1$, and suppose that these
are well described by the Weyl invariant action (2.3)-$(A.2)$.
Now suppose we want to know the effective action at some lower
energy scale $k_2$. As discussed in section 4,
this is given by a functional integral over all
fluctuations of the fields which lie in the momentum shell between
$k_1$ and $k_2$. But how is this momentum shell defined? Since we
are postulating that the geometry of spacetime is given by the
dimensionless metric $g$, the shell is defined by
$k_2<g^{\mu\nu}q_\mu q_\nu<k_1$ (for simplicity we are assuming
here a sharp cutoff, but this is by no means essential).
One sees that the definition of the shell introduces in the definition
of the effective action $S_{k_2}$ a dependence on $g$
which is not compensated by a dependence on $\rho$. Unavoidably,
$S_k$ will not be a function of the combinations (2.4) alone.
It will be a genuine functional of $g$, $\igamma$ and $\rho$,
and will not be invariant under infinitesimal Weyl transformations.

To be more specific, suppose that we want to compute only the
scale-dependent effective potential $V_k(\rho)$, which is given by
$S_k=\int d^4x \sqrt{\det g} \,V_k$, for a field of the form (3.1).
As we have mentioned above, Weyl invariance requires the potential
to be purely quartic.
If we assume that $V_{k_1}$ is purely quartic, then unavoidably
$V_{k_2}$ will not be anymore, because the integration
procedure breaks scale invariance.
It is instructive to see this in detail in an explicit calculation.
Consider a single wave operator of the form
${\cal O}=q^2+c\rho^2$ (in principle (3.3) can be written as a sum of
such terms), and suppose the momentum shell is defined by sharp
UV and IR cutoffs. Then we have
$$
\eqalign{
\Delta V=& V_{k_1}-V_{k_2}=
-{1\over2}\int\limits_{q^2=k_2^2}^{\ \ k_1^2}
{d^4q\over(2\pi)^4}\ln\left(q^2+c\rho^2\over q^2+c\rho_{k_1}^2\right)\cr
=&\ {1\over64\pi^2}\biggl[
k_2^4\ln\left({k_2^2+c\rho^2\over k_2^2+c\rho_{k_1}^2}\right)
-k_1^4\ln\left({k_1^2+c\rho^2\over k_1^2+c\rho_{k_1}^2}\right)\cr
&
+c^2\rho^4\ln\left({k_1^2+c\rho^2\over k_2^2+c\rho^2}\right)
-c^2\rho_{k_1}^4\ln\left({k_1^2+c\rho_{k_1}^2\over
k_2^2+c\rho_{k_1}^2}\right)
+c(\rho^2-\rho_{k_1}^2)(k_2^2-k_1^2)\biggr]\ ,\cr}
\eqno(A.6)
$$
which is obviously no longer purely quartic.
It is still a function of $\rho^2$ only.

Let us approximate the potential $V_k$ by a quartic polynomial
of the form
$$
V_k(\rho)={1\over2}m_k^2\rho^2+\lambda_k\rho^4\ .\eqno(A.7)
$$
The constants $m_k^2$ and $\lambda_k$ are to be thought of as
the coefficients of the Taylor expansion of $V_k$ around the origin:
$$
m_k^2={d^2 V_k\over d\rho^2}\Big|_0\ \ ,\qquad
\lambda_k={1\over 24}{d^4 V_k\over d\rho^4}\Big|_0\ .\eqno(A.8)
$$
Then we have
$$
\eqalignno{
\Delta m_k^2=m_{k_1}^2-m_{k_2}^2=&{c\over32\pi^2}(k_2^2-k_1^2)\
,&(A.9a)\cr
\Delta\lambda_k=\lambda_{k_1}-\lambda_{k_2}=&{c^2\over64\pi^2}
\ln{k_1^2\over k_2^2}\ .&(A.9b)\cr}
$$

The most important conclusion of this discussion is that the
potential cannot be consistently assumed to be purely quartic at all scales:
if $m^2$ is zero at some scale, it will be nonzero as soon as one
begins to integrate.
Weyl invariance is broken and a mass term is generated.
There follows that if we want to study the renormalization group
flow of the effective potential taking into account what we called
the ``renormalization group improvement'', we have to assume from the
outset that the potential is not purely quartic.
Of course, if we were to study the whole effective action rather than
just the effective potential we would find many more terms that are
not present in the starting action. For example, instead of
the factors $\rho^2$ in front of the torsion terms there will now
be general functions of $\rho$. In principle, this will also have an
effect on the running of the potential, but we
neglect this effect.

\bigskip
\goodbreak

\leftline{\bf Appendix B: Lorentz decomposition}
\medskip
\noindent
In the main text we have taken as independent coupling constants
the coefficients of all possible contractions of two curvature or
torsion tensors. However, these contractions do not carry any
special geometrical significance. One may expect that the final
results would look simpler if they were expressed in terms of
another set of parameters, which are related to geometrically
significant quantities.
Let us consider the irreducible parts of the torsion and curvature
tensors with respect to the Lorentz group [23]:
$$
T_{\lambda\mu\nu}
=\sum\limits_{i=1}^3{} T^{(i)}_{\lambda\mu\nu}\ \ \ ,
\qquad
R_{\mu\nu\rho\sigma}
=\sum\limits_{i=1}^6{}R^{(i)}_{\mu\nu\rho\sigma}\ ,
\eqno(B.1)
$$
where
$$
\eqalignno{
&T^{(1)}_{\lambda\mu\nu}=
{1\over3}\left(2T_{\lambda\mu\nu}
-T_{\nu\lambda\mu}+T_{\nu\mu\lambda}\right)
+{1\over 3}\left(g_{\lambda\mu}T_\nu
-g_{\nu\mu}T_\lambda\right)\ ,&(B.2a)\cr
&T^{(2)}_{\lambda\mu\nu}=
-{1\over 3}\left(g_{\lambda\mu}T_\nu
-g_{\nu\mu}T_\lambda\right)\ ,&(B.2b)\cr
&T^{(3)}_{\lambda\mu\nu}=
{1\over3}\left(T_{\lambda\mu\nu}
+T_{\nu\lambda\mu}+T_{\mu\nu\lambda}\right)\ ,&(B.2c)\cr}
$$
and
$$
\eqalignno{
&R^{(1)}_{\mu\nu\rho\sigma}=
{1\over6}\left(2R_{\mu\nu\rho\sigma}+2R_{\rho\sigma\mu\nu}
-R_{\mu\rho\sigma\nu}+R_{\mu\sigma\rho\nu}
+R_{\nu\rho\sigma\mu}-R_{\sigma\nu\rho\mu}\right)\cr
&\qquad\quad-{1\over2}\left(g_{\mu\rho}{R^{\rm (ST)}}_{\nu\sigma}
-g_{\nu\rho}{R^{\rm (ST)}}_{\mu\sigma}
-g_{\mu\sigma}{R^{\rm (ST)}}_{\nu\rho}
+g_{\nu\sigma}{R^{\rm (ST)}}_{\mu\rho}\right)\cr
&\hskip 7cm -{1\over12}\left(g_{\mu\rho}g_{\nu\sigma}-
g_{\mu\sigma}g_{\nu\rho}\right)R
\ ,\qquad&(B.3a)\cr
&R^{(2)}_{\mu\nu\rho\sigma}=
{1\over2}\left(R_{\mu\nu\rho\sigma}-R_{\rho\sigma\mu\nu}\right)
-{1\over2}\Bigl(g_{\mu\rho}{R^{\rm (A)}}_{\nu\sigma}
-g_{\nu\rho}{R^{\rm (A)}}_{\mu\sigma}
-g_{\mu\sigma}{R^{\rm (A)}}_{\nu\rho}\cr
&\hskip 9cm +g_{\nu\sigma}{R^{\rm (A)}}_{\mu\rho}\Bigr)\ ,&(B.3b)\cr
&R^{(3)}_{\mu\nu\rho\sigma}=
{1\over6}\left(R_{\mu\nu\rho\sigma}+R_{\rho\sigma\mu\nu}
+R_{\mu\sigma\nu\rho}+R_{\nu\rho\mu\sigma}
+R_{\mu\rho\sigma\nu}+R_{\sigma\nu\mu\rho}\right)
\ ,&(B.3c)\cr
&R^{(4)}_{\mu\nu\rho\sigma}=
{1\over2}\left(g_{\mu\rho}{R^{\rm (ST)}}_{\nu\sigma}
-g_{\nu\rho}{F^{\rm (ST)}}_{\mu\sigma}
-g_{\mu\sigma}{R^{\rm (ST)}}_{\nu\rho}
+g_{\nu\sigma}{R^{\rm (ST)}}_{\mu\rho}\right)\cr
& \hskip 7cm -{1\over12}\left(g_{\mu\rho}g_{\nu\sigma}
-g_{\mu\sigma}g_{\nu\rho}\right)R
\ ,&(B.3d)\cr
&R^{(5)}_{\mu\nu\rho\sigma}=
{1\over2}\left(g_{\mu\rho}{R^{\rm (A)}}_{\nu\sigma}
-g_{\nu\rho}{R^{\rm (A)}}_{\mu\sigma}
-g_{\mu\sigma}{R^{\rm (A)}}_{\nu\rho}
+g_{\nu\sigma}{R^{\rm (A)}}_{\mu\rho}\right)
\ ,&(B.3e)\cr
&R^{(6)}_{\mu\nu\rho\sigma}=
{1\over12}\left(g_{\mu\rho}g_{\nu\sigma}-g_{\mu\sigma}g_{\nu\rho}\right)F
\ ,&(B.3f)\cr}
$$
where ${R^{\rm (ST)}}_{\mu\nu}={1\over2}\left(R_{\mu\nu}+R_{\nu\mu}\right)
-{1\over4}g_{\mu\nu}F$ and
${R^{\rm (A)}}_{\mu\nu}={1\over2}\left(R_{\mu\nu}-R_{\nu\mu}\right)$
are the symmetric traceless and antisymmetric parts of the Ricci
tensor.
These decompositions are orthogonal with respect to the inner
products $(A,B)=\int d^4x\sqrt{\det g}g^{\mu\rho}g^{\nu\sigma}
g^{\lambda\tau}\cdots A_{\mu\nu\lambda\ldots}B_{\rho\sigma\tau\ldots}$.
The quantity (2.6) can then be rewritten in the form
$$
{1\over2}\sum\limits_{i=1}^6
\tilde g_i R^{(i)}_{\mu\nu\rho\sigma}R^{(i)\mu\nu\rho\sigma}
+{1\over2}\rho^2\sum\limits_{i=1}^3
\tilde a_i T^{(i)}_{\lambda\mu\nu}T^{(i)\lambda\mu\nu}
\ .\eqno(B.4)
$$
where $\tilde a_i$ and $\tilde g_i$ is a new set of coupling constants,
related to the coupling constants
$a_i$, $g_i$ appearing in the action (2.2) by the linear transformation
$$
\left[\matrix{\tilde g_1\cr\tilde g_2\cr\tilde g_3\cr
\tilde g_4\cr\tilde g_5\cr\tilde g_6\cr}\right]=
\left[\matrix{
1 &1 &{1/2} &0 &0 &0 \cr
1 &-1 &0 &0 &0 &0 \cr
1 &1 &-1 &0 &0 &0 \cr
1 &2 &{1/2} &{1/2} &{1/2} &0 \cr
1 &-1 &0 &{1/2} &-{1/2} &0 \cr
1 &1 &{1/2} &{3/2} &{3/2} &6 \cr}\right]
\left[\matrix{ g_1\cr g_2\cr g_3\cr
g_4\cr g_5\cr\ g_6\cr}\right]
\ \ \ ,\qquad
\left[\matrix{\tilde a_1\cr\tilde a_2\cr\tilde a_3\cr}\right]=
\left[\matrix{
2 &1 &0 \cr
2 &1 &3 \cr
2 &-2 &0 \cr}\right]
\left[\matrix{a_1\cr a_2\cr a_3\cr}\right]
\eqno(B.5)
$$
Most formulae are more compactly written in terms of the new coupling
constants. This is evident from the following table, which gives
the relation between the coefficients $G_i$ and $B_i$
appearing in (3.11) and the parameters appearing in the Lagrangian:
\medskip
\vbox{
\settabs\+\indent\quad&$4g_1+2g_2+4g_3+4g_4+4g_5+12 g_6$\ & \cr %sample line
\+$G_1$=&$4g_1+2g_2+4g_3+g_4+g_5$&$=(1/2)\tilde g_1+\tilde g_4\ ,$\cr
\+$G_2$=&$4g_1+g_2$&$=\tilde g_1+\tilde g_2\ ,$\cr
\+$G_3$=&$4g_1-4g_3+g_4-g_5$&$=\tilde g_2+\tilde g_5\ ,$\cr
\+$G_4$=&$4g_1+g_2+2g_4$&$=\tilde g_4+\tilde g_5\ ,$\cr
\+$G_5$=&$4g_1+2g_2+4g_3+4g_4+4g_5+12 g_6
    $&$=2\tilde g_6+8\tilde g_4-5\tilde g_1\ ,$\cr
\+$G_6$=&$4g_1-2g_2$&$=(1/3)\tilde g_1+\tilde g_2\ ,$\cr
\+$B_1$=&$2a_1+a_2+g_0$&$=\tilde a_1+g_0\ ,$\cr
\+$B_2$=&$4a_1+2a_2$&$=2\tilde a_1\ ,$\cr
\+$B_3$=&$6a_1-5a_2-g_0$&$=(1/3)(\tilde a_1+8\tilde a_3)-g_0\ ,$\cr
\+$B_4$=&$2a_1-3a_2-g_0$&$=(1/3)(4\tilde a_3-\tilde a_1)-g_0\ ,$\cr
\+$B_5$=&$4a_1-2a_2$&$=(1/3)(2\tilde a_1+4\tilde a_3)\ ,$\cr
\+$B_6$=&$2a_1+a_2+2a_3-g_0$&$=(1/3)(\tilde a_1+2\tilde a_2)-g_0\ ,$\cr
\+$B_7$=&$a_3-g_0$&$=(1/3)(\tilde a_2-\tilde a_1)-g_0\ ,$\cr
\+$B_8$=&$2a_1+a_2+a_3$&$=(1/3)(2\tilde a_1+\tilde a_2)\ ,$\cr
\+$B_9$=&$2a_1+a_2+3a_3-2g_0$&$=\tilde a_2-2g_0\ ,$\cr
\+$B_{10}$=&$4a_1+2a_2+6a_3$&$=2\tilde a_2\ ,$\cr
\+$B_{11}$=&$b_1-2g_0$&$=b_1-2g_0\ ,$\cr
\+$B_{12}$=&$2b_1$&$=2b_1\ ,$\cr
\+$B_{13}$=&$8a_1-8a_2-2g_0$&$=4\tilde a_3-2g_0\ .$\cr
}
\vfil
\eject

\leftline{\bf Appendix C: Spin-projector operators}
\medskip
\noindent
For completeness, we list in this appendix the explicit expressions
of the spin-projector operators $P_{ij}^{AB}(J^{\cal P})$
that have been used to rewrite the action (3.3) into the form (3.10).
There are in fact some differences with respect to those in [10],
due to the different spin-parity content of the fields appearing here.

For fixed spin $J$ and parity $\cal P$, these operators are labelled
by the indices $A$, $B$ that identify the fields $\omega$, $\varphi$ and
$\sigma$, and by $i$, $j$ that identify
isomorphic Lorentz representations occurring more than once.
For example for spin-parity $2^+$,
$i=1,2$; for $1^-$, $i=1,2,3$ {\it etc}. The operators
$P_{ii}^{AA}(J^{\cal P})$ project out of a field a given
irreducible representation, while the intertwiners $P_{ij}^{AB}(J^{\cal P})$
(with $i\neq j$) give isomorphisms between the different
representations occurring more than once. (Note that the indices $A$,
$B$ in these projectors are redundant since $i$, $j$ already label
the representations. It is nevertheless convenient to keep them in order
to remember by what field a certain representation is carried,
{\it e.g.} for $J^{\cal P}=1^-$ the representations $i=1$ and $i=2$
are carried by $\omega$ and $i=3$ is carried by $\varphi$).
The operators $P_{ij}^{AB}(J^{\cal P})$
are orthonormal and complete:
$$
\eqalign{&P_{ij}^{AB}(J^{\cal P})\cdot P_{kl}^{CD}(I^{\cal Q})=
\delta_{IJ}\,\delta_{\cal PQ}\,\delta_{jk}\,\delta_{BC}\
P_{il}^{AD}(J^{\cal P})\ ,\cr
            &\sum_{J,{\cal P},A,i} P_{ii}^{AA}(J^{\cal P})={\bf 1}\ .}
\eqno(C.1)
$$
It is useful to introduce the following notations:
$$ {\hat q}^\mu=q^\mu/\sqrt{q^2}\ ,\qquad \l_\mu^\nu={\hat q}_\mu{\hat q}^\nu\
,
\qquad \t_\mu^\nu=\delta_\mu^\nu-\l_\mu^\nu\ ,\eqno(C.2)$$
obeying the relations:
$$ \l^\mu_\nu\, \t_\mu^\rho=\, 0\ ,\qquad \t_\mu^\nu \t_\nu^\rho=\l_\mu^\rho\ ,
\qquad \l_\mu^\nu \l_\nu^\rho=\l_\mu^\rho\ .\eqno(C.3)$$
In terms of ${\hat q}^\mu$, $\l_\mu^\nu$, and $\t_\mu^\nu$, one finds:
$$ \eqalignno{
&\left[P(2^+)\right]=\left[\matrix{
\left[P_{11}^{\omega\omega}(2^+)\right]_{\tau\rho\sigma}{}^{\alpha\beta\gamma}
&
\left[P_{12}^{\omega\varphi}(2^+)\right]_{\tau\rho\sigma}{}^{\alpha\beta}\cr
\left[P_{21}^{\varphi\omega}(2^+)\right]_{\rho\sigma}{}^{\alpha\beta\gamma} &
\left[P_{22}^{\varphi\varphi}(2^+)\right]_{\rho\sigma}
{}^{\alpha\beta}}\right]\ ,\cr
&\null\cr
& \left[P_{11}^{\omega\omega}(2^+)\right]_{\tau\rho
                                       \sigma}{}^{\alpha\beta\gamma}=
  \t_\tau^\alpha\, \t_{[\rho}^{[\beta}\, \l_{\sigma]}^{\gamma]}+
  \t_\tau^{[\gamma}\, \l_{[\rho}^{\beta]}\, \t_{\sigma]}^\alpha-
  {2\over3}\, \t_{\tau[\rho}\, \l_{\sigma]}^{[\gamma}\, \t^{\beta]\alpha}\ ,\cr
& \left[P_{12}^{\omega\varphi}(2^+)\right]_{\tau\rho\sigma}{}^{\alpha\beta}=
  \sqrt{2}\, \t_\tau^{(\alpha}\, \t_{[\rho}^{\beta)}\, \q_{\sigma]}-
  {\sqrt{2}\over3}\, \t^{\alpha\beta}\, \t_{\tau[\rho}\, \q_{\sigma]}\ ,\cr
& \left[P_{21}^{\varphi\omega}(2^+)\right]_{\rho\sigma}{}^{\alpha\beta\gamma}=
  \sqrt{2}\, \t_{(\rho}^\alpha\,\t_{\sigma)}^{[\beta}\, \q^{\gamma]}-
  {\sqrt{2}\over3}\, \t_{\rho\sigma}\, \t^{\alpha[\beta}\, \q^{\gamma]}\ ,\cr
& \left[P_{22}^{\varphi\varphi}(2^+)\right]_{\rho\sigma}{}^{\alpha\beta}=
  \t_{(\rho}^{(\alpha}\, \t_{\sigma)}^{\beta)}-{1\over3}\, \t_{\rho\sigma}\,
  \t^{\alpha\beta}\ ,\cr
&\null\cr
&\null\cr
&\left[P^{\omega\omega}(2^-)\right]_{\tau\rho\sigma}{}^{\alpha\beta\gamma}=
 {2\over3}\,\t_\tau^\alpha\, \t_{[\rho}^{[\beta}\, \t_{\sigma]}^{\gamma]}+
 {2\over3}\,\t_\tau^{[\gamma}\, \t_{[\rho}^{\beta]}\, \t_{\sigma]}^\alpha-
 \t_{\tau[\rho}\, \t_{\sigma]}^{[\gamma}\, \t^{\beta]\alpha}\ ,\cr
}$$
\vfill
\eject
%
%&\null\cr
%&\null\cr
%
$$\eqalign{& \left[P(1^+)\right]=\left[\matrix{
\left[P^{\omega\omega}_{11}(1^+)\right]_{\tau\rho\sigma}{}^{\alpha\beta\gamma}&
\left[P^{\omega\omega}_{12}(1^+)\right]_{\tau\rho\sigma}{}^{\alpha\beta\gamma}
\cr
\left[P^{\omega\omega}_{21}(1^+)\right]_{\tau\rho\sigma}{}^{\alpha\beta\gamma}&
\left[P^{\omega\omega}_{22}(1^+)\right]_{\tau\rho\sigma}{}^{\alpha\beta\gamma}
\cr}\right]
\ ,\cr
&\null\cr
&\left[P^{\omega\omega}_{11}(1^+)\right]_{\tau\rho\sigma}{}^{\alpha\beta\gamma}
 =\t_\tau^\alpha\,\t_{[\rho}^{[\beta}\,\l_{\sigma]}^{\gamma]}-
  \t_\tau^{[\gamma}\,\l_{[\rho}^{\beta]}\,\t_{\sigma]}^\alpha\ ,\cr
&\left[P^{\omega\omega}_{12}(1^+)\right]_{\tau\rho\sigma}{}^{\alpha\beta\gamma}
 =\sqrt{2}\,\t_\tau^{[\gamma}\,\t _{[\rho}^{\beta]}\, \l_{\sigma]}^\alpha\ ,\cr
&\left[P^{\omega\omega}_{21}(1^+)\right]_{\tau\rho\sigma}{}^{\alpha\beta\gamma}
 =\sqrt{2}\,\l_\tau^{[\gamma}\,\t _{[\rho}^{\beta]}\, \t_{\sigma]}^\alpha\ ,\cr
&\left[P^{\omega\omega}_{22}(1^+)\right]_{\tau\rho\sigma}{}^{\alpha\beta\gamma}
 =\l_\tau^\alpha\,\t_{[\rho}^{[\beta}\,\t_{\sigma]}^{\gamma]}\ ,\cr
&\null\cr
&\null\cr
& \left[P(1^-)\right]=\left[\matrix{
\left[P^{\omega\omega}_{11}(1^-)\right]_{\tau\rho\sigma}{}^{\alpha\beta\gamma}&
\left[P^{\omega\omega}_{12}(1^-)\right]_{\tau\rho\sigma}{}^{\alpha\beta\gamma}&
\left[P^{\omega\varphi}_{13}(1^-)\right]_{\tau\rho\sigma}{}^{\alpha\beta}\cr
\left[P^{\omega\omega}_{21}(1^-)\right]_{\tau\rho\sigma}{}^{\alpha\beta\gamma}&
\left[P^{\omega\omega}_{22}(1^-)\right]_{\tau\rho\sigma}{}^{\alpha\beta\gamma}&
\left[P^{\omega\varphi}_{23}(1^-)\right]_{\tau\rho\sigma}{}^{\alpha\beta}\cr
\left[P^{\varphi\omega}_{31}(1^-)\right]_{\rho\sigma}{}^{\alpha\beta\gamma}&
\left[P^{\varphi\omega}_{32}(1^-)\right]_{\rho\sigma}{}^{\alpha\beta\gamma}&
\left[P^{\varphi\varphi}_{33}(1^-)\right]_{\rho\sigma}{}^{\alpha\beta}\cr}
\right]\ ,\cr
&\null\cr
&\left[P^{\omega\omega}_{11}(1^-)\right]_{\tau\rho\sigma}{}^{\alpha\beta\gamma}
 =\t_{\tau[\rho}\,\t_{\sigma]}^{[\gamma}\,\t^{\beta]\alpha}\ ,\cr
&\left[P^{\omega\omega}_{12}(1^-)\right]_{\tau\rho\sigma}{}^{\alpha\beta\gamma}
 =\sqrt{2}\,\l^{\alpha[\beta}\,\t_{\sigma]}^{[\gamma}\,\t_{\rho]\tau}\ ,\cr
&\left[P^{\omega\varphi}_{13}(1^-)\right]_{\tau\rho\sigma}{}^{\alpha\beta}=
 \sqrt{2}\,\t_{\tau[\rho}\,\t_{\sigma]}^{(\alpha}\,\q^{\beta)}\ ,\cr
&\left[P^{\omega\omega}_{21}(1^-)\right]_{\tau\rho\sigma}{}^{\alpha\beta\gamma}
 =\sqrt{2}\,\l_{\tau[\rho}\,\t_{\sigma]}^{[\gamma}\,\t^{\beta]\alpha}\ ,\cr
&\left[P^{\omega\omega}_{22}(1^-)\right]_{\tau\rho\sigma}{}^{\alpha\beta\gamma}
 =2\,\l_\tau^\alpha\,\l_{[\rho}^{[\beta}\,\t_{\sigma]}^{\gamma]}\ ,\cr
&\left[P^{\omega\varphi}_{23}(1^-)\right]_{\tau\rho\sigma}{}^{\alpha\beta}
 =2\, \q_\tau\, \l_{[\rho}^{(\alpha}\,\t_{\sigma]}^{\beta)}\ ,\cr
&\left[P^{\varphi\omega}_{31}(1^-)\right]_{\rho\sigma}{}^{\alpha\beta\gamma}=
 \sqrt{2}\,\t^{\alpha[\beta}\,\t_{(\rho}^{\gamma]}\,\q_{\sigma)}\ ,\cr
&\left[P^{\varphi\omega}_{32}(1^-)\right]_{\rho\sigma}{}^{\alpha\beta\gamma}=
 2\,\q^\alpha\,\l_{(\rho}^{[\beta}\,\t_{\sigma)}^{\gamma]}\ ,\cr
&\left[P^{\varphi\varphi}_{33}(1^-)\right]_{\rho\sigma}{}^{\alpha\beta}=
 2\, \t_{(\rho}^{(\alpha}\,\l_{\sigma)}^{\beta)}\ ,\cr
}$$
\vfill
\eject
%
%&\null\cr
%&\null\cr
%
$$\eqalign{& \left[P(0^+)\right]=\left[\matrix{
\left[P^{\omega\omega}_{11}(0^+)\right]_{\tau\rho\sigma}{}^{\alpha\beta\gamma}
&
\left[P^{\omega\varphi}_{12}(0^+)\right]_{\tau\rho\sigma}{}^{\alpha\beta} &
\left[P^{\omega\varphi}_{13}(0^+)\right]_{\tau\rho\sigma}{}^{\alpha\beta} &
\left[P^{\omega\sigma}_{14}(0^+)\right]_{\tau\rho\sigma} \cr
\left[P^{\varphi\omega}_{21}(0^+)\right]_{\rho\sigma}{}^{\alpha\beta\gamma} &
\left[P^{\varphi\varphi}_{22}(0^+)\right]_{\rho\sigma}{}^{\alpha\beta} &
\left[P^{\varphi\varphi}_{23}(0^+)\right]_{\rho\sigma}{}^{\alpha\beta} &
\left[P^{\varphi\sigma}_{24}(0^+)\right]_{\rho\sigma} \cr
\left[P^{\varphi\omega}_{31}(0^+)\right]_{\rho\sigma}{}^{\alpha\beta\gamma} &
\left[P^{\varphi\varphi}_{32}(0^+)\right]_{\rho\sigma}{}^{\alpha\beta} &
\left[P^{\varphi\varphi}_{33}(0^+)\right]_{\rho\sigma}{}^{\alpha\beta} &
\left[P^{\varphi\sigma}_{34}(0^+)\right]_{\rho\sigma}\cr
\left[P^{\sigma\omega}_{41}(0^+)\right]^{\alpha\beta\gamma} &
\left[P^{\sigma\varphi}_{42}(0^+)\right]^{\alpha\beta} &
\left[P^{\sigma\varphi}_{43}(0^+)\right]^{\alpha\beta} &
\left[P^{\sigma\sigma}_{44}(0^+)\right] \cr}
\right]\ ,\cr
&\null\cr
&\left[P^{\omega\omega}_{11}(0^+)\right]_{\tau\rho\sigma}{}^{\alpha\beta\gamma}
 ={2\over3}\, \t_{\tau[\rho}\, \l_{\sigma]}^{[\gamma}\, \t^{\beta]\alpha}\ ,\cr
&\left[P^{\omega\varphi}_{12}(0^+)\right]_{\tau\rho\sigma}{}^{\alpha\beta}=
 {\sqrt{2}\over3}\, \t^{\alpha\beta}\, \t_{\tau[\rho}\,\q_{\sigma]}\ ,\cr
&\left[P^{\omega\varphi}_{13}(0^+)\right]_{\tau\rho\sigma}{}^{\alpha\beta}=
 \sqrt{2\over3}\,\l^{\alpha\beta}\,\t_{\tau[\rho}\,\q_{\sigma]}\ ,\cr
&\left[P^{\omega\sigma}_{14}(0^+)\right]_{\tau\rho\sigma}=
 \sqrt{2\over3}\,\q_{[\sigma}\,\t_{\rho]\tau}\ ,\cr
&\left[P^{\varphi\omega}_{21}(0^+)\right]_{\rho\sigma}{}^{\alpha\beta\gamma}=
 {\sqrt{2}\over3}\, \t_{\rho\sigma}\,\t^{\alpha[\beta}\,\q^{\gamma]}\ ,\cr
&\left[P^{\varphi\varphi}_{22}(0^+)\right]_{\rho\sigma}{}^{\alpha\beta}=
 {1\over3}\,\t_{\rho\sigma}\,\t^{\alpha\beta}\ ,\cr
&\left[P^{\varphi\varphi}_{23}(0^+)\right]_{\rho\sigma}{}^{\alpha\beta}=
 {1\over\sqrt{3}}\, \t_{\rho\sigma}\,\l^{\alpha\beta}\ ,\cr
&\left[P^{\varphi\sigma}_{24}(0^+)\right]_{\rho\sigma}=
 {1\over\sqrt{3}}\, \t_{\rho\sigma}\ ,\cr
&\left[P^{\varphi\omega}_{31}(0^+)\right]_{\rho\sigma}{}^{\alpha\beta\gamma}=
 \sqrt{2\over3}\,\l_{\rho\sigma}\, \t^{\alpha[\beta}\, \q^{\gamma]}\ ,\cr
&\left[P^{\varphi\varphi}_{32}(0^+)\right]_{\rho\sigma}{}^{\alpha\beta}=
 {1\over\sqrt{3}}\, \l_{\rho\sigma}\, \t^{\alpha\beta}\ ,\cr
&\left[P^{\varphi\varphi}_{33}(0^+)\right]_{\rho\sigma}{}^{\alpha\beta}=
 \l_{\rho\sigma}\, \l^{\alpha\beta}\ ,\cr
&\left[P^{\varphi\sigma}_{34}(0^+)\right]_{\rho\sigma}=
 \l_{\rho\sigma}\ ,\cr
&\left[P^{\sigma\omega}_{41}(0^+)\right]^{\alpha\beta\gamma}=
 \sqrt{2\over3}\,\q^{[\gamma}\,\t^{\beta]\alpha}\ ,\cr
&\left[P^{\sigma\varphi}_{42}(0^+)\right]^{\alpha\beta}=
 {1\over\sqrt{3}}\, \t^{\alpha\beta}\ ,\cr
&\left[P^{\sigma\varphi}_{43}(0^+)\right]^{\alpha\beta}=
 \l^{\alpha\beta}\ ,\cr
&\left[P^{\sigma\sigma}_{44}(0^+)\right]=1 \ ,\cr
&\null\cr
&\null\cr
&\left[P^{\omega\omega}(0^-)\right]_{\tau\rho\sigma}{}^{\alpha\beta\gamma}=
{1\over3}\,\t_\tau^\alpha\, \t_{[\rho}^{[\beta}\, \t_{\sigma]}^{\gamma]}-
{2\over3}\,\t_\tau^{[\gamma}\, \t_{[\rho}^{\beta]}\, \t_{\sigma]}^\alpha\equiv
\t_{[\tau}^{[\alpha}\, \t_\rho^\beta\, \t_{\sigma]}^{\gamma]}\ ,\cr
}$$

\centerline{\bf References}
\bigskip
\noindent
\item{1.} S. Weinberg, in ``General Relativity: an Einstein Centenary
Survey'', ed. S. Hawking and W. Israel, Cambridge University Press
(1986);\hfil\break
J.F. Donoghue, ``General relativity as an effective
field theory: the leading quantum correction'', Univ. Mass.
UMHEP-408 (gr-qc/9405057).
\smallskip
\item{2.} A.D. Sakharov, Dokl. Akad. Nauk, SSR {\bf 177}, 70 (1967);
\hfil\break
K. Akama, Y. Chikashige, T. Matsuki and H. Terazawa, Progr. Theor.
Phys. {\bf 60}, 868 (1978); \hfil\break
B. Hasslacher and E. Mottola, Phys. Lett. {\bf 95 B}, 237 (1980);
\hfil\break
S. Adler, Rev. Mod. Phys. {\bf 54}, 729 (1982);\hfil\break
D. Amati and G. Veneziano, Nucl. Phys. {\bf B 204}, 451
(1982); \hfil\break
I.L. Buchbinder and S.D. Odintsov, Class and Quantum Grav. {\bf 2},
721 (1985);\hfil\break
H. Terazawa, in the A.D. Sakharov memorial volume, ed. L.V. Keldysh et
al., Nauka, Moscow (1991);\hfil\break
S.D. Odintsov and I.L. Shapiro, Class. and Quantum Grav. {\bf 9},
(1992).
\smallskip
\item{3.} J. Gasser and H. Leutwyler, Ann. of Phys. (NY) {\bf 158},
142 (1984);\hfil\break
H. Leutwyler, Ann. of Phys. (NY) {\bf 235}, 165 (1994).
\smallskip
\item{4.} K.S. Stelle, Phys. Rev. {\bf D 16}, 953 (1977).
\smallskip
\item{5.} I.L. Buchbinder, S.D. Odintsov and I.L. Shapiro,
{\it Effective Action in Quantum Gravity}, (IOP, Bristol (UK), 1992).
\smallskip
\item{6.} R. Utiyama, Phys. Rev. {\bf 101}, 1597 (1956); Progr. Theor.
Phys. {\bf 64}, 2207 (1980);\hfil\break
T.W.B. Kibble, J. Math. Phys. {\bf 2}, 212 (1961);\hfil\break
D.W. Sciama, in ``Recent developments in General Relativity'',
Infeld Festschrift, Pergamon, Oxford (1962);\hfil\break
F.W. Hehl, P. von der Heyde, G.D. Kerlick and J.M. Nester,
Rev. Mod. Phys. {\bf 48}, 393 (1976);\hfil\break
L.L. Smalley, Phys. Lett. {\bf 61 A}, 436 (1977);\hfil\break
K. Hayashi and Y. Shirafuji, Prog. Theor. Phys. {\bf 64}, 866, 883,
1435, 2222 (1980);\hfil\break
D. Ivanenko and G. Sardanashvily, Phys. Rep. {\bf 94}, 1 (1983).
\smallskip
\item{7.} R. Percacci, Phys. Lett. {\bf 144 B}, 37 (1984);
Nucl. Phys. {\bf B 353}, 271 (1991).
\smallskip
\item{8.} E. Sezgin and P. van Nieuwenhuizen, Phys. Rev. D {\bf
21}, 3269 (1980);\hfil\break
E. Sezgin, Phys. Rev. D {\bf 24}, 1677 (1981);\hfil\break
R. Kuhfuss and J. Nitsch, Gen. Rel. and Grav. {\bf 18}, 947 (1986).
\smallskip
\item{9.} N. Nakanishi and I. Ojima, {\it Covariant operator formalism
of gauge theory and quantum gravity}, (World Scientific, Singapore,
1990).
\smallskip
\item{10.} R. Floreanini and R. Percacci, ``Average effective
potential for the conformal factor'', Nucl. Phys. {\bf B}, to appear.
\smallskip
\item{11.} R. Floreanini and R. Percacci, ``A multiplicative background
field method'', in the volume in honor of D. Ivanenko,
V. Koloskov, ed., to appear.
\smallskip
\item{12.} K.J. Barnes, Ph.D. Thesis (1963) unpublished;\hfil\break
R.J. Rivers, Nuovo Cim. {\bf 34} 387 (1964).
\smallskip
\item{13.} K.G. Wilson and J.B. Kogut, Phys. Rep. {\bf 12C}, 75
(1974);\hfil\break
K.G. Wilson, Rev. Mod. Phys. {\bf 47}, 774 (1975).
\smallskip
\item{14.} J. Polchinski, Nucl. Phys. {\bf B 231}, 269 (1984).
\smallskip
\item{15.} B. Warr, Ann. of Phys (NY) {\bf 183}, 1 (1988) \hfil\break
C. Becchi, ``On the construction of renormalized quantum field theory
using renormalization group techniques'', in {\it Elementary
Particles, Field Theory and Statistical Mechanics},
M. Bonini, G. Marchesini and E. Onofri, eds.,
University of Parma (1993);\hfil\break
M. Bonini, M. D'Attanasio and G. Marchesini, Nucl. Phys. {\bf B 418},
81 (1994); {\it ibid.} {\bf B 421},
429 (1994); ``BRS symmetry for Yang--Mills theory with exact
renormalization group'', Parma Preprint, UPRF 94-412 (1994).
\smallskip
\item{16.} C. Wetterich, Nucl. Phys. {\bf B 334}, 506 (1990);
{\it ibid.} {\bf B 352}, 529 (1991);
Z. Phys. C {\bf 57}, 451 (1993); {\it ibid.} {\bf C 60}, 461 (1993);
\hfil\break
M. Reuter and C. Wetterich, Nucl. Phys. {\bf B 391},
147 (1993).
\smallskip
\item{17.} M. Reuter and C. Wetterich, Nucl. Phys. {\bf B 417}, 181 (1994);
{\it ibid.} {\bf B 427}, 291 (1994);
\smallskip
\item{18.} C. Ford, D.R.T. Jones, P.W. Stephenson and
M.B. Einhorn, Nucl. Phys. {\bf B 395}, 17 (1993).
\smallskip
\item{19.} F. Wegner and A. Houghton, Phys. Rev. {\bf A 8},
401 (1973);\hfil\break
A. Hasenfratz and P. Hasenfratz, Nucl. Phys. {\bf B 270}, 685 (1986);
\hfil\break
C. Wetterich, Phys. Lett. {\bf B 301}, 90 (1993);\hfil\break
S.B. Liao and J. Polonyi, ``Renormalization group and universality'',
Duke TH-94-64.
\smallskip
\item{20.} J. Julve and M. Tonin, Nuovo Cimento {\bf 46 B}, 137
(1978);
\hfil\break
A. Salam and J. Strathdee, Phys. Rev. D {\bf 18}, 4480 (1978).
\smallskip
\item{21.} L. Griguolo and R. Percacci, in preparation.
\smallskip
\item{22.} N.C. Tsamis and R.P. Woodard, Ann. of Phys. {\bf 168},
457 (1986);\hfil\break
E.T. Tomboulis, Nucl. Phys. {\bf B 329}, 410 (1990).
\smallskip
\item{23.} P. Baekler, F.W. Hehl and H.J. Lenzen, in {\it Proceedings
of the third Marcel Grossmann meeting on General Relativity},
H. Ning, ed., (North Holland, Amsterdam, 1983).

\bye

\def\tr{{\rm tr}}
\def\thetacl{\theta_{\rm (cl)}{}}
\def\acl{A_{\rm (cl)}{}}
\def\fcl{F_{\rm (cl)}{}}
\def\mcl{M_{\rm (cl)}{}}
\def\xcl{X_{\rm (cl)}{}}
\def\ncl{N_{\rm (cl)}{}}
\def\o{{\cal O}}
\def\btheta{\bar\theta}
\def\igamma{{\mit\Gamma}}
\font\titlefont=cmbx10 scaled\magstep1
\magnification=\magstep1

\null
\vskip 1.5cm
\centerline{\titlefont MEAN-FIELD QUANTUM GRAVITY}
\smallskip
\vskip 1.5cm
\centerline{\bf R. Floreanini}
\smallskip
\centerline{Istituto Nazionale di Fisica Nucleare, Sezione di Trieste}
\centerline{Dipartimento di Fisica Teorica, Universit\`a di Trieste}
\centerline{Strada Costiera 11, 34014 Trieste, Italy}
\bigskip\smallskip
\centerline{\bf R. Percacci}
\smallskip
\centerline{International School for Advanced Studies, Trieste, Italy}
\centerline{and}
\centerline{Istituto Nazionale di Fisica Nucleare, Sezione di Trieste}
\vskip 1.8cm
\centerline{\bf Abstract}
\smallskip\midinsert\narrower\narrower\noindent
We describe a new approach to quantum gravity, based on a kind of mean-field
approximation. The action, which we choose to be
quadratic in curvature and torsion, is made
polynomial by replacing the inverse vierbein by its mean value.
This action is used to compute the effective action for the vierbein
and hence its vacuum expectation value.
Self-consistency is then enforced by requiring
that this vacuum expectation value be proportional to the mean field.
We have explicitly carried out this self-consistent procedure
at one-loop in the case of a mean field corresponding to Minkowski space,
de Sitter space and in the long wavelength limit
for a generic space. General Relativity is recovered as a low energy
approximation.
\endinsert
\vskip 1cm
\vfil\eject

\beginsection {1. INTRODUCTION}

{}From the point of view of Elementary Particle Physics, Einstein's
theory of gravity has many features in common with non-linear chiral
models of QCD: both theories have derivative couplings, nonpolynomial
interactions, a dimensionful coupling constant, are not renormalizable a.s.o.
[1].
All this originates from more basic similarities at the
kinematical level. In fact, it has been known for a long time that the
metric carries a nonlinear realization of the group $GL(4)$, linear
with respect to the Lorentz subgroup $O(1,3)$, and therefore
is similar to the field variables of a nonlinear sigma model with
values in the coset space $GL(4)/O(1,3)$ [2].
Another fruitful analogy is the one between gravity and Yang-Mills theory.
This is particularly striking in the vierbein formalism and in first
order formulations, where the Lorentz connection is an independent
dynamical variable and the theory is invariant under local Lorentz
transformations, in addition to coordinate transformations [3].

At first sight these two analogies have little in common and one may
think that only one of them can be pursued at the time. However, this
is not so. In fact, one can give a locally-$GL(4)$-invariant reformulation
of General Relativity which makes both analogies apparent [4-6]. The best
analogy is then between General Relativity and a chiral model in which
the flavor group has been gauged. In this formulation,
in addition to a $GL(4)$ connection, there are two nonlinear fields:
the soldering form and an internal Lorentzian metric.
The nonlinearity arises from the constraints that the soldering form
be nondegenerate and the internal metric has Lorentzian signature.
Either one of these fields (but not both at the same time) can be gauged away,
leaving us with General Relativity either in metric or in vierbein
formulation. Without the soldering form, the theory
would describe a gauged $GL(4)/O(1,3)$-valued nonlinear sigma model.
So this is more than just an analogy: one may say that gravity is very
literally a soldered gauged sigma model.

If one takes this point of view seriously, one is led to believe that
General Relativity should not be quantized, but rather be regarded
as a low energy limit of some more fundamental theory. Now, the
dynamical variables of QCD (quarks and gluons) are different from
those of the chiral models (mesons): the former are described by
spinor and vector fields, the latter by non-linear scalar fields.
In the same way it may well be that the fundamental variables
underlying gravity could have little or nothing to do with the metric,
vierbein or connection. This possibility has been discussed for some
time in the so-called ``induced gravity'' programme, where the Einstein
action was seen as part of the effective action of some matter fields
[7,8]. String and membrane theories also go in this direction.
Our attempt here will be less radical: we shall assume that the familiar
objects which appear in General Relativity (metric, vierbein, connection)
are indeed fundamental variables. However, we shall try to go beyond
the picture of an effective induced theory.

A nonlinear sigma model can always be regarded as a linear Higgs model
with some constraint of the form $\Phi^a\Phi^a={\rm const}$,
which forces the field to lie in an orbit of the gauge group. This
constraint can be regarded as the effect of a Higgs potential in the
strong coupling limit. A similar picture can be applied also to gravity,
but with an important difference.
In the usual models considered in particle physics, the constraints
are holonomic and therefore the orbits of the
gauge group are lower dimensional submanifolds of the space of
the linear Higgs fields.
On the contrary in gravity the constraints are of the anholonomic
type and therefore the orbits corresponding to the nonlinear
bosons are open subsets in certain tensor spaces.
This geometrical fact means in more physical terms that
while in particle physics the linear theory has more degrees of
freedom than the nonlinear one, in the case of gravity the linear
(unconstrained) theory has the same number of degrees of freedom of
the nonlinear (constrained) theory.

Thus, if we were able to construct a theory of gravity in
which the soldering form and the metric were not constrained a priori
to be nondegenerate, then without having to introduce new degrees of freedom
this theory would be more akin to a Higgs model than to a nonlinear
sigma model [9]. Its quantum properties would presumably
be improved and even if was not to be regarded as a fundamental
theory, it would probably have the same status of the standard model
of Particle physics. As an additional bonus, a theory of this type
would provide a natural framework for the unification of gravity with
the other interactions [10,6].

Our aim in this paper is to discuss the construction and quantization
of such a ``Higgs-like'' or ``unconstrained'' theory of gravity, and its
relation to the usual theory of General Relativity. In order to
simplify the discussion we will assume that the internal metric is
nondegenerate and choose the $GL(4)$ gauge so that it is equal to the
Minkowski metric $\eta_{ab}$. Furthermore we will also assume that the
$GL(4)$ connection is metric, i.e. reduces to an $O(1,3)$ connection.
Thus we will effectively work within the framework of the vierbein
formulation, but we will not assume a priori that the vierbein is
nondegenerate, nor that torsion is zero.

In the construction of our model we try to follow as much as
possible the example of the Higgs model as used in elementary particle
and condensed matter physics.
In doing so we encounter two different but related difficulties.
First of all if the metric is allowed to become degenerate, then it is
impossible to define its inverse, which is needed in the Lagrangian
for contracting covariant indices (e.g. on derivatives).
The other difficulty has to do with the construction of the potential,
which is needed to guarantee that the vacuum expectation value
of the Higgs fields (here the soldering form) is not zero.
One can easily convince himself that it is impossible to write a nontrivial
potential for the vierbein. For example, suppose we try to write
down a term containing two vierbeins,
no derivatives and invariant under coordinate and local Lorentz
transformations. The only possibility is
$g^{\mu\nu}\theta^a{}_\mu\theta^b{}_\nu\eta_{ab}$,
where $g_{\mu\nu}$ is the space-time metric. If $\theta$ was unrelated
to $g$, this would be a true mass-term for $\theta$. However,
in gravity one has the relation
$g_{\mu\nu}=\, \theta^a{}_\mu\, \theta^b{}_\nu\, \eta_{ab}$
and so the term written above is not really quadratic: it is
independent of $\theta$ and equal to 4. For the same reason one cannot
write potential terms of higher order.

The origin of both these difficulties can be traced to the double role
which is played in the gravitational Lagrangian by the metric (or vierbein).
In any field theory one needs a metric to contract indices in the Lagrangian
and to define the volume element; in this role the metric provides
the geometrical standard according to which lengths and angles are
measured. In addition in the theory of gravity the metric also plays
the role of dynamical variable. This duality
is the source of the beautiful geometrical interpretation of the
classical theory, where the two roles coexist peacefully. However,
it is also at the root of most difficulties, both conceptual and
practical, which are encountered in the quantization of gravity.
For example, the nonpolynomiality of gravitational Lagrangians
is due to the fact that covariant indices must
be contracted with the contravariant (inverse) metric.
Also, the fact that the theory must contain a dimensionful coupling
constant can be traced to the fact that geometrical and field theoretic
arguments lead to different dimensions for the metric.

We overcome the two difficulties mentioned above by considering a
mean-field quantum theory of gravity in which the two roles of the
metric are kept separate:
we assume that lengths and angles are not be measured with the
dynamical, fluctuating metric, but rather with its vacuum expectation
value (which we will refer to as ``mean value''), assumed nondegenerate.
On the other hand, the dynamical metric (or vierbein) can fluctuate
without constraints and evolves on the background provided by its own
mean value.

We modify the gravitational action by replacing the inverse metric
$g^{\mu\nu}$ by the inverse mean metric $\bar g^{\mu\nu}$.
Since the mean metric is assumed nondegenerate, the first difficulty
does not arise anymore. The second difficulty is also avoided
because one can now write terms like
$\bar g^{\mu\nu}\theta^a{}_\mu\theta^b{}_\nu\eta_{ab}$,
which are part of a genuine potential for the vierbein.
The action will then look like a Higgs model action in a fixed background
metric. One can use this action to perform quantum calculations in
which the mean metric is kept fixed. Among other things one can
compute the vacuum expectation value of the composite operator
$g_{\mu\nu}=\, \theta^a{}_\mu\, \theta^b{}_\nu\, \eta_{ab}$.
The whole scheme is self-consistent if one finds that this
vacuum expectation value is equal (up to a dimensionful multiplicative
constant $\ell^{-2}$) to the mean metric that one had postulated
in the beginning.

For technical reasons it is easier to compute the vacuum expectation
value of $\theta$ rather than that of $g$. In this case
self-consistency means that the vacuum expectation value of $\theta$
is equal (up to a dimensionful multiplicative constant $\ell^{-1}$)
to the mean vierbein $\btheta$, where
$\bar g_{\mu\nu}=\, \btheta^a{}_\mu\, \btheta^b{}_\nu\, \eta_{ab}$.
Note that since $\langle g_{\mu\nu}\rangle \neq\,
\langle\theta^a{}_\mu\rangle\langle\theta^b{}_\nu\rangle\eta_{ab}$,
the two procedures will lead to different values for the fundamental
length $\ell$, although qualitatively the results should be the same.

In Section 2 we will introduce a particularly simple action which
incorporates the ideas illustrated above and discuss in more
detail the general outline of the mean field approach.

In Section 3 we discuss the case in which the mean field corresponds to
the flat Minkowski metric. In this particular case one can use Fourier
analysis and the computations follow the familiar pattern from Elementary
Particle models. We are led to an effective potential for the
classical vierbein which is of the Coleman-Weinberg form.
The minima of this potential occur at multiples of the unit matrix, thus
ensuring selfconsistency.

In Section 4 we begin to take into account curvature effects by considering
a de Sitter mean field. In this case one can still compute the one-loop
effective action exactly. A closed form for the minimum can be given
in the limit of large de Sitter radius, and in this approximation
self-consistency can be explicitly checked.

In Section 5 we consider a generic mean field $\bar g$, and evaluate the
first three terms in the long wavelength, low momentum expansion
of the effective action.
Evaluating this action at its minimum yields an effective action for the
mean field which contains the Einstein term. Thus, at large distances
General Relativity is recovered as an induced effect.

Finally, Section 6 contains further remarks and conclusions.

\beginsection {2. A SIMPLE MODEL}

In order to make the previous discussion more concrete, we will illustrate
the mean field approach to quantum gravity by discussing a particular model.
We work within the context of the vierbein formulation of gravity,
taking as independent dynamical variables the vierbein $\theta^a{}_\mu$
and an $O(1,3)$ gauge field $A_\mu{}^a{}_b$ (here $a,b=0,1,2,3$ are internal
indices and $\mu,\,\nu=0,1,2,3$ are spacetime indices).
The spacetime metric is given by
$$g_{\mu\nu}=\, \theta^a{}_\mu\, \theta^b{}_\nu\, \eta_{ab}\ ,
\eqno(2.1)$$
where $\eta_{ab}={\rm diag}(-1,1,1,1)$.
The analogy with the Higgs model of elementary particle physics suggests
an action quadratic in the curvature of the $O(1,3)$ gauge field,
$F_{\mu\nu}{}^a{}_b=\partial_\mu A_\nu{}^a{}_b-\partial_\nu A_\mu{}^a{}_b
+g(A_\mu{}^a{}_c A_\nu{}^c{}_b- A_\nu{}^a{}_c A_\mu{}^c{}_b)$
and in the covariant derivative of the order parameter,
$\nabla_\mu\theta^a{}_\nu=\partial_\mu\theta^a{}_\nu
+gA_\mu{}^a{}_b\theta^b{}_\nu
-\igamma_\mu{}^\lambda{}_\nu\theta^a{}_\lambda$
($g$ is the gauge coupling constant and $\mit\Gamma$ are the
Christoffel symbols of the composite metric (2.1)).

The simplest such action has the form
$$S(\theta,A)=\int d^4x\ \sqrt{|\det g|}\ \left[-{1\over4}
g^{\mu\rho}g^{\nu\sigma}\eta_{ac}\eta^{bd}
F_{\mu\nu}{}^a{}_b \, F_{\rho\sigma}{}^c{}_d \
-{1\over2} \, g^{\mu\rho} g^{\nu\sigma}\eta_{ab}
\nabla_\mu\theta^a{}_\nu \nabla_\rho\theta^b{}_\sigma\ \right].\eqno(2.2)$$
It is manifestly invariant under local Lorentz and general coordinate
transformations.
A more general action of this type would contain several other terms
in which the indices are contracted in different ways, each term
weighted with a different coefficient. For the purposes of this paper it
will be sufficient to consider the particular case (2.2).

As discussed in the Introduction, the vierbein has two roles in the
theory: it defines the geometry of spacetime and at the same time it
is a dynamical variable. One can identify occurrences of the vierbein
in the action where it plays the role of geometrical standard, and
others where it plays the role of dynamical field.
If one compares (2.2) with the action of the Higgs model,
the only place where $\theta$ plays the role of dynamical variable
is under the covariant derivative.
The mean field approach consists in replacing $\theta$ by a ``mean
vierbein'' $\btheta$ in all other places. Due to the particular form
of the action, this is equivalent to replacing everywhere the
composite metric $g_{\mu\nu}$ by
$\bar g_{\mu\nu}=\btheta^a{}_\mu\btheta^b{}_\nu\eta_{ab}$.

As already mentioned, at this stage it becomes also possible to add to
the action a potential term.
In fact we will see later that to ensure renormalizability one has to
consider terms containing arbitrary powers of the curvature tensor
of $\bar g$ and up to four powers of the field $\theta$. Since we will
restrict ourselves to
one-loop calculations, it will be sufficient to consider terms linear
in curvature and quadratic in $\theta$, and terms quadratic in
curvature (which we do not write).
Thus our starting action will be of the form
$$\eqalign{\bar S(\theta,A;\bar g)
=\int d^4x\ \sqrt{|\det \bar g|}\
\Bigl[-{1\over4}\bar g^{\mu\rho}\bar g^{\nu\sigma}&\eta_{ac}\eta^{bd}
F_{\mu\nu}{}^a{}_b \, F_{\rho\sigma}{}^c{}_d \
-{1\over2} \, \bar g^{\mu\rho} \bar g^{\nu\sigma}\eta_{ab}
\bar\nabla_\mu\theta^a{}_\nu \bar\nabla_\rho\theta^b{}_\sigma \cr
&-{1\over 2} (m^2+\xi\bar R)\, \tr X -{\lambda_1\over 4}(\tr X)^2
-{\lambda_2\over 4}\tr(X^2)
\Bigr]\ ,\cr}\eqno(2.3)$$
where $X^a{}_b=\bar g^{\mu\nu}\theta^a{}_\mu\theta^c{}_\nu\eta_{cb}$,
$\bar R$ is the scalar curvature of $\bar g$,
$\bar\nabla$ denotes the covariant derivative constructed
with $A$ and the Christoffel symbols of $\bar g$ and
$m^2$, $\xi$, $\lambda_1$, $\lambda_2$ are coupling constants.
We shall refer to the second line in (2.3), with opposite sign,
as to the tree-level potential $V^{(0)}$.

The step from the action (2.2) to the action (2.3) is not
entirely unambiguous. The second term in (2.2) can be rewritten in
terms of the torsion
$\mit\Theta_\mu{}^a{}_\nu=
\nabla_\mu\theta^a{}_\nu-\nabla_\nu\theta^a{}_\mu$, and using the identity
$\nabla_\mu\theta^a{}_\nu={1\over2}(\Theta_\mu{}^a{}_\nu
- g^{\lambda\rho}\eta_{bc}\,\theta^a{}_\lambda\,
\theta^b{}_\nu\Theta_\mu{}^c{}_\rho
- g^{\lambda\rho}\eta_{bc}\,\theta^a{}_\lambda\,
\theta^b{}_\mu\Theta_\nu{}^c{}_\rho)$
one can rewrite
$$-{1\over2} \, g^{\mu\rho} g^{\nu\sigma}\eta_{ab}
\nabla_\mu\theta^a{}_\nu \nabla_\rho\theta^b{}_\sigma=
-{3\over8}g^{\mu\rho} g^{\nu\sigma}\eta_{ab}
\Theta_\mu{}^a{}_\nu\Theta_\rho{}^b{}_\sigma
-{1\over4}g^{\mu\rho}\theta^{-1}{}_a{}^\sigma\theta^{-1}{}_b{}^\nu
\Theta_\mu{}^a{}_\nu\Theta_\rho{}^b{}_\sigma\ .$$
Had we started from the action $S$ written in this alternative way,
we would have arrived at an action $\bar S$
with a different kinetic term for $\theta$.
One could eliminate this ambiguity by restricting the possible forms
of the kinetic term for $\theta$.
Anyway, the preceding discussion is meant as
a motivation for, not as a derivation of the action $\bar S$.
Our main reason for chosing the action (2.3) is that
it leads to a simple form of the propagator.

Taking the coordinates to have dimensions of length and the
``geometric'' metric $\bar g$ to be dimensionless, the dynamical
fields $A$ and $\theta$ have canonical dimension of inverse
length. Thus $m^2$ has the dimension of squared mass and the coupling
constants $\lambda_1$, $\lambda_2$ and $\xi$ are dimensionless, as usual.
Note that if we assumed the composite metric (2.1) to be
dimensionless, as required by geometric considerations, then for the
field $\theta$ to have canonical dimension one would have to introduce
in (2.1) a dimensionful constant $\ell$. This is the approach that was
adopted in [6,11]. Here we will let the composite metric (2.1) have
dimension of mass squared, as required by the canonical dimension of
$\theta$, and $\bar g$ be dimensionless. The constant $\ell$ will then
reappear in the self-consistency conditions which relate $g$ to $\bar g$
(or $\theta$ to $\bar\theta$).

We emphasize that nothing goes wrong in the action $\bar S$ when the
dynamical field $\theta^a{}_\mu$ becomes degenerate or even becomes
identically zero.
So in quantizing the theory with action (2.3) one need not
worry at all about this aspect and one can functionally integrate
over $\theta$ without constraints.
Unlike the original action $S$, $\bar S$ is polynomial and contains
interaction terms which are at most quartic in the dynamical fields;
it has exactly the same type of interactions of the usual Higgs model.
Thus the theory defined by (2.3) is power-counting
renormalizable in flat space.

Since in the action $\bar S$ the metric $\bar g$ has to be treated as
a fixed background, general coordinate invariance is lost.
The choice of the mean vierbein $\btheta$ would also break local
Lorentz invariance, but since $\btheta$ appears only through the
combination
$\bar g_{\mu\nu}=\btheta^a{}_\mu\btheta^b{}_\nu\eta_{ab}$,
the action (2.3) is invariant under local Lorentz transformations.
Furthermore, if we allow the transformations to act also on the
background $\bar g$, the action (2.3) has the same invariances of the
action (2.2).

We will evaluate the one-loop effective potential
for $\theta^a{}_\mu$ using the saddle point approximation,
treating $\bar g$ as a fixed background.
As is well known this reduces to the calculation of a functional
determinant.
We first expand $\bar S$ up to second order around a classical
solution $A_{\rm (cl)}$, $\thetacl$ of the field equations.
In the expansion of
$\bar S$ terms linear in the fluctuations are then absent.
Defining
$$\eqalignno{\theta^a{}_\mu&=\ \thetacl{}^a{}_\mu+\varphi^a{}_\mu\
,&(2.4a)\cr
A_\mu{}^a{}_b&=\ \acl_\mu{}^a{}_b+\omega_\mu{}^a{}_b\
,&(2.4b)\cr}$$
the quadratic action has the form
$$\eqalignno{\bar S^{(2)}(\varphi,\omega;\thetacl,\acl;\bar g)=\
&{1\over2}\int d^4x \sqrt{|\det \bar g|}\
\tr\left[\left(\omega\ \varphi\right)
\pmatrix{{\cal O}_{[\omega\omega]}&{\cal O}_{[\omega\varphi]}\cr
{\cal O}_{[\varphi\omega]}&{\cal O}_{[\varphi\varphi]}\cr}
\pmatrix{\omega\cr\varphi\cr}\right]\ &(2.5a) \cr
&\null\cr
=\ &{1\over2}\int d^4x \sqrt{|\det \bar g|}\
\Bigl[\omega_\mu{}^a{}_b\,
{\cal O}_{[\omega\omega]}{}^\mu{}_a{}^{b\nu}{}_c{}^d\, \omega_\nu{}^c{}_d\cr
&\quad+2\,\varphi^a{}_\mu\,
{\cal O}_{[\varphi \omega]}{}_a{}^\mu{}^\nu{}_c{}^d{}\, \omega_\nu{}^c{}_d
+\varphi^a{}_\mu\, {\cal O}_{[\varphi\varphi]}{}_a{}^\mu{}_b{}^\nu\,
\varphi^b{}_\nu \Bigr]\ .&(2.5b)\cr}$$
This linearized action is invariant under the linearized gauge transformations:
the fields
$$A_\mu{}^a{}_b={1\over g}\bar\nabla_\mu\epsilon^a{}_b\ ,\qquad
\varphi^a{}_\mu=-\epsilon^a{}_b\thetacl{}^b{}_\mu \eqno(2.6)$$
are null vectors for the operator ${\cal O}$ which is the block matrix
operator appearing in (2.5a). We therefore have to fix the gauge.
We choose the t'Hooft gauge and add to the linearized action the
gauge-fixing term
$$-{1\over2\alpha}\int d^4x\ \sqrt{|\det \bar g|}\
\Bigl[\bar g^{\mu\nu}(\bar\nabla_\mu \omega_\nu{}^a{}_b+\alpha\, g\,
\eta_{cb}\,\thetacl^c{}_\mu\,\varphi^a{}_\nu)\Bigr]^2\ .\eqno(2.7)$$
Collecting all terms, the operators governing the dynamics of small
fluctuations are
$$\eqalignno{&{\cal O}_{[\omega\omega]}{}_{\mu ab}{}^{\nu cd}=\
\delta^{[c}_{[a}\,\delta^{d]}_{b]}\ \left[\delta_\mu^\nu\,
\bar\nabla_\lambda\bar\nabla^\lambda-
\Bigl(1-{1\over\alpha}\Bigr)\bar\nabla_\mu\bar\nabla^\nu
-\bar R_\mu{}^\nu\right]\cr
&{\hskip 5cm}+4g\,\delta_{[a}^{[c}\, \fcl_\mu{}^\nu{}_{b]}{}^{d]}
-\delta_\mu^\nu\ \mcl_{ab}{}^{cd}\ ,&(2.8a)\cr
&{\cal O}_{[\varphi \omega]}{}_{a\mu}{}^{\nu cd}
=\,2g\,\delta_a^{[c}\,\bar\nabla^\nu\thetacl^{d]}{}_\mu\ ,&(2.8b)\cr
&{\cal O}_{[\varphi\varphi]}{}_{a\mu}{}^{b\nu}=
\ \delta_a^b\delta_\mu^\nu\
\left[\bar\nabla_\lambda\bar\nabla^\lambda-\bigl(m^2+\xi\,\bar R\bigr)\right]
-\ncl{}_{a\mu}{}^{b\nu}
\ ,&(2.8c)\cr}$$
where
$$\eqalignno{\mcl_{ab}{}^{cd}=&
\ g^2\,\delta_{[a}^{[c}\, \xcl_{b]}{}^{d]}=\ g^2\, \delta_{[a}^{[c}\,
\thetacl_{b]}{}^\mu\thetacl^{d]}{}_\mu\ ,&(2.9a)\cr
\ncl_{a\mu}{}^{b\nu}=
&\ \alpha\, g^2\delta_a^b\, \xcl_\mu{}^\nu
+\lambda_1\left[\xcl_c{}^c\ \delta_a^b\delta_\mu^\nu
+2\,\thetacl_{a\mu}\,\thetacl^{b\nu}\right]\cr
&\qquad\quad+\lambda_2\left[\xcl_a{}^b\ \delta_\mu^\nu+\delta_a^b\
\xcl_\mu{}^\nu+\thetacl_a{}^\nu\thetacl^b{}_\mu\right]\ .&(2.9b)\cr}$$
Here and in the following indices are raised and lowered
with $\eta_{ab}$ and $\bar g_{\mu\nu}$ and transformed from latin to greek
by means of $\btheta$. The ghost operator is
$${\cal O}_{\rm [gh]}{}_{ab}{}^{cd}=
\delta^{[c}_{[a}\,\delta^{d]}_{b]}\ \bar\nabla_\lambda\bar\nabla^\lambda
-\alpha\,\mcl_{ab}{}^{cd}\ .\eqno(2.10)$$
In order to compute the functional determinants, one has to
analytically continue the differential operators to the Euclidean
sector. This amounts simply to changing the overall sign of the
operators (2.8) and (2.10). The one-loop effective action is then formally
given by
$$\Gamma^{(1)}(\thetacl,\acl;\bar g)
={1\over2}\ln\det{\cal O}-\ln\det{\cal O_{[\rm gh]}}\ .
\eqno(2.11)$$
The explicit evaluation of these determinants requires a
regularization due to the presence of ultraviolet divergences.
In the subsequent three sections we will use a simple cutoff, zeta
function and heat kernel regularization for the cases when the mean
field is flat space, de Sitter space and a generic space, respectively.
These methods will be found to give entirely consistent results.

If the regularization procedure respects general
coordinate invariance, as is the case with the method we adopt in
Section 5, the effective action (2.11) will have
the same invariance properties of the classical action $\bar S$.
In particular, it will be invariant under coordinate transformations
when all fields, including $\bar g$, are transformed.

The vacuum expectation values $\langle\theta\rangle$ and
$\langle A\rangle$ are obtained by minimizing the total effective action
$\Gamma=\bar S+\Gamma^{(1)}$ with respect to $\thetacl$ and $\acl$.
These minima depend upon $\bar g$.
Since we want to interpret the background vierbein $\btheta$ as the
vacuum expectation value of $\theta$, the theory will be self-consistent if
$\langle \theta\rangle=\ell^{-1}\btheta$, where $\ell$ is a constant
with the dimension of length. In the next Sections we shall see that
self-consistency can be achieved in the case of a flat or de Sitter
background, and, in the long wavelength limit, for a generic
background.

\beginsection {3. FLAT SPACE}

By the equivalence principle, any macroscopic metric can be approximated
in sufficiently small regions by a flat metric.
On the other hand, quantum gravity is supposed to supersede the classical
theory precisely at short distances. Thus it is perfectly appropriate that
quantum gravity should begin by explaining flat space.
Note that in view of the analogy with the Higgs model, the Minkowski
metric is already a nontrivial background: it is the same as having a
constant nonzero Higgs field.

We take then $\bar g_{\mu\nu}=\eta_{\mu\nu}$ and
$\btheta^a{}_\mu=\delta^a_\mu$. For the classical fields we choose
$A_{\rm (cl)}=0$ and $\thetacl={\rm const}$.
The operators governing the dynamics of small fluctuations become,
in momentum space,
$$\eqalignno{&{\cal O}_{[\omega\omega]}{}_{\mu ab}{}^{\nu cd}=\
\delta^{[c}_{[a}\,\delta^{d]}_{b]}\,\left[-\delta_\mu^\nu\
k_\lambda k^\lambda
+\Bigl(1-{1\over\alpha}\Bigr)k_\mu k^\nu\right]
-\, \delta_\mu^\nu\,\mcl_{ab}{}^{cd}\ ,&(3.1a)\cr
&{\cal O}_{[\varphi \omega]}{}_{a\mu}{}^{\nu cd}
=\ 0\ ,&(3.1b)\cr
&{\cal O}_{[\varphi\varphi]}{}_{a\mu}{}^{b\nu}=\ \delta_a^b\delta_\mu^\nu\
\left(-k_\lambda k^\lambda-m^2\right)
-\ncl_{a\mu}{}^{b\nu}\ ,&(3.1c)\cr
&{\cal O}_{[\rm gh]}{}_{ab}{}^{cd}=
-\delta^{[c}_{[a}\,\delta^{d]}_{b]}\ k_\lambda k^\lambda
-\alpha\,\mcl{}_{ab}{}^{cd}\ .&(3.1d)\cr}$$
Due to $(3.1b)$, the one-loop effective action reduces to
$$\Gamma^{(1)}={1\over2}\ln\det{\cal O}_{[\omega\omega]}+
{1\over2}\ln\det{\cal O}_{[\varphi\varphi]}-\ln\det{\cal O}_{\rm
[gh]}\ .\eqno(3.2)$$

Since the classical fields are constant, the effective action is the
spacetime integral of the effective potential $V$.
For every $k$, the matrices in (3.1) can be explicitly diagonalized.
This is achieved by first bringing $\thetacl$ to
diagonal form by means of independent global Lorentz transformations on the
internal and spacetime indices.
Up to an irrelevant multiplicative factor, the eigenvalues of every
operator $\o$ are of the form $-(k^2+\lambda_i)$.
Then, in the Euclidean regime,
$\ln\det {\cal O}={1\over(2\pi)^2}\sum_i\int^\Lambda d^4k
\ln(k^2+\lambda_i)$, where $\Lambda$ is an ultraviolet cutoff.
The integral over $k$ can be performed explicitly and one finds,
for $\Lambda$ large,
$$\eqalign{V^{(1)}=&\ {3\over64\pi^2}\left[2\Lambda^2\tr \mcl+
\,\tr \left(\mcl^2\left(\ln{ \mcl\over\Lambda^2}
-{1\over2}\right)\right)\right]\cr
&-{1\over64\pi^2}\left[2\alpha\,\Lambda^2\tr \mcl+\alpha^2
\,\tr \left(\mcl^2\left(\ln{\alpha \mcl\over\Lambda^2}
-{1\over2}\right)\right)\right]\cr
&+{1\over64\pi^2}\biggl[2\Lambda^2\,\tr \tilde\ncl+
\,\tr \biggl(\tilde\ncl^2\biggl(\ln{ \tilde\ncl\over\Lambda^2}
-{1\over2}\biggr)\biggr)\biggr]\ ,\cr}\eqno(3.3)$$
where $\tilde\ncl=\, m^2+\ncl$, the traces are over double
indices and an infinite, field-independent constant has been dropped.
In (3.3) the first line is the contribution of the transverse
components of $\omega$, the second line comes from the longitudinal components
of $\omega$ and from the ghosts, the last line is the $\varphi$ contribution.
This effective potential contains divergences proportional to
$\tr \mcl$, $\tr(\mcl^2)$ and $\tr\tilde\ncl$, $\tr(\tilde\ncl^2)$.
These terms are proportional to $\tr\xcl$ and to suitable combinations
of $(\tr\xcl)^2$ and $\tr(\xcl^2)$, and thus of the same form of the
potential terms in the starting action. The infinities in (3.3) can then
be absorbed in a renormalization of the coupling constants of the tree-level
potential $V^{(0)}$. In fact, we see that the addition of the
tree-level potential was necessary to ensure renormalizability of the theory.
This is very similar to what happens in scalar electrodynamics,
where renormalizability of the meson-photon interaction demands
the presence of a quartic self-interaction of the scalars.

With a suitable choice of the renormalization scale
$\mu$ the total renormalized effective potential takes the form
$$\eqalign{ V=V^{(0)}+V^{(1)}
&=\ {1\over2}m^2\tr \xcl+{\lambda_1\over4}(\tr \xcl)^2
+{\lambda_2\over4}\tr (\xcl^2)\cr
&+{1\over64\pi^2}\,\left[(3-\alpha^2)\tr \left(\mcl^2\left(
\ln{\mcl\over\mu^2}-{3\over2}\right)\right)
-\alpha^2\ln\alpha\ \tr\mcl^2\right]\cr
&+{1\over64\pi^2}\,\tr \biggl(\tilde\ncl^2\biggl(\ln
{\tilde\ncl\over\mu^2}-{3\over2}\biggr)\biggr)\ ,\cr}\eqno(3.4)$$
where the coupling constants are the renormalized ones.
Notice that for $\alpha=\ 0$, the t'Hooft gauge reduces to
the Landau gauge; in that case the potential (3.4) reduces, modulo
a finite redefinition of the renormalization scale $\mu$, to that
already computed in [11].

The minimum of the effective potential (3.4) cannot easily be written
in analytic form. To simplify calculations we follow
[12] in setting $m^2=\,0$, and assume that the coupling constants
$\lambda_1$ and $\lambda_2$ are of order $g^4$. Then, in the last term of (3.4)
the dependence on $\lambda_1$ and $\lambda_2$
can be neglected. For a physically acceptable range of the parameters
$\lambda_1$, $\lambda_2$ and $g$, the absolute minimum of the effective
potential (3.4) occurs for
$$\thetacl^a{}_\mu={\mu\over g} \exp\left[{1\over2}
-{16\pi^2\over9+5\alpha^2}\left(
{4\lambda_1+\lambda_2\over g^4}+{5\over32\pi^2}\alpha^2\, \ln\alpha\right)
\right]\ \delta_\mu^a\ ,\eqno(3.5)$$
and global Lorentz transformations thereof.
Thus, the quantum dynamics of the model drives the vacuum
expectation value of the vierbein to be nondegenerate.
This is a gravitational analog of the so called Coleman-Weinberg
mechanism [12].
In fact, the vacuum expectation value of $\theta$ is proportional to
the mean vierbein $\btheta$, with proportionality constant $\ell^{-1}$
of the order of the renormalization mass $\mu$. We will in Section 5
that $\ell^2$ can be identified with Newton's constant.

The effective potential (3.4) and its minimum (3.5) depend explicitly
on the gauge parameter $\alpha$. This is a common feature of effective
potential calculations [13]. In order to obtain a gauge-independent
result one could compute the Vilkovisky-de Witt effective action for our
problem [14]. We do not expect this to yield qualitatively different
results. In any event, notice that the existence of a
non-degenerate absolute minimum is guaranteed for any finite value of
$\alpha$ (an infinite value for $\alpha$ is in any case excluded
because it would give rise to an effective potential unbounded from
below). Flat space is then a self-consistent
solution of the theory in any gauge.

\beginsection {4. DE SITTER SPACE}

In this Section we start taking into account effects due to the
curvature of the mean metric. We choose $\bar g$ to be the de Sitter
metric, which enables us to compute the one-loop effective action
exactly. The results we obtain will also be used as an independent
check of more general calculations in the next section.
One loop effective actions in de Sitter space have been computed
before in a variety of contexts [15-20].
Since calculations are performed in the Euclidean sector,
de Sitter space is just a four-dimensional sphere.
We will not need to choose a coordinate system to write down
the metric explicitly. The only properties that we will need are
$$\bar R_{\mu\nu\rho\sigma}={1\over r^2}\,(\bar g_{\mu\rho}\bar
g_{\nu\sigma}-\bar g_{\mu\sigma}\bar g_{\nu\rho})\ ,\qquad
\bar R_{\mu\nu}={3\over r^2}\,\bar g_{\mu\nu}\ ,\qquad
\bar R={12\over r^2}\ , \eqno(4.1)$$
where $r$ is the radius of the de Sitter space (related to the
cosmological constant $\lambda_0$ by $r^2={3/\lambda_0}$).
For the classical fields we take
$$\thetacl^a{}_\mu=\rho\,\bar\theta^a{}_\mu\ ,\qquad
g\acl_\lambda{}^a{}_b=\btheta^a{}_\mu\,\bar\igamma_\lambda{}^\mu{}_\nu\,
\btheta^{-1}{}_b{}^\nu+\btheta^a{}_\mu\partial_\lambda\,
\btheta^{-1}{}_b{}^\mu\ ,\eqno(4.2)$$
with $\rho$ a constant with the dimensions of mass. We thus reduce the
freedom in the classical vierbein to a single constant $\rho$
and our task in this Section will be to show that the quantum dynamics
of the theory requires $\rho$ to be nonzero.
The curvature of $\acl$ is related now to the Riemann tensor of the
mean metric by
$g \fcl_{\mu\nu}{}^a{}_b=\btheta^a{}_\rho\,
\bar R_{\mu\nu}{}^\rho{}_\sigma\, \btheta^{-1}{}_b{}^\sigma$.
Using this, one can verify that (4.2) solve the classical equations of motion.
We will now compute the one-loop effective action,
which will depend on the proportionality constant $\rho$ and,
parametrically, on the radius $r$.

With the assumptions above, the differential operators (2.8) and (2.10)
simplify; in particular
$$\eqalignno{\mcl_{ab}{}^{cd}=&\,g^2\rho^2\
\delta^{[c}_{[a}\,\delta^{d]}_{b]}\ ,&(4.3a)\cr
\ncl_{a\mu}{}^{b\nu}=&\left(\alpha\,
g^2+2\,(2\lambda_1+\lambda_2)\right)\rho^2\,
\delta_a^b\delta_\mu^\nu
+2\lambda_1\,\rho^2\,\btheta_{a\mu}\btheta^{b\nu}
+\lambda_2\rho^2\btheta_a{}^\nu\btheta^b{}_\mu\ .&(4.3b)\cr}$$
Also, ${\cal O}_{[\omega\varphi]}=\, 0$, so that $\Gamma^{(1)}$ is again
given by (3.2).

In order to deal with the nonminimal term
$\bar\nabla_\mu\bar\nabla^\nu$ in ${\cal O}_{[\omega\omega]}$,
it is convenient to decompose $\omega$ in its transverse and
longitudinal parts, satisfying respectively
$\bar\nabla_\mu\omega^T{}^{\mu}{}_{ab}=0$ and
$\omega^L{}_{\mu ab}=\bar\nabla_\mu\epsilon_{ab}$.
The operator $\o_{[\omega\omega]}$ maps transverse fields to transverse
fields and longitudinal fields to longitudinal fields.
Therefore, $\ln\det{\cal O}_{[\omega\omega]}=
\ln\det^T{\cal O}_{[\omega\omega]}+\ln\det^L{\cal O}_{[\omega\omega]}$.
Using the formula
$${\cal O}_{[\omega\omega]}{}_{\mu ab}{}^{\nu cd}\ \bar\nabla_\nu\epsilon_{cd}
=\bar\nabla_\mu\left({\cal O}^\prime{}_{ab}{}^{cd}\,\epsilon_{cd}\right)
\ ,\eqno(4.4)$$
where ${\cal O}^\prime{}_{ab}{}^{cd}={1\over\alpha}\,
{\cal O}_{[gh]}{}_{ab}{}^{cd}$,
we can rewrite: $\ln\det^L{\cal O}_{[\omega\omega]}=\ln\det{\cal O}^\prime$.
On the other hand when the operator ${\cal O}_{[\omega\omega]}$ acts
on transverse fields the nonminimal term drops out.

The spectrum of the operator ${\cal O}_{[\omega\omega]}$ on transverse
fields, and the spectra of
${\cal O}^\prime$, ${\cal O}_{[\varphi\varphi]}$ and
${\cal O}_{\rm [gh]}$ can be determined
explicitly using group-theoretic arguments. We begin by decomposing
the fields $\omega$ and $\varphi$ in their irreducible components with
respect to the group $O(4)$. Transforming all indices to latin by
means of $\btheta$, we have
$$\eqalignno{\omega_{abc}=&\ {2\over3}(t_{abc}-t_{acb})+{1\over3}
(\eta_{ab}\,v_c-\eta_{ac}\,v_b)+\varepsilon_{abcd}\,w^d\ ,&(4.5a)\cr
\varphi_{ab}=&\ \psi_{ab}+\chi_{ab}+\eta_{ab}\,\tau\ ,&(4.5b)\cr}$$
where $t_{abc}={1\over2}(\omega_{abc}+\omega_{bac})
+{1\over6}(\eta_{ac}\,\omega^d{}_{db}+\eta_{bc}\,\omega^d{}_{da})
-{1\over3}\eta_{ab}\omega^d{}_{dc}$ carries the 16-dimensional representation
$\left({3\over2},{1\over2}\right)\oplus\left({1\over2},{3\over2}\right)$,
$v_a=\omega^b{}_{ba}$ and
$w_a={1\over6}\varepsilon_{abcd}\,\omega^{bcd}$ carry the 4-dimensional
representation $\left({1\over2},{1\over2}\right)$,
$\psi_{ab}={1\over2}(\varphi_{ab}+\varphi_{ba})-{1\over4}\eta_{ab}\,
\varphi^c{}_c$ transforms according to the 9-dimensional
representation $(1,1)$,
$\chi_{ab}={1\over2}(\varphi_{ab}-\varphi_{ba})$
transforms according to the 6-dimensional representation
$(1,0)\oplus(0,1)$ and $\tau={1\over4}\varphi^a{}_a$ transforms
according to the 1-dimensional representation $(0,0)$ [21].

It can be shown that the transverse part of $\omega$
is given by the tensor $t$ and the longitudinal parts $v^L$ and $w^L$
of the vectors $v$ and $w$, while the longitudinal part of $\omega$
is given by the transverse parts of $v$ and $w$.
This is made plausible by observing that the 18 degrees of freedom of
$\omega^T$ match the 16 of $t$ plus one each for $v^L$ and $w^L$;
similarly $\omega^L$ has 6 degrees of freedom, matching the ones of $v^T$
and $w^T$.  A rigorous proof of this result is given in the Appendix.
When acting on transverse fields the operator ${\cal O}_{[\omega\omega]}$
respects the $O(4)$ decomposition. Symbolically,
$$\omega^T{\cal O}_{[\omega\omega]}\omega^T=
{4\over3}\,t\,{\cal O}_{[tt]}\,t+{2\over3}\,v^L\,{\cal O}_{[v^Lv^L]}\,v^L
+6\,w^L\,{\cal O}_{[w^Lw^L]}\,w^L\ ,\eqno(4.6)$$
where
$$\eqalignno{&{\cal O}_{[tt]}=
-\bar\nabla_\lambda\bar\nabla^\lambda+{5\over r^2}+z\ ,&(4.7a)\cr
&{\cal O}_{[v^Lv^L]}={\cal O}_{[w^Lw^L]}=
-\bar\nabla_\lambda\bar\nabla^\lambda-{1\over r^2}+z\ ,&(4.7b)\cr}$$
with $z=\,g^2\rho^2$
(tensor indices have been suppressed since these operators are
multiples of the identity in the appropriate tensor space).
Similarly one finds
$$\varphi\,{\cal O}_{[\varphi\varphi]}\,\varphi=
\psi\,{\cal O}_{[\psi\psi]}\,\psi+\chi\,{\cal O}_{[\chi\chi]}\,\chi
+4\,\tau\,{\cal O}_{[\tau\tau]}\,\tau\ ,\eqno(4.8)$$
where
$$\eqalignno{{\cal O}_{[\psi\psi]}=&
-\bar\nabla_\lambda\bar\nabla^\lambda+z_\psi\ , \qquad
z_\psi=m^2+{12\over r^2}\xi
+\rho^2\,(4\lambda_1+3\lambda_2+g^2\alpha)\ ,&(4.9a)\cr
{\cal O}_{[\chi\chi]}=&
-\bar\nabla_\lambda\bar\nabla^\lambda+z_\chi\ ,\qquad
z_\chi=m^2+{12\over r^2}\xi
+\rho^2\,(4\lambda_1+\lambda_2+g^2\alpha)\ ,&(4.9b)\cr
{\cal O}_{[\tau\tau]}=&
-\bar\nabla_\lambda\bar\nabla^\lambda+z_\tau\ ,\qquad
z_\tau=m^2+{12\over r^2}\xi
+\rho^2\,(12\lambda_1+3\lambda_2+g^2\alpha)\ .&(4.9c)\cr}$$
Finally, ${\cal O}_{\rm [gh]}$ and ${\cal O}^\prime$ act on
the antisymmetric tensor representation $(1,0)\oplus(0,1)$.

The spectra of all these operators can be determined using the method
explained in [22]. The eigenvalues $\lambda_n$ and the
corresponding multiplicities $d_n$ are given in the following table.
\bigskip
\newbox\sstrutbox
\setbox\sstrutbox=\hbox{\vrule height12pt depth8pt width0pt}
\def\bigstrut{\relax\ifmmode\copy\sstrutbox\else\unhcopy\sstrutbox\fi}
\relax
\vbox{\tabskip=0pt \offinterlineskip
\def\tablerule{\noalign{\hrule}}
\halign to 16truecm {\strut#& \vrule#\tabskip=1em plus 2em&
\hfil#& \vrule#& \hfil#\hfil& \vrule#&
\hfil#\hfil& \vrule#&
\hfil#\hfil& \vrule#
\tabskip=0pt \cr\tablerule
&&\multispan7\hfil\bigstrut Eigenvalues and multiplicities \hfil&\cr\tablerule
&&\omit\hidewidth\bigstrut  operator \hidewidth&&
\omit\hidewidth $\lambda_n$ \hidewidth&&
\omit\hidewidth $d_n$ \hidewidth&&
\omit\hfil &\cr\tablerule
&&\bigstrut $\phantom{\Bigg|^|}\o_{[tt]}\phantom{\Bigg|_|}$
&& $\eqalign{&r^{-2}(n^2+3n+2)+ z\cr &r^{-2}(n^2+3n-2)+z\cr}$
&& $\eqalign{{\hbox{$5\over4$}}(n&+4)(n-1)(2n+3)\cr &n(n+3)(2n+3)\cr}$
&& $\eqalign{&n\geq2\cr &n\geq2\cr}$
&\cr
\tablerule
&&\bigstrut $\o_{[v^Lv^L]}$
&& $r^{-2}(n^2+3n-4)+ z$
&& ${1\over6}(n+1)(n+2)(2n+3)$
&& $n\geq1 $
&\cr\tablerule
&&\bigstrut ${\phantom{\Bigg|^{\Big|}}}^{}\o_{[\psi\psi]_{\phantom{\bigg|}}}$
&& $\eqalign{&r^{-2}(n^2+3n-2)+z_\psi\cr
             &r^{-2}(n^2+3n-6)+z_\psi\cr
             &r^{-2}(n^2+3n-8)+z_\psi\cr}$
&& $\eqalign{\hbox{${5\over6}$}(n&-1)(n+4)(2n+3)\cr
             &\hbox{${1\over2}$}n(n+3)(2n+3)\cr
             \hbox{${1\over6}$}(n&+1)(n+2)(2n+3)\cr}$
&& $\eqalign{&n\geq2\cr
             &n\geq2\cr
             &n\geq2\cr}$
&\cr\tablerule
&&\bigstrut $\o_{[\chi\chi]}$
&& $r^{-2}(n^2+3n-2)+z_\chi$
&& $n(n+3)(2n+3)$
&& $n\geq1$
&\cr\tablerule
&&\bigstrut $\o_{[\tau\tau]}$
&& $r^{-2}(n^2+3n)+z_\tau$
&& ${1\over6}(n+1)(n+2)(2n+3)$
&& $n\geq0$
&\cr\tablerule
&&\bigstrut $\o_{[\rm gh]}$
&& $r^{-2}(n^2+3n-2)+\alpha\,z$
&& $n(n+3)(2n+3)$
&& $n\geq1$
&\cr\tablerule
%\noalign{\smallskip}
\hfil\cr}}
\relax
\bigskip
We note that for $\rho=\,0$ the operators ${\cal O}_{[v^Lv^L]}$ and
${\cal O}_{[w^Lw^L]}$ have five zero eigenvalues each. This is because for
$\rho=\,0$ the linearized Lagrangian (2.5) is the one of pure Yang-Mills
theory for the group $SO(4)=SU(2)\times SU(2)$, and the classical
field $\acl$ is the direct sum of an instanton and an anti-instanton.
Thus, these are the familiar zero-modes due to the $O(5)$-invariance
of the instanton background [15]. When $m^2=\,0$, $\xi=\,0$, the operator
${\cal O}_{[\tau\tau]}$ also has a zero eigenvalue for $\rho=\,0$.

The complete one-loop effective action is given by
$$\eqalign{\Gamma^{(1)}=&\ {1\over2}\ln\det{\cal O}_{[tt]}
+{1\over2}\ln\det{\cal O}_{[v^Lv^L]}
+{1\over2}\ln\det{\cal O}_{[w^Lw^L]}
+{1\over2}\ln\det{\cal O}^\prime \cr
+&{1\over2}\ln\det{\cal O}_{[\psi\psi]}
+{1\over2}\ln\det{\cal O}_{[\chi\chi]}
+{1\over2}\ln\det{\cal O}_{[\tau\tau]}
-\ln\det{\cal O}_{\rm [gh]}\ .\cr}\eqno(4.10)$$
For each operator ${\cal O}$, we define the dimensionless zeta-function
$\zeta(\o,s)=\sum_n d_n (r^2\lambda_n)^{-s}$ [23].
Then, $\ln\left(\det{\o/\mu^2}\right)
=\sum_n d_n \ln\left({\lambda_n/\mu^2}\right)
=-\zeta^\prime(\o,0)-\ln(\mu^2 r^2)\zeta(\o,0)$,
where the prime signifies derivative with respect to $s$ and
$\mu$ is the renormalization scale.
These zeta functions can be evaluated exactly in terms of digamma functions
[18,20]. The complete expression is complicated and not very revealing.
Instead we will present the approximated form of $\Gamma^{(1)}$ for
$r \rho$ large and $\alpha\not=0$:
$$\eqalign{\Gamma^{(1)}(\rho;r)&=
{3\over4}r^4z^2\left(\ln{z\over\mu^2}-{3\over2}\right)
+3r^2z\left(\ln{z\over\mu^2}-1\right)
+{59\over10}\ln{z\over\mu^2}\cr
&+{3\over8}r^4z_\psi^2\left(\ln{z_\psi\over\mu^2}-{3\over2}\right)
-{3\over2}r^2z_\psi\left(\ln{z_\psi\over\mu^2}-1\right)
+{9\over20}\ln{z_\psi\over\mu^2}\cr
&+{1\over4}r^4z_\chi^2\left(\ln{z_\chi\over\mu^2}-{3\over2}\right)
-r^2z_\chi\left(\ln{z_\chi\over\mu^2}-1\right)
+{19\over30}\ln{z_\chi\over\mu^2}\cr
&+{1\over24}r^4z_\tau^2\left(\ln{z_\tau\over\mu^2}-{3\over2}\right)
-{1\over6}r^2z_\tau\left(\ln{z_\tau\over\mu^2}-1\right)
+{29\over180}\ln{z_\tau\over\mu^2}\cr
&-{1\over4}r^4\alpha^2z^2\left(\ln{\alpha z\over\mu^2}-{3\over2}\right)
+r^2\alpha z\left(\ln{\alpha z\over\mu^2}-1\right)
-{19\over30}\ln{\alpha z\over\mu^2}\ .\cr}\eqno(4.11)$$
We assume again $m^2=0$, $\xi=0$ and $\lambda_1\approx\lambda_2\approx g^4$.
The one-loop effective potential $V^{(1)}$ is defined by
$\Gamma^{(1)}(\rho;r)=
\Omega(r) V^{(1)}(\rho;r)$, where $\Omega(r)={8\over3}\pi^2 r^4$
is the volume of de Sitter space and $\rho$ is constant.
The effective potential is given to order $g^4$ by
$$\eqalignno{V^{(1)}=&\, {1\over32\pi^2}\Biggl\{\Bigl(9+5\alpha^2\Bigr)
z^2\Bigl(\ln{z\over\mu^2}-{3\over2}\Bigr)+{5\over3}z^2\alpha^2\ln\alpha\cr
&+{4\over r^2}\Bigl[\Bigl(9-5\alpha-18{2\lambda_1+\lambda_2\over g^2}\Bigr)
z\ln{z\over\mu^2}
-\!\left(9-5\alpha+\ln\alpha\left(5\alpha+18{2\lambda_1+\lambda_2\over g^2}
\right)\right)z \Bigr]\cr
&\qquad\qquad\qquad\qquad\qquad
+{560\over3}{1\over r^4}\ln{z\over\mu^2}\Biggr\}\ ;&(4.12)\cr}$$
an additive $\rho$-independent costant has been dropped.
Notice that in the flat space limit $r\rightarrow\infty$ only the
first line remains. It coincides with the one-loop effective potential in
flat space discussed in Section 3, when we set
$\thetacl=\rho\bar\theta$. Therefore the method of the zeta function
and the method of the cutoff give the same result for the effective
action in flat space.

The minimum of the total effective potential $V=V^{(0)}+V^{(1)}$ occurs,
for any finite nonzero value of the gauge parameter $\alpha$, at a non-zero
value of $z$, whose explicit expression can not be given analytically.
However, expanding around the flat-space minimum
$\thetacl^a{}_\mu=\rho_0\delta^a_\mu$ given in (3.5)
and keeping only terms of lowest order in $r^{-2}$, one gets:
$$\eqalign{\rho^2=\ \rho_0^2-{1\over r^2
g^2}{2\over 9+5\alpha^2}&\Bigl[
\left(9-5\alpha-18{2\lambda_1+\lambda_2\over g^2}\right)
\ln\left({g^2\,\rho_0^2\over \mu^2}\right)\cr
& \qquad\qquad -5\alpha\,\ln\alpha
-18(1+\ln\alpha){2\lambda_1+\lambda_2\over g^2}\Bigr]
+O({1\over r^4})\ .\cr}\eqno(4.13)$$
For large $\mu\, r$ the minimum of $V$
is in the region for which the expression (4.12) can be trusted.

For small values of $\rho$ the exact expression of the one-loop
effective potential is found to be logarithmically divergent. This is not
surprising since, as observed, the operators $\o_{[v^Lv^L]}$ and
$\o_{[w^Lw^L]}$ have five zero-modes each at $\rho=\ 0$. These clearly
give the dominant contribution to the effective potential, which then
diverges like $\ln(\rho/\mu)$ for $\rho$ close to zero.
Of course this behaviour can not be trusted within the one-loop approximation.
Indeed as briefly discussed in the following, by resumming an
infinite number of loop contributions due to the
zero-modes one gets a sensible (finite) result.
Similar behaviour had been observed before [24].
Notice that (4.11) and (4.12)
are also logarithimically divergent for $z$ close to zero; however,
these infinities are not significant since they occur for values of $\rho$ for
which those expressions are no longer valid.

In order to compute higher-loop contributions to the effective action,
it is necessary to go beyond the quadratic approximation in the expansion of
the action (2.3), taking into account also interaction terms in the
fluctuating fields $\omega$ and $\varphi$.
Since it is $v^L$ and $w^L$ that have zero modes, we shall concentrate
on the quartic self-interactions
${g^2\over27}(v^L{}_\mu\, v^L{}^\mu)^2$ and $3g^2(w^L{}_\mu\, w^L{}^\mu)^2$
and add them to the quadratic part of the action,
symbolically given by the last two terms in (4.6).
Other quartic interaction terms can be added, but the ones written above
will be sufficient for our purposes.

We shall now compute, in the limit of small $\rho$, the
contributions to the effective action given by graphs with an arbitrary
number of loops generated by these new interaction terms.
We use the method of the auxiliary field, as explained for example in
[25]. We introduce two auxiliary fields $\phi_1$ and $\phi_2$ and
rewrite the action for $v^L$ and $w^L$ as
$$\eqalign{ \int d^4x\ \sqrt{|\det\bar g|}
\Bigl[{2\over3}v^L\,\tilde\o_{[v^Lv^L]}&\,
v^L+6\,w^L\,\tilde\o_{[w^Lw^L]}\,w^L\cr
&-{3\over g^2}(\phi_1-g^2\,\rho^2)^2
-{3\over g^2}(\phi_2-g^2\,\rho^2)^2\Bigr]\ ,\cr}\eqno(4.14)$$
where $\tilde\o_{[v^Lv^L]}$ and $\tilde\o_{[w^Lw^L]}$ are given by
$(4.7b)$ in which $z$ is replaced by $\phi_1$ and $\phi_2$, respectively.
Then the one-loop effective action, parametrically depending on $\phi_1$
and $\phi_2$, is formally given by
$$ \eqalign{\Gamma(\rho,\phi_1,\phi_2;r)=\ {1\over2}\ln&\det\tilde\o_{[v^Lv^L]}
+{1\over2}\ln\det\tilde\o_{[w^Lw^L]}\cr
&-{3\over g^2} \int d^4x\ \sqrt{|\det\bar g|}\left[(\phi_1-g^2\,\rho^2)^2
+(\phi_2-g^2\,\rho^2)^2\right]\ .\cr}\eqno(4.15)$$
As already observed, only zero-mode contributions need to be included in
the computation of the determinants, so that one finds
$$\Gamma(\rho,\phi_1,\phi_2;r)=\ {5\over2}\ln{\phi_1\over\mu^2}
-{3\, \Omega\over g^2}(\phi_1-g^2\,\rho^2)^2+{5\over2}\ln{\phi_2\over\mu^2}
-{3\, \Omega\over g^2}(\phi_2-g^2\,\rho^2)^2\ ,\eqno(4.16)$$
where $\phi_1$ and $\phi_2$ are now constant.
The dependence of $\Gamma$ on the auxiliary variables $\phi_1$ and $\phi_2$
can be eliminated by using their corresponding equations of motion,
$$\phi_i=g^2\,\rho^2+{5\, g^2\over 12\, \Omega}{1\over\phi_i}\ ,
\qquad\qquad i=\, 1,2\ .\eqno(4.17)$$
Inserting the solutions of (4.17) back in (4.16) one obtains the expression
of the effective action which includes the contributions of the zero-modes
to all loop-orders. The first terms in the expansion for small $\rho$
of the total effective potential can be easily determined
$$V=-{15\over16\pi^2\, r^4}\left[1+\ln\left({32\pi^2\mu^4\, r^4\over5\, g^2}
\right)\right]+\sqrt{5\over2}{3\, g\over\pi\, r^2}\ \rho^2
-(3\, g^2-4\lambda_1-\lambda_2)\ \rho^4+\ldots\ .\eqno(4.18)$$
where again we have set $m^2= 0$ and $\xi= 0$.
The resummation of a certain class of graphs (the so called ``daisy
graphs'') has thus eliminated from the effective potential the logarithmic
divergence of the one-loop contribution, producing a regular power-law
behaviour.
%Notice that, according to (4.18), $\rho= 0$ is a minimum of
%the effective potential; it is however a local minimum, since the
%coefficient of $\rho^4$ in (4.18) is negative for $\lambda_1\approx\lambda_2
%\approx g^4$ (the gauge coupling constant $g$ is naturally small).

%At this stage, one can evaluate the exact total effective action at its
%absolute minimum, approximatively given by (4.13). In this way, one gets
%an effective action $\Gamma(r)$ for the radius of the de Sitter space.
%This expression has quartic, quadratic and logarithmic divergences
%when $r\rightarrow\infty$. The quartic divergence corresponds to the
%value of the potential at its flat space minimum, multiplied by the
%volume of flat space. One can normalize the effective potential so
%that its minimum value in flat space is zero. This corresponds to
%discarding the quartically divergent terms. The next leading terms,
%for reasonable values of the coupling constants and every value of
%$\alpha$, tend to $-\infty$ for large values of $r$.
%Appealing to standard arguments from quantum cosmology [24] one can
%then conclude that the probability distribution for $r$,
%given by $e^{-\Gamma(r)}$, is exponentially peaked around flat space.

In conclusion we have found that for large radius of the background de Sitter
metric $\bar g$, the effective potential has its absolute minimum for
some nonzero value of $\rho$ and hence de Sitter space gives a
self-consistent solution of the theory. The constant $\ell$ appearing
in the self-consistency condition $\thetacl=\ell^{-1}\btheta$ is given
just by $\ell=\rho_{\rm (min)}^{-1}$.
For reasonable values of the coupling constants the second term on the
r.h.s. of (4.13) is negative, so $\ell^{-1}$ decreases for decreasing
values of $r$ (increasing curvature). We expect that, as in similar
models [17,18], $\rho$ will go to zero for some sufficiently
small value of $r$. This is usually taken as a signal of a phase
transition. In the present model it also signals a breakdown of
selfconsistency. We then come the remarkable conclusion that for given
values of the coupling constants, only a certain range of values of
the radius may be permitted.

\beginsection {5. GRAVITY-INDUCED GRAVITY}

The fact that the dynamical variables of our theory have the right
tensorial structure is not enough to qualify it as a theory of
gravity. This comes from the identification of $\langle g \rangle$,
via $\bar g$, with the classical, macroscopic metric.
Further indications come from studying the effective dynamics for $\bar g$.
Such a dynamics can be obtained by evaluating the effective action
$\Gamma(\thetacl,\acl;\bar g)$ at its minimum values for $\thetacl$
and $\acl$. It is of course impossible to compute exactly the one-loop
effective action for a generic $\bar g$, but one can study its behaviour
at large distances. Note that this is really all that is needed,
since $\bar g$ is supposed to represent the macroscopic metric.

In computing the Euclidean one-loop effective action (2.11)
for an arbitrary $\bar g$ we will use the method of the heat kernel.
Given an operator $\cal O$, acting on a space of tensors possibly
carrying also internal indices, we define its determinant through the formula:
$$ \ln\det{\cal O}=-\int_{1\over\Lambda^2}^\infty ds\, s^{-1}
\int\, d^4x \sqrt{|\det\bar g|}\ \tr K(x,x;s)\ ,\eqno(5.1)$$
where $\Lambda$ is an ultraviolet cutoff and $K$ is the heat kernel of
the operator ${\cal O}$, satisfying
${d\over ds}K+{\cal O}K=\,0$ and $\tr$ means trace over both tensorial and
internal indices. For small $s$, the trace of the heat kernel
has the well-known asymptotic expansion
$$\int\, d^4x \sqrt{|\det\bar g|}\ \tr K(x,x;s)\approx
B_0 s^{-2}+B_2 s^{-1}+ B_4 +O(s)\ ,\eqno(5.2)$$
where $B_n=\int\, d^4x \sqrt{|\det\bar g|}\ \tr\, b_n(x)$.
For an operator of the form
${\cal O}= -\bar\nabla_\lambda\bar\nabla^\lambda+Z$, one has [26]
$$\eqalignno{&b_0={1\over(4\pi)^2} {\bf 1} &(5.3a)\cr
             &b_2={1\over(4\pi)^2}\left({\bar R\over6}{\bf 1}-Z\right)
&(5.3b)\cr
             &b_4={1\over(4\pi)^2}\biggl[
\left({1\over180}\bar R_{\mu\nu\rho\sigma}
\bar R^{\mu\nu\rho\sigma}-{1\over180}\bar R_{\mu\nu}\bar R^{\mu\nu}
+{1\over72}\bar R^2+{1\over30}\bar\nabla_\mu\bar\nabla^\mu\bar R \right){\bf 1}
\cr &\qquad\qquad\qquad\qquad
-{1\over6}\bar R Z-{1\over6}\bar\nabla_\mu\bar\nabla^\mu Z+{1\over2}Z^2
+{1\over12}{\cal F}_{\mu\nu}{\cal F}^{\mu\nu}\biggr]\ ,&(5.3c)\cr}$$
where $\bf 1$ is the unity in the appropriate internal space and
${\cal F}$ acts both on spacetime and internal indices and is defined by
$[\bar\nabla_\mu,\bar\nabla_\nu]={\cal F}_{\mu\nu}$.
In order to be able to apply these formulae to the operators ${\cal O}$
and ${\cal O}_{[\rm gh]}$ given in (2.8) and (2.10), we choose henceforth the
Feynman-t'Hooft gauge for which $\alpha=1$. To extract the dependence
of $\Gamma^{(1)}$ on $\thetacl$, we split ${\cal O}=\tilde{\cal O}+Q$,
where $Q$ contains all the terms of $Z$ quadratic in $\thetacl$
and $\tilde{\cal O}= -\bar\nabla_\lambda\bar\nabla^\lambda+\tilde Z$.
Then,
$$\eqalign{\int\, d^4x& \sqrt{|\det\bar g|}\ \tr\left(b_0({\cal O}) s^{-3}
+b_2(\o) s^{-2}+b_4(\o) s^{-1}+\ldots\right)\cr
&=\int\, d^4x \sqrt{|\det\bar g|}\tr\left(b_0(\tilde\o)e^{-sQ}s^{-3}
+b_2(\tilde\o)e^{-sQ}s^{-2}+b_4(\tilde\o)e^{-sQ}s^{-1}+\ldots\right)\ .\cr}
\eqno(5.4)$$
Inserting in (5.1), the integration over $s$ can be performed explicitly.
The result is
$$\eqalign{\ln\det\o=&\int\, d^4x \sqrt{|\det\bar g|}\ \tr\Biggl[
-{\Lambda^4\over2}b_0(\o)-\Lambda^2 b_2(\o)-\left(\ln{\Lambda^2\over\mu^2}
-\gamma\right)b_4(\o)\cr
&+{1\over2}b_0(\tilde\o)\,Q^2\left(\ln{Q\over\mu^2}-{3\over2}\right)
-b_2(\tilde\o)\,Q\left(\ln{Q\over\mu^2}-1\right)
+b_4(\tilde\o)\,\ln{Q\over\mu^2}\Biggr]\ ,\cr}\eqno(5.5)$$
where $\gamma$ is Euler's constant.
The first line contains all the divergent parts, while the second contains
only finite terms. We will use a renormalization procedure which
amounts to taking the second line as the definition of the finite part
of $\ln\det \o$.

In the case of the operator $\o$ given in (2.8), in the gauge $\alpha=1$,
$Z=\tilde Z+Q$ where
$$\tilde Z=
\left[\matrix{\tilde Z_{[\omega\omega]}{}_{\mu ab}{}^{\nu cd}&
\tilde Z_{[\omega\varphi]}{}_{\mu ab}{}^{c\nu}\cr
\tilde Z_{[\varphi\omega]}{}_{a\mu}{}^{\nu cd}&
\tilde Z_{[\varphi\varphi]}{}_{a\mu}{}^{c\nu}\cr}\right]=
\left[\matrix{\delta^{[c}_{[a}\,\delta^{d]}_{b]}\bar R_\mu{}^\nu
-4g\,\fcl_\mu{}^\nu{}_{[a}{}^{[c}\delta_{b]}^{d]}
& -2g\,\delta_{[a}^c\bar\nabla_\mu\thetacl_{b]}{}^\nu \cr
-2g\,\delta_a^{[c}\bar\nabla^\nu\thetacl^{d]}{}_\mu
& (m^2+\xi\, \bar R)\, \delta_a^c\delta_\mu^\nu }\right]\eqno(5.6)$$
$$Q=\left[\matrix{Q_{[\omega\omega]}{}_{\mu ab}{}^{\nu cd}&
Q_{[\omega\varphi]}{}_{\mu ab}{}^{c\nu}\cr
Q_{[\varphi\omega]}{}_{a\mu}{}^{\nu cd}&
Q_{[\varphi\varphi]}{}_{a\mu}{}^{c\nu}\cr}\right]
=\left[\matrix{\delta_\mu^\nu\,\mcl_{ab}{}^{cd}
& 0 \cr 0 & \ncl_{a\mu}{}^{c\nu}\cr}\right]\eqno(5.7)$$
For the ghost operator, $\tilde Z_{[{\rm gh}]}=0$ and
$Q_{\rm [gh]}=\alpha\,\mcl$. It is now a straightforward task to insert
these formulae in (5.3), (5.5) and compute the traces.
The part of the one-loop effective action (2.11) that contains the
divergent contributions is given by
$$\eqalign{\Gamma^{(1)}_{\rm div}&={1\over(4\pi)^2}\int d^4x\
\sqrt{|\det\bar g|}\Biggl\{-7\,\Lambda^4+\Lambda^2\Bigl[8\, m^2+
8\bigl(\xi-{1\over6}\bigr)\,\bar R+\tr\mcl+{1\over2}\tr\ncl\Bigr]\cr
&-\left(\ln{\Lambda^2\over\mu^2}-\gamma\right)\biggr[{1\over2}\tr\mcl^2
+{1\over4}\tr\ncl^2+{\bar R\over3}\, \tr\mcl+{1\over2}\Bigl(m^2+\bigl(\xi
-{1\over6}\bigr)\bar R\Bigr)\tr\ncl\cr
&+4 m^2\Bigl(1+2 m^2\bigr(\xi-{1\over6}\bigl)\bar R\Bigl)
-{61\over180}\bar R_{\mu\nu\rho\sigma}\bar R^{\mu\nu\rho\sigma}
+{64\over45}\bar R_{\mu\nu}\bar R^{\mu\nu}-\Bigl({11\over24}-4\bigl(
\xi-{1\over6}\bigr)^2\Bigr)\, \bar R^2\cr
&+{5\, g^2\over3}\, \fcl_{\mu\nu ab}\, \fcl^{\mu\nu ab}+
3\, g^2\, \bar\nabla_\lambda\thetacl^a{}_\mu\bar\nabla^\lambda\thetacl_a{}^\mu
\biggr]\Biggr\}\ .\cr}\eqno(5.8)$$
It has terms that are either of the same form of the starting action
(2.3), or quadratic in the curvature of $\btheta$. The former can be
eliminated by a suitable renormalization of the coupling constants
of (2.3), while the latter are cancelled by adding suitable counterterms
(we did not write these terms in the action (2.3) because they are
independent of the dynamical variables $\theta$ and $A$).
The remaining finite part, written in terms of the renormalized
coupling constants, reads
$$\eqalign{\Gamma^{(1)}&={1\over2}{1\over(4\pi)^2}\!
\int d^4x\ \sqrt{|\det\bar g|}
\Biggl\{\tr\!\left(\!\mcl^2\!
\left(\ln{\mcl\over\mu^2}-{3\over2}\right)\!\right)\!
+{1\over2}\tr\!\left(\!\ncl^2\!
\left(\ln{\ncl\over\mu^2}-{3\over2}\right)\!\right)\cr
&+{2\over3}\,\bar R\,\tr\left(\mcl\!
\left(\ln{\mcl\over\mu^2}-1\right)\!\right)
+\Bigl(m^2+\bigl(\xi-{1\over6}\bigr)\bar R\Bigr)
\tr\left(\ncl\!\left(\ln{\ncl\over\mu^2}-1\right)\!\right)\cr
&+\left(-{13\over180}\bar R_{\mu\nu\rho\sigma}\bar R^{\mu\nu\rho\sigma}
+{22\over45}\bar R_{\mu\nu}\bar R^{\mu\nu}-{5\over36}\bar R^2
-{1\over10}\bar\nabla_\lambda\bar\nabla^{\lambda}\bar R\right)
\tr\left(\ln{\mcl\over\mu^2}\right)\cr
&+\left(2g^2\delta_a^c\bar\nabla_\mu\thetacl_b{}^\nu
\bar\nabla^\mu\thetacl^d{}_\nu-{22\over3}g^2\, \delta_a^{[c}\,
\fcl_{\mu\nu b}{}^{e]} \fcl^{\mu\nu}{}_e{}^d\right)
{\left(\ln{\mcl\over\mu^2}\right)}{}_{cd}{}^{ab}\cr
&+\biggr({1\over180}\Bigl(\bar R_{\mu\nu\rho\sigma}\bar R^{\mu\nu\rho\sigma}
-\bar R_{\mu\nu}\bar R^{\mu\nu}\Bigr)
+{1\over2}\bigl(\xi-{1\over6}\bigr)^2\bar R^2
-{1\over30}(5\xi-1)\bar\nabla_\lambda\bar\nabla^{\lambda}\bar R \cr
&+{m^4\over2}+\bigl(\xi-{1\over6}\bigr)\, m^2\bar R\biggr)\,
\tr\left(\ln{\ncl\over\mu^2}\right)
+\biggl( 2g^2\, \delta_a^{[b}\,\bar\nabla_\lambda\thetacl^{c]}{}_\mu
\bar\nabla^\lambda\thetacl_c{}^\nu\cr
&-\!{1\over12}\Bigl(\delta_a^b\, \bar R_{\lambda\rho\sigma\mu}
\bar R^{\lambda\rho\sigma\nu}
\!-\!2g\,\bar R_{\rho\sigma\mu}{}^\nu\, \fcl^{\rho\sigma}{}_a{}^b
\!-\!g^2\,\delta_\mu^\nu\, \fcl_{\rho\sigma a}{}^c\,
\fcl^{\rho\sigma}{}_c{}^b\Bigr)\!\biggr)
{\left(\ln{\ncl\over\mu^2}\right)}{}_{\!b\nu}^{\ a\mu}\Biggr\}\cr}
\eqno(5.9)$$
This formula reduces exactly to (4.11) in the case of de Sitter space
and for $\thetacl=\rho\btheta$. This shows two things. Firstly,
the expansion in inverse powers of the de Sitter radius used in (4.11)
coincides with the expansion in powers of curvature used in (5.9).
Moreover, the renormalization scheme we have adopted here to extract
the finite part of the effective action is
equivalent to the renormalization procedure which is implicit in the
zeta function approach.

To proceed further one needs to compute the vacuum expectation value of
the fields $\theta$ and $A$. These are the solutions of the equations
of motion obtained by varying the total effective action
$\Gamma=\bar S+\Gamma^{(1)}$ with respect to $\thetacl$ and $\acl$.
For simplicity, we shall keep in $\Gamma$ only the dominant parts in the
long wavelength expansion, those containing at most terms linear in the
curvature of $\btheta$ and without derivatives of the classical fields.
As in the previous Sections, we set $m^2=\ 0$, assume
$\lambda_1\approx\lambda_2\approx g^4$
and choose $\xi={1\over6}$ to simplify a bit the expressions.
The total effective action $\Gamma$ is then given by the potential terms
in (2.3) plus the first three terms in (5.9).

We notice that to this order $\Gamma$ is independent on $\acl$ so that
the vacuum expectation value $\langle A\rangle$ is left arbitrary.
This slightly unpleasant fact depends on our choice of action (2.3)
and is not a general consequence of the mean-field approach.
If we added to the action (2.3) a term linear in the curvature,
such as the second term in (6.1) below, this term would appear
in $\Gamma$ to the order we are discussing and by itself would
give vanishing torsion as an equation of motion.

The variation of $\Gamma$ with respect to $\thetacl$ gives the equation
for $\langle\theta\rangle$. Since we have to impose self-consistency
on the theory anyway, we shall look for solutions for which
$\langle\theta\rangle$ is proportional to $\btheta$, and check that
the minimum of $\Gamma$ occurs for a non-zero value of the proportionality
constant $\rho$. Indeed, one finds that in our approximation
the absolute minimum of $\Gamma$ occurs for
$$\rho^2={\mu^2\over g^2}e^{1-{16\pi^2\over7}\left({4\lambda_1+\lambda_2
\over g^4}\right)}-{\bar R\over7 g^2}\left[1+{8\pi^2\over3 g^2}
-{16\pi^2\over7}{4\lambda_1+\lambda_2\over g^4}\right]\ .
\eqno(5.10)$$
Note that since $\rho$ has to be a constant, self-consistency requires
$\bar R={\rm constant}$. In particular this condition is satisfied
by all solutions of Einstein's vacuum equations.

One can now obtain an effective action for $\bar g$
by evaluating $\Gamma$ at its minimum (5.10). The explicit
computation gives
$$ \Gamma(\bar g)=\int  d^4x\ \sqrt{|\det\bar g|}\left\{
{\mu^2\over3 g^2}e^{1-{16\pi^2\over7}\left({4\lambda_1+\lambda_2
\over g^4}\right)}\ \left[1-{6\over7}{4\lambda_1+\lambda_2
\over g^4}\right]\,\bar R+\ldots\right\}\ ,\eqno(5.11)$$
where $\Gamma$ has been normalized such that it vanishes in flat space.
The Einstein-Hilbert action is thus recovered as the action that
governs gravity at large distances. The mechanism by which this
happens is very similar to the one discussed in [27]. There are also
close ties with the ``induced gravity'' programme [7,8] and with
the ideas in [28], where the method of the effective
potential was applied to the gravitational field.
Newton's constant $G_N$, which
appears in the Einstein-Hilbert action in the form
$-{1\over 16\pi G_N}\int d^4x\bar R$, is seen to be of the order of
the renormalization point $\mu$, which in turn appears in the theory
as an arbitrary dimensionful constant. Newton's constant therefore
appears in this theory to arise through a sort of dimensional
transmutation.

\beginsection{6. CONCLUSIONS}

The original motivation for this work was the recognition that the
metric and/or vierbein play in the theory of gravity the role of order
parameter. In particular one can see a kind of Higgs phenomenon
occurring already in the standard formulation of General Relativity [6].
Since the Higgs phenomenon plays such an important role in the description
of Elementary Particle physics, it is tempting to try and construct a
theory of gravity following the same lines. The questions that the
Higgs model is designed to answer are: Why is the order
parameter nonzero? What is the origin of the mass of the gauge fields?
In the context of gravity, analougous questions are:
Why are the metric and/or vierbein nondegenerate? Why is the
connection metric and torsionfree? These are the questions that we
have tried to answer with our mean-field model of quantum gravity.

In the traditional approach to quantum gravity it is implicitly assumed
that the geometry of spacetime is determined by the quantum metric or,
in the path integral framework, by the fluctuating metric.
One could try to give some conceptual foundation to the mean-field
approach by postulating that lengths and angles should not be
measured with the quantum (fluctuating) metric but rather with its vacuum
expectation value. Having thus two metrics at our disposal,
we can write the action $\bar S$ given in (2.3) in which the
mean metric and the fluctuating metric play different roles.
In several respects this action lends itself better to treatment by
traditional field-theoretic methods than an action of the form (2.2).
All quantum calculations are to be performed by keeping
$\bar g$ fixed, so the technical aspects of our approach are identical
to those of quantum field theory in a fixed curved background metric,
a well-studied subject [29].
The difference with more traditional approaches to quantum gravity
lies in the appearance of the vacuum state in the action through $\bar g$,
so that the mean-field theory is not a quantum field theory in the
traditional sense. In practice, this is reflected in the necessity of
verifying, at the end of the day, the self-consistency conditions
$\langle\theta\rangle=\ell^{-1}\btheta$.

We have found that Minkowski space is a
self-consistent solution of the one-loop quantum dynamics of the theory.
Furthermore, insofar as quantum field theory in flat space preserves
global Lorentz invariance, it seems very likely that Minkowski space
will be a self-consistent solution at all orders of perturbation
theory. In the case of de Sitter space we have found that for given
values of the coupling constants self-consistency may be achieved at
least for a certain range of values of the de Sitter radius.
When one considers a more general mean field, it becomes much
more difficult to establish conditions of self-consistency. Our
approach has been to deal with this problem order by order in an
expansion in powers of momentum. It appears
that to lowest order in such an expansion any solution of Einstein's
equations in vacuum will give a self-consistent solution of the
theory. At the next order, terms quadratic in the curvature appear,
suggesting that some kind of Yang-Mills-type equation will become
relevant at short distances.

Any theory of quantum gravity has to reproduce General Relativity in
the classical limit. In our approach, General Relativity appears as
an effective low-energy theory, much as in certain ``induced
gravity'' schemes: the effective action depends on $\thetacl$ and
$\btheta$, so when it is evaluated at its minimum with respect to
$\thetacl$ one remains with an effective action for $\btheta$,
which contains the Einstein-Hilbert term.
In particular, Newton's constant appears as
the vacuum expectation value of the vierbein, thus providing a
concrete realization of an old idea [30].

The small fluctuations of the fields $\theta$ and $A$ have masses of
the order of the vacuum expectation value of the vierbein,
{\it i.e.} of Planck's mass.
This is in accordance with the empirical observation that there are
no massless spin-one bosons related to gravity, and is also related to
the absence of torsion and nonmetricity at low energy.
The graviton only appears in the low-energy effective theory
and its masslessness is a consequence of the invariances of the theory.

We have tried to present the essential ideas of the mean field
approach keeping technical complications at a minimum. There are
several directions in which our work can be improved and generalized,
some of which are rather straightforward while others require
substantial work.

A simple generalization would consist in replacing the gauge group
$O(1,3)$ by $O(1,N-1)$. In this case the internal (latin) indices
run from 0 to $N-1$, while the spacetime (greek) indices remain as before.
In this generalized theory, gravity is unified with an $O(N-4)$
Yang-Mills theory, the form $\theta^a{}_\mu$ playing the role of
order parameter [10,6]. At least in the case of Minkowski space,
the calculation of the effective potential proceeds as in Section 3
and the results are the same [11].

Another possibility is to add to the action $\bar S$
a ``cosmological'' term and an ``Einstein'' term, written in polynomial
form:
$${\lambda_0\over 24}\int d^4x\, \varepsilon_{abcd}
\theta^a{}_\mu\theta^b{}_\nu\theta^c{}_\rho\theta^d{}_\sigma
\varepsilon^{\mu\nu\rho\sigma}
+\kappa\int d^4x\ \varepsilon_{abcd}
\theta^a{}_\mu\theta^b{}_\nu F_{\rho\sigma}{}^{cd}
\varepsilon^{\mu\nu\rho\sigma}\ .\eqno(6.1)$$
Note that the ``cosmological'' term should be regarded as part of the quartic
potential, while the ``Einstein'' term describes cubic and quartic
interactions between $\theta$ and $A$.
We observe that, perhaps surprisingly, the addition of the
``cosmological'' term to $\bar S$ does not ruin the results of section 3:
Minkowski space is still a minimum of the effective
action, except for a $\lambda_0$-dependent modification of (3.5).

The action (2.3), while giving a simple form for the propagators,
suffers from a serious drawback: the massive modes $A_\mu{}^0{}_i$ and
$A_\mu{}^i{}_0$ have propagators with negative residues at the poles.
This is because the first term in the action
(2.3) is a Yang-Mills action for a noncompact gauge group.
It is possible to circumvent this problem by choosing a more
complicated Lagrangian, containing terms with all possible
contractions of the curvatures and torsion.
One could also argue that since the ghosts occur at the Planck mass,
where the theory probably ceases to be meaningful anyway, their
presence may not be fatal. This possibility has to be
investigated in greater detail.

We have chosen to study the vacuum expectation value of the gauge-variant
order parameter $\theta$. This is analogous to studying the vacuum
expectation value of the Higgs field $\Phi$ in the Standard Model.
There, a more rigorous procedure would be to compute the vacuum
expectation value of the gauge-invariant operator $\tr(\Phi^2)$.
Similarly, in our case, one could try to compute $\langle g\rangle$,
which is invariant under local Lorentz transformations, although
not under general coordinate transformations. Our result can be regarded
as a calculation of $\langle g\rangle$ in the approximation
in which $\langle\theta^2\rangle=\langle\theta\rangle^2$.
As we have already mentioned in Section 2, a direct determination of
$\langle g\rangle$ is technically more complicated,
since $g$ has to be treated as a composite operator,
but we expect it to yield essentially equivalent results.

Also, we are aware that from a rigorous point of view the
effective potential has to be convex and hence what we have said
can only be true in a metaphoric sense. We believe that a correct
treatment of this point can be given, for instance using the
concept of constrained effective potential [31]. The use of this idea seems
to be quite natural to our problem, where self-consistency demands
$\langle\theta\rangle$ to take a fixed value.

It may seem that since the fluctuating metric is no longer used as the
geometrical standard of lengths and angles, the theory has lost some of its
geometrical flavor. In fact, this is only partly so.
The geometric nature of gravity lies therein, that the geometry of
spacetime is dynamically determined. This is still true in our approach,
since the vacuum expectation value of the metric, which we take as the
standard of lengths and angles, is determined selfconsistently by the
quantum dynamics of the theory. What is gone is the idea of quantum
fluctuations of the geometry (and consequently also of the topology).
While this idea may be fascinating, it is also at the origin
of most difficulties of quantum gravity and a more conservative
approach like the one we are proposing may have better chances
of success. Whether this will be the case requires much more
work to establish.

\vskip 1.2cm
\centerline{\bf Acknowledgements}
\smallskip\noindent
We thank S. Randjbar-Daemi for explaining the spectra on the sphere to
us and E. Spallucci for much help and advice through all stages of this work.

\vfil\eject

\beginsection{APPENDIX}

In Section 3 we have introduced two decompositions in the space
of all tensors $\omega_{\mu ab}$, antisymmetric in the ``internal''
indices $a,b$.
The first was the Hodge decomposition of $\omega$, regarded as a
Lie-algebra valued one-form, into its exact (longitudinal) and coexact
(transverse) part (recall that the first cohomology group $H^1(S^4)=0$
and there are no harmonic one-forms on $S^4$).
The second was the decomposition (4.5a) into $O(4)$-invariant parts.
We will refer to the three terms on the r.h.s. of (4.5a) as the tensor,
vector and axial vector parts of $\omega$.
In this appendix we examine the relationship between these two
decompositions. We prove the following theorem:
\item{1)} {\sl the tensor part of $\omega$ is coexact};
\item{2)} {\sl the vector part of $\omega$ is coexact iff $v$ is exact
and is exact iff $v$ is coexact};
\item{3)} {\sl the axial vector part of $\omega$ is coexact iff $w$ is
exact and is exact iff $w$ is coexact}.

\noindent
In the proofs we will use $\btheta$ to transform all indices from
latin to greek and vice-versa when convenient. These operations can be
performed freely under the covariant derivatives because, due to the
second condition in (4.2), $\bar\nabla\btheta=0$.

We begin by proving 3). Assume that
$\omega_{\lambda\mu\nu}=\varepsilon_{\lambda\mu\nu\rho}w^\rho$.
Clearly if $w=df$ for some function $f$,
$\bar\nabla^\lambda\omega_{\lambda\mu\nu}=0$. Conversely if
$\bar\nabla^\lambda\omega_{\lambda\mu\nu}=0$, then
$\bar\nabla_\lambda w_\rho-\bar\nabla_\rho w_\lambda=0$, and therefore
$w=df$ for some function $f$. Thus the axial vector part of $\omega$
is coexact if and only if $w$ is exact.
Next suppose that $\omega$ is exact, {\it i.e.} that
$\omega_{\lambda ab}=\bar\nabla_\lambda \epsilon_{ab}$ for some
antisymmetric tensor $\epsilon_{ab}$. Then we have
$\bar\nabla_\lambda
w^\lambda=\bar\nabla^\mu{1\over\sqrt{\bar g}}
\varepsilon^{\mu\nu\rho\sigma}\bar\nabla_\nu\epsilon_{\rho\sigma}
={1\over\sqrt{\bar g}}\varepsilon^{\mu\nu\rho\sigma}
[\bar\nabla_\mu,\bar\nabla_\nu]\epsilon_{\rho\sigma}=0$,
{\it i.e.} $w$ is coexact.
Conversely, if $w$ is coexact, one sees immediately
that as a three-form $\omega$ is closed, and therefore there exists
a two-form $\epsilon_{\rho\sigma}$ such that $\omega=d\epsilon$.
Since the axial vector part of $\omega$ is totally antisymmetric,
this is equivalent to saying that
$\omega_{\lambda ab}=\bar\nabla_\lambda\epsilon_{ab}$.

Next to prove 2), assume that
$\omega_{\lambda\mu\nu}={1\over3}(\bar g_{\lambda\mu}v_\nu
-\bar g_{\lambda\nu}v_\mu)$.
Clearly if $v_\mu=\partial_\mu f$ for some function $f$,
$\bar\nabla^\mu\omega_{\mu\nu\rho}=0$. Conversely if
$\bar\nabla^\mu\omega_{\mu\nu\rho}=0$, then
$(\bar\nabla_\nu v_\rho-\bar\nabla_\rho v_\nu)=0$, i.e. $v$ is closed.
But every closed one-form on the sphere is exact, so $v_\mu=\partial_\mu f$.
Thus the vector part of $\omega$ is coexact if and only if $v$ is exact.
Next suppose that $\omega$ is exact, {\it i.e.} that
$\omega_{\lambda ab}=\bar\nabla_\lambda \epsilon_{ab}$ for some
antisymmetric tensor $\epsilon_{ab}$.
We have
$v_\lambda=\omega^\mu{}_{\mu\lambda}=\bar\nabla^\mu\epsilon_{\mu\lambda}$,
so $v$ is coexact.
Conversely if $v$ is coexact, {\it i.e.} there exists an antisymmetric
tensor $\epsilon$ such that $v_\mu=\bar\nabla^\nu\epsilon_{\nu\mu}$,
then $\omega_{\lambda\mu\nu}=
{1\over3}(\bar g_{\lambda\mu}\bar\nabla^\tau\epsilon_{\tau\nu}
-\bar g_{\lambda\nu}\bar\nabla^\tau\epsilon_{\tau\mu})$.
This is precisely the vector part of
$\bar\nabla_\lambda\epsilon_{\mu\nu}$, and since by assumption
$\omega$ was purely vectorial, it must be itself of this form.

Finally to prove 1), assume
$\omega_{\lambda\mu\nu}={2\over3}(t_{\lambda\mu\nu}-t_{\lambda\nu\mu})$.
We show that $\omega$ cannot be exact as a one-form. In fact, if
$\omega_{\mu ab}=\bar\nabla_\mu\epsilon_{ab}$, since $t$ is symmetric
in the first pair of indices, $0=\varepsilon^{\lambda\mu\nu\rho}
\omega_{\lambda\mu\nu}=\varepsilon^{\lambda\mu\nu\rho}
\bar\nabla_\lambda\epsilon_{\mu\nu}$. Regarding $\epsilon$ as a
two-form, this implies $d\epsilon=0$ and since the second cohomology group
$H^2(S^4)=0$, $\epsilon=du$ for some
one-form $u$. On the other hand, since $t$ is traceless, also the
tensor part of $\omega$ is traceless, so
$\bar\nabla^\mu\epsilon_{\mu\nu}=0$. This implies $\delta du=0$.
The one-form $u$ can also be decomposed into exact and coexact parts.
The exact part does not contribute to $\epsilon$, so we can assume
without loss of generality that $\delta u=0$. But then
$(d\delta+\delta d)u=0$. Since there are no harmonic one-forms on the
sphere, $u=0$. So $\omega$ must be coexact as a one-form.

\vfil\eject

\centerline{\bf References}
\bigskip
\noindent
\item{1.} S. Weinberg, in ``General Relativity: an Einstein centenary
Survey'', ed. S. Hawking and W. Israel, Cambridge University Press (1986).
\smallskip
\item{2.} C. Isham, A. Salam and J. Strathdee, Ann. of Physics {\bf 62}, 98
(1971);\hfil\break
K. Cahill, Phys. Rev. {\bf D 18}, 2930 (1978);\hfil\break
D. Popovi\'c, Phys. Rev. {\bf D 34},1764 (1986);\hfil\break
J. Dell, J.L. deLyra and L. Smolin, Phys. Rev. {\bf D 34}, 3012 (1986).
\smallskip
\item{3.} R. Utiyama, Phys. Rev. {\bf 101}, 1597 (1956); Progr. Theor.
Phys. {\bf 64}, 2207 (1980);\hfil\break
T.W.B. Kibble, J. Math. Phys. {\bf 2}, 212 (1961);\hfil\break
D.W. Sciama, in ``Recent developments in General Relativity'',
Infeld Festschrift, Pergamon, Oxford (1962).
\smallskip
\item{4.} R. Percacci, ``Geometry of Nonlinear Field Theories'',
World Scientific (1986);\hfil\break
R. Percacci, in ``XIII Conference on Differential
Geometric Methods in Theoretical Physics'', World Scientific (1984).
\smallskip
\item{5.}R. Floreanini and R. Percacci, Class. and Quantum Grav. {\bf 7},
975 (1990); {\it ibidem} {\bf 7}, 1805 (1990);
{\it ibidem} {\bf 8}, 273 (1991).
\smallskip
\item{6.} R. Percacci, Nucl. Phys. {\bf B 353}, 271 (1991).
\smallskip
\item{7.}
A. d'Adda, Phys. Lett. {\bf 119 B}, 334 (1982); {\it ibidem}
{\bf 152 B}, 63 (1985);\hfil\break
D. Amati and J. Russo, Phys. Lett. {\bf B 248}, 44 (1990).
\smallskip
\item{9.} A. Hanson and T. Regge, in Proceedings of the Integrative
Conference on Group Theory and Mathematical Physics, Austin (Texas),
1978.
\smallskip
\item{10.} R. Percacci, Phys. Lett. {\bf 144 B}, 37 (1984).
\smallskip
\item{11.} R. Floreanini, E. Spallucci and R. Percacci,
``Coleman-Weinberg effect in Quantum Gravity'', SISSA preprint.
\smallskip
\item{12.} S. Coleman and E. Weinberg, Phys. Rev. D {\bf 7}, 888 (1973).
\smallskip
\item{13.} R. Jackiw, Phys. Rev. {\bf D 9}, 1686 (1974);\hfil\break
G. Kunstatter, in ``Super field theories'', ed. H.C. Lee et
al., Plenum, New York (1987); ``Geometrical approach to the effective action''
in the Proceedings of the Banff Summer School on
Gravitation, Banff, August 1990.
\smallskip
\item{14.} G.A. Vilkovisky, Nucl. Phys. {\bf B 234}, 125 (1984);
in ``Quantum theory of Gravity'' ed. S.M. Christensen, Adam Hilger,
Bristol (1984);\hfil\break
B. deWitt, in ``Quantum field theory and quantum statistics; essays in
honor of the Sixtieth birthday of E.S. Fradkin'', ed. I.A. Batalin et
al., Adam Hilger, Bristol (1987);
in the Proceedings of the Johns Hopkins workshop on current problems
in Particle Physics, ed. G. Domokos and S. Kovesi-Domokos,
World Scientific, Singapore (1989).
\smallskip
\item{15.} G. 't Hooft, Phys. Rev. {\bf D14}, 3432 (1976);\hfil\break
A.A. Belavin and A.M. Polyakov, Nucl. Phys. {\bf B 123},
429 (1977);\hfil\break
F. Ore, Phys. Rev. {\bf D 16}, 2577 (1977);\hfil\break
S. Chadha, A. d'Adda, P. di Vecchia and F. Nicodemi,
Phys. Lett. {\bf 72 B}, 103 (1977).
\smallskip
\item{16.} G. Gibbons and M.J. Perry, Nucl. Phys. {\bf B 146}, 90
(1978);\hfil\break
S.M. Christensen and M.J. Duff, Nucl. Phys. {\bf B170}, 480 (1980).
\smallskip
\item{17.} G.M. Shore, Ann. of Phys. {\bf 128}, 376 (1980).
\smallskip
\item{18.} B. Allen, Nucl. Phys. {\bf B 226}, 228 (1983);
Ann. of Physics {\bf 161}, 152 (1985).
\item{19.} S.W. Hawking and I. Moss, Nucl. Phys. {\bf B 224}, 180
(1983).
\smallskip
\item{20.} E.S. Fradkin and A.A. Tseytlin, Nucl. Phys. {\bf B 234},
472 (1984).
\smallskip
\item{21.} K. Hayashi and T. Shirafuji, Prog. Theor. Phys. {\bf 64}, 866
(1980).
\smallskip
\item{22.} S. Randjbar-Daemi, Abdus Salam and J. Strathdee,
Nucl. Phys. {\bf B 242}, 447 (1984).
\smallskip
\item{23.} S. Hawking, Comm. Math. Phys. {\bf 55}, 133 (1977).
\smallskip
\item{24.} B.L. Hu and D.J. O'Connor,
Phys. Rev. Lett. {\bf 56}, 1613 (1986);
Phys. Rev. {\bf D 36}, 1701 (1987).
\smallskip
\item{25.} S. Coleman, R. Jackiw and H.D. Politzer, Phys. Rev.
{\bf D 10}, 2491 (1974);\hfil\break
R.W. Haymaker, ``Variational methods for composite operators'',
LSU-HE 108, to appear in Rivista del Nuovo Cimento.
\smallskip
\item{26.} B. de Witt, ``Dynamical theory of groups and fields'',
in ``Relativity, Groups and Topology'', ed. C. deWitt and B. deWitt,
Gordon and Breach, New York (1964);\hfil\break
R.T. Seeley, Proc. Symp. Pure Math. Am. Math. Soc. {\bf 10}, 288
(1967);\hfil\break
P.B. Gilkey, J. Diff. Geom. {\bf 10}, 601 (1975).
\smallskip
\item{27.} L. Smolin, Nucl. Phys. {\bf B 160}, 253 (1979);\hfil\break
A. Zee, Phys. Rev. Lett. {\bf 42}, 417 (1979);
Phys. Rev. Lett. {\bf 44}, 703 (1980);
Phys. Rev. {\bf D 23}, 858 (1981).
\smallskip
\item{28.} L. Smolin, Phys. Lett. {\bf 93 B}, 95 (1980).
\smallskip
\item{29.} N.D. Birrell and P.C.W. Davies, ``Quantum fields in curved
space'', Cambridge University press (1982).
\smallskip
\item{30.} Y. Fujii, Phys. Rev. {\bf D 9}, 874 (1974);\hfil\break
V. de Alfaro, S. Fubini and G. Furlan, Il Nuovo Cimento
{\bf A 50}, 523 (1979); {\it ibidem} {\bf B 57}, 227 (1980);
Phys. Lett. {\bf B 97}, 67 (1980);\hfil\break
R. Floreanini and \"O. Oguz, Il Nuovo Cimento {\bf 71 A}, 534 (1982);
\hfil\break
N. Matsuo, ``Einstein gravity as spontaneously broken Weyl Gravity'',
OU-HET 111 (1987).
\smallskip
\item{31.} L. O'Raifeartaigh, A. Wipf and H. Yoneyama, Nucl. Phys.
{\bf B 271}, 653 (1986).

\bye